\documentclass[review]{elsarticle}

\usepackage{lineno,hyperref}
\modulolinenumbers[5]

\usepackage{graphicx}
\usepackage{amsmath}
\usepackage{amssymb}

\usepackage{float}

\usepackage[noend]{algpseudocode}
\usepackage{algorithmicx,algorithm}
\journal{ }

\begin{document}

\begin{frontmatter}

\title{Misfit function for full waveform inversion based on Earth Mover's Distance with dynamic formulation}


\author[mymainaddress]{Peng Yong}
\ead{yongpeng2015@s.upc.edu.cn}
\author[mysecondaryaddress]{Wenyuan Liao}
\ead{wliao@ucalgary.ca}
\author[mymainaddress]{Jianping Huang}
\ead{jphuang@upc.edu.cn}
\author[mymainaddress]{Zhenchun Li}
\ead{leonli@upc.edu.cn}
\author[mysecondaryaddress]{Yaoting Lin}
\ead{yaoting.lin@ucalgary.ca}
\address[mymainaddress]{Department of Geophysics, China University of Petroleum (East China), Qingdao, China}
\address[mysecondaryaddress]{Department of Mathematics and Statistics, University of Calgary, Calgary, Canada}

\begin{abstract}
Conventional full waveform inversion (FWI) using least square distance (LSD) between the observed and predicted seismograms suffers from local minima. Recently, earth mover's distance (EMD) has been introduced to FWI to compute the misfit between two seismograms. Instead of comparisons bin by bin, EMD allows to compare signal intensities across different coordinates. This measure has great potential to account for time and space shifts of events within seismograms. However, there are two main challenges in application of EMD to FWI. The first one is that the compared signals need to satisfy nonnegativity and mass conservation assumptions. The second one is that the computation of EMD between two seismograms is a computationally expensive problem.  In this paper, a strategy is used to satisfy the two assumptions via decomposition and recombination of original seismic data. In addition, the computation of EMD  based on dynamic formulation is formulated as a convex optimization problem. A primal-dual hybrid gradient method with linesearch has been developed to solve this large-scale optimization problem on GPU device. The advantages of the new method are that it is easy to implement and has high computational efficiency. Compared to LSD based FWI, the computation time of the proposed method will approximately increase by $11\%$ in our case studies. A 1D time-shift signals case study has indicated that EMD is more effective in capturing time shift and makes the misfit function more convex. Two applications to synthetic data using transmissive and reflective recording geometries have demonstrated the effectiveness of EMD in mitigating cycle-skipping issues. We have also applied the proposed method to SEG 2014 benchmark data, which has further demonstrated that EMD can mitigate local minima and provide reliable velocity estimations without using low frequency information in the recorded data. 
\end{abstract}

\end{frontmatter}

\section{Introduction}
Full waveform inversion (FWI) is an indirect inversion method, which attempts to obtain high-resolution estimations of subsurface parameters by minimizing 
the misfit between the observed and calculated data \citep{lailly1983seismic,tarantola1984inversion,  tarantola2005inverse,virieux2016direct}. Different from tomography method which matches  the traveltimes only, FWI uses full wavefield data for inversion. Hence, FWI has great potential to extract quantitative information from seismograms \citep{ sirgue2010thematic,zhu2015seismic,zelt2016frequency,pan2018interparameter}. Due to the high computational cost of FWI, gradient based local optimization methods are usually applied to solve this large-scale PDE-constrained optimization problem \citep{epanomeritakis2008newton,metivier2013full,van2015penalty,virieux2017introduction}. In general, the misfit function is defined by  least square distance (LSD), which is computed as the $L_2$ norm of the difference between the observed and calculated seismograms. Studies have demonstrated that high wavenumber perturbations are responsible for amplitude of seismic data, while the low wavenumber variations of the velocity mainly  affect the  traveltime of the seismic events \citep{jannane1989wavelengths}.  From an inverse problem point of view, it is supposed to first invert the smooth background by matching these traveltime shifts and then inject high wavenumber perturbations gradually \citep{wu1987diffraction,mora1989inversion,alkhalifah2014scattering}. However, the LSD based on bin by bin
comparisons is not suitable to capture the time shifts between two oscillatory seismic signals. To converge towards the global minimum, FWI requires an
initial model accurate enough to make the predicted
data match the observed data within half a phase \citep{virieux2009overview,metivier2016measuring}. 
 
In practice, such an accurate initial model may not always be  available \citep{virieux2009overview,virieux2017introduction}. To mitigate the local minima problem, a series of inversion strategies
have been proposed. In the time domain, a multi-scale strategy is presented by Bunks to expand the radius of convergence by inversion starting with low-frequency contents and gradually increases to high-frequency contents, since low-frequency contents are less sensitive to cycle-skipping \cite{bunks1995multiscale}. However, in realistic seismic data, the low-frequency band is always contaminated by noise. Method has been studied to recover low-frequency contents from high-frequency contents in synthetic data \citep{li2016full}.   Different from waveform inversion in data domain, migration velocity analysis aims to expand the search space  by introducing subsurface offsets and time shifts in image domain \citep{symes2008migration,symes2009seismic}. The high computational  cost result from the construction of extended image volumes seems to have precluded their use in 3D configurations up to now. To make use of the time shifts in seismograms, 
misfits based on cross-correlation \citep{luo1991wave,van2010correlation} and later on warping techniques \citep{ma2013wave} have been proposed to automatically measure the time shifts between seismograms. One great challenge of these methods is to effectively and accurately obtain traveltime residuals especially when wavefields are complex.
Reflection FWI alternately updates the velocities of the smooth backgrounds and
unsmooth perturbations with wavefield decomposition to mitigate local minima \citep{xu2012inversion,wu2015simultaneous,zhou2015full}. However, Reflection FWI requires a good reflectivity model as a secondary source to construct
a backscattering wavefield, the effectiveness of reflection FWI is affected by the complexity of the velocity model  \citep{brossier2015velocity}. Following the idea of designing more convex objective functions, various types of dataset comparison and misfit design have been proposed to mitigate local minima, such as envelope inversion \citep{bozdaug2011misfit,wu2014seismic}, adaptive waveform inversion \citep{warner2014adaptive} and wavefield reconstruct inversion \citep{van2015penalty}.  For ill-posed waveform inverse problems, regularization techniques have been applied to effectively overcome local minima in large-contrast salt inversion \citep{esser2018total,peters2017constraints,qiu2016full,yong2018total}.

Recently, an optimal transport distance (OTD) has been introduced  to measure the misfit in FWI \citep{engquist2013application}.  The OTD also known as earth mover's distance (EMD) has received significant attention in many research areas such as image processing, computer vision and statistics, because of its capability to compare signal intensities across different signal/image coordinates \citep{kolouri2016transport}. The main motivation of its application to FWI is to take advantage of the capability to capture time shifts between signals. Despite their appealing theoretical properties, two underlying assumptions of the standard EMD are that the compared signals should be nonnegative, and that no energy is lost in the process of mapping one signal to the other \citep{engquist2013application}. For seismic data, these two assumptions are not satisfied. In this paper, a strategy proposed by Mainini to deal with signed signals is adopted to overcome these difficulties \citep{mainini2012description}. This strategy has been implicitly used in the previous studies \citep{metivier2016measuring,metivier2016optimal}. Note that EMD of seismic data is also a large-scale problem.  
The proposition from \cite{engquist2013application} is to use Monge's formulation of the OT problem, a nonlinear system of partial-differential equations, which can be solved using finite-difference based method \citep{benamou2014numerical}. 
However, it is expensive to obtain the misfit of EMD by numerically solve the Monge's equation. For 2D seismic seismogram, the total inversion using EMD with Monge's formulation takes 3 to 4 times the time of the FWI with LSD misfit \citep{yang2017analysis,yang2018application}.
The methodology proposed by \citep{metivier2016optimal} is based on a modified dual Kantorovich problem \citep{villani2008optimal} and it is solved with Simultaneous Descent Method of Multipliers (SDMM), in which a linear system corresponding to a second-order finite-differences discretization of Poisson's problem has to be solved at each iteration of the SDMM algorithm. Compared to the classical FWI, the computation cost of FWI with OTD will
increase by 20$\%$-70$\%$ \citep{metivier2016measuring}. 
In this paper, we will introduce a state-of-the-art method to efficiently compute EMD based on a dynamic formulation \citep{benamou2002monge}. Here, the computation of EMD is recast as an  $L_1$ type convex optimization problem \citep{li2017parallel} and efficiently solved by a primal-dual hybrid gradient method (PDHG) with linesearch \citep{malitsky2018first}, which is widely used in compressed sensing and image processing \citep{esser2010general,chambolle2011first,goldstein2013adaptive}. Compared to the methods which have been employed in existing application of OTD to FWI, the new method is very simple to code and easy to parallelize with the GPU device. Moreover, there is  no need to solve the complex nonlinear partial-differential equations \citep{engquist2013application} or the large-scale linear system of equations \citep{metivier2016measuring, metivier2016optimal}. The numerical case studies of Marmousi 2 model and Chevron 2014 data have shown that, compared to the classical FWI method, the computation time of the proposed method approximately increases by 11$\%$. 

The rest of the paper is organized as the follows. In Section 2, we first give a brief introduction of several forms of OT problem and introduce a  PDHG method with linesearch to efficiently compute EMD based on the dynamic formulation. The application of EMD to FWI problem is also presented in this section.  Section 3 gives four different case studies to emphasize the main properties of FWI based on EMD. Finally,  discussion and  conclusion are drawn in the two last
sections. The main contributions of this paper are the detailed derivation of the PDHG method to efficiently compute EMD for seismic inversion and the numerical experimental validation of the proposed method to mitigate local minima in FWI.

\section{Theory}

\subsection{Definition of the Earth Mover's Distance}
The optimal mass transport problem seeks the most efficient way to transform one distribution of mass to another, relative to a given cost function. Consider two nonnegative measures $\mu(x)$ and $\nu(y)$ defined on the spaces $X\subset{\mathbf{R}^d}$ and $Y\subset{\mathbf{R}^d}$. Monge's optimal transportation problem is to minimize the total transportation cost \citep{monge1781memoire,benamou2014numerical}
\begin{equation}
\begin{aligned}
& M(\mu,\nu)=\inf_{{T}}\int_X{||x-T(x)||^p\mu(x)dx}   \\
& \mbox{  s.t. } \nu(T(x))\mbox{det}(\nabla{T(x)})=\mu(x),
\end{aligned}
\end{equation}
where $p\ge{1}$ and $||\cdot||^p$ denotes $L_p$ norm on $\mathbf{R}^d$. $T(x)$ is a map from $X$ to $Y$ that rearranges the measure $\mu$ into the measure $\nu$. The first application of EMD to FWI is based on this non-linear PDE formulation \citep{engquist2013application}. Because of the high computational cost,  EMD of 2D seismogram is usually computed trace by trace using 1D algorithm \citep{yang2018application}. However, macro-scale variations of the background velocity shifts the seismic events not only along the time axis but also along the receiver (space) axis \citep{metivier2016measuring}.    

Kantorovich formulated the transportation problem by finding an optimal transport plan $\gamma\in{\mathbf{R}^{d\times{d}}}$ through minimizing the transportation cost \citep{kantorovich1960mathematical,cuturi2013sinkhorn,benamou2015iterative} 
\begin{equation}
\begin{aligned}
& K(\mu,\nu)=\min_{\gamma}\int_{X\times{Y}}{c(x,y)d\gamma(x,y)}  \\
& \mbox{  s.t.  } \int_{X}\gamma(x,y)dx=\nu(y) \mbox{ and } \int_{Y}\gamma(x,y)dy=\mu(x),
\end{aligned}
\end{equation}
where $c(x,y)$ is the cost of transporting one unit of mass from $x$ to $y$. In most studies, $c(x,y)$ is defined as $||x-y||^p$, with $p=1,2$. When $p=2$, the optimal plan $\gamma$ is unique and this map is a gradient of a convex function. Note that when $p=1$, the cost is not strictly convex, an optimal map exists but is not unique \citep{cuturi2013sinkhorn}. The infimum is also known
as Wasserstein distance. The scale of $\gamma\in{\mathbf{R}^{d\times{d}}}$ seems to obstruct the application to 2D or 3D seismic data. In the particular case where $X=Y$ and the $L_p$ Wasserstein distance between $\mu$ and $\nu$ can be calculated with a dual Kantorovich formulation  \citep{villani2008optimal,santambrogio2015optimal,metivier2016measuring}     
\begin{equation}
DK(\mu,\nu)=\max_{\phi\in\mathbf{R}^d}\int_{\mathbf{R}^d}\phi(x)d(\mu(x)-\nu(x)),  \phi(x)\in{Lip_1},
\end{equation}
where $Lip_1$ is the space of 1-Lipschitz functions. By once again considering the dual, we
readily obtain an equivalent dynamic formulation of Monge-Kantorovich problem \citep{benamou2002monge,santambrogio2015optimal,kolouri2016transport}, which will be employed in this paper to compute the optimal transport distance and can be written as \citep{benamou2002monge,li2017parallel}
\begin{equation}
\begin{aligned}
& OT(\mu,\nu)=\min{\int_{\Omega}}||\mathbf{u}(x)||^pdx \\
& \mbox{  s.t. } \nabla{\cdot}\mathbf{u}+\mu-\nu=0 \mbox{ in }\Omega, \frac{\partial\mathbf{u}}{\partial{n}} =0 \mbox{ on } \partial{\Omega}, 
\end{aligned}
\end{equation}
where $\Omega\subset{\mathbf{R}^d}$ is the  closure of a Lipschitz domain. The optimization variable $\mathbf{u}$ is a flux vector which satisfies the zero flux boundary condition. This problem shares a similar structure of total variation norm in image processing, and can be
solved by some efficient numerical methods \citep{goldstein2013adaptive,esser2010general,chambolle2011first}.  It is necessary to point out that four formulations of EMD above are equivalent in mathematics \citep{villani2008optimal,kolouri2016transport}. The difference is the   numerical solutions to these problems, which have different computation efficiencies.

FWI attempts to adjust the model parameters to minimize the distance between the observed and calculated common shot gathers ($P_{obs}(x_r,t)$ and $P_{cal}[m](x_r,t)$). Here, the variable $x_r$ denotes the receiver position and the variable $t$ is time. $P_{cal}[m](x_r,t)$  depends on the given model parameters $m$. While the optimal transport distance has many desirable properties, there remain challenges to compute optimal transport distance of the seismic signals \citep{engquist2013application}. The first one is that signals are supposed to be nonnegative, which is typically not the case with seismic signals. The second one is mass conservation that requires $\int_X{\mu(x)}dx=\int_Y{\nu(y)}dy$. Generally, the second assumption cannot be guaranteed in seismic imaging as well. 
 
To overcome these difficulties, we adopt Mainini strategy  \citep{mainini2012description} to deal with the signed seismic data, which has been  implicitly used in the previous studies \citep{metivier2016measuring,metivier2016optimal}.  Firstly, seismic signals are decomposed into positive and negative components.  Then, we recombine the data with the positive and negative parts to compare positive measures with mass conservation.  
\begin{equation}
OT(P_{cal},P_{obs})=OT(P_{cal}^{+}+P_{obs}^{-},P_{obs}^{+}+P_{cal}^{-})
\end{equation}
with $P_{cal}=P_{cal}^{+}-P_{cal}^{-}$ and $P_{obs}=P_{obs}^{+}-P_{obs}^{-}$. For seismic data, we usually have 
\begin{equation}
\int{P_{cal}^{+}-P_{cal}^{-}}=\int{P_{obs}^{+}-P_{obs}^{-}}.
\end{equation}
Hence, we arrive at
\begin{equation}
\int{P_{cal}^{+}+P_{obs}^{-}}=\int{P_{obs}^{+}+P_{cal}^{-}}.
\end{equation}
Thus, both nonnegativity and mass conservation are satisfied with Mainini strategy. Now the optimal transport distance between the observed and calculated data can be expressed as 
\begin{equation}
\label{eq:OT_seismic}
\begin{aligned}
& f_{OT}(P_{cal}[m](x_r,t),P_{obs}(x_r,t)) =\min_{\mathbf{u}}{\int_{\Omega}}||\mathbf{u}(x_r,t)||^pdx_{r}dt  \\
& \mbox{  s.t. } \nabla{\cdot}\mathbf{u}(x_r,t)+P_{cal}-P_{obs}=0 \mbox{ in }\Omega,
\frac{\partial\mathbf{u}}{\partial{n}} =0 \mbox{ on } \partial{\Omega}. 
\end{aligned}
\end{equation}

Note that, EMD used in this paper is equivalent to the modified dual Kantorovich distance used by \citep{metivier2016measuring,metivier2016optimal} in mathematics. It is also necessary to point out that, we do not need to decompose the signal into the positive and negative parts in practice from the formulation (\ref{eq:OT_seismic}) , which avoids discontinuity  when the EMD is computed. Please note that misfit function proposed here are indeed computing the  optimal transport between $P_{cal}^{+}+P_{obs}^{-}$ and $P_{obs}^{+}+P_{cal}^{-}$. We will discuss the limitation of this strategy in the following section. 
 
For comparison, the conventional $L_2$ misfit function is given as
\begin{equation}
f_{L_2}(P_{cal}[m](x_r,t),P_{obs}(x_r,t))=\frac{1}{2}{\int_{\Omega}}(P_{cal}-P_{obs})^2dx_{r}dt.
\end{equation}

\subsection{Fast numerical method for Earth Mover's Distance} 
For the sake of simplicity of programming, we assume that 2D common shot gathers  $P_{cal}(x_r,t)$ and $P_{obs}(x_r,t)$ are defined on $\Omega=[0,T]\times[0,X]$ with a regular Cartesian gird of size ${nr}\times {nt}$ and the spacing are  $\Delta{x_r}=\frac{X}{nr-1}$ and $\Delta{t}=\frac{T}{nt-1}$, respectively.  Let $(i,j)$ denote the indices of the discrete locations $((i-1)\times{\Delta{x_r}}, (j-1)\times{\Delta{t}})$ for $i= 1,2, \ldots, nr$, $j= 1,2, \ldots, nt$. Thus, we have the calculated data $P_{cal}\in{\mathbf{R}^{nr\times{nt}}}$, the observed data $P_{obs}\in{\mathbf{R}^{nr\times{nt}}}$ and the flux vector $\mathbf{u}=(u^x,u^t)\in{\mathbf{R}^{nr\times{nt}\times{2}}}$, where $u^x\in{\mathbf{R}^{nr\times{nt}}}$ and $u^t\in{\mathbf{R}^{nr\times{nt}}}$ represent the $x$- and $t$- direction components, respectively.

The divergence operator $\nabla{\cdot}\mathbf{u}\in{\mathbf{R}^{nr\times{nt}}}$ can be written as
\begin{equation}
(\nabla{\cdot}\mathbf{u})_{i,j}=(\partial_x{u^x})_{i,j} + (\partial_t{u^t})_{i,j}.
\end{equation}
To discretize the divergence operator in equation (\ref{eq:OT_seismic}), we use forward finite differences with Neumann boundary condition, which is given by

\begin{equation}
(\partial_x{u^x})_{i,j}  = \left\{
\begin{array}{rl}
\frac{1}{\Delta{x_r}}(u_{i+1,j}^x-u_{i,j}^x)  & \mbox{if } i< nr,   \\ 
 0 & \mbox{if } i = nr.
\end{array} \right.
\end{equation}

\begin{equation}
(\partial_t{u^t})_{i,j}  = \left\{
\begin{array}{rl}
\frac{1}{\Delta{t}}(u_{i,j+1}^t-u_{i,j}^t)  & \mbox{if } j< nt,   \\ 
 0 & \mbox{if } j = nt.
\end{array} \right.
\end{equation}
We apply backward finite differences with Neumann boundary condition to discretize gradient operator 
$
(\nabla{\varphi})_{i,j} =  \left(
\begin{array}{c}
(\partial_x{\varphi})_{i,j} \\
(\partial_t{\varphi})_{i,j}
\end{array}
\right),  
$
which is given by 
\begin{equation}
(\partial_x{\varphi})_{i,j}  = \left\{
\begin{array}{rl}
\frac{\varphi_{i,j}-\varphi_{i-1,j}}{\Delta{x_r}}  & \mbox{if } i > 1,   \\ 
 0 & \mbox{if } i = 1.
\end{array} \right.
\end{equation}

\begin{equation} 
(\partial_t{\varphi})_{i,j}  = \left\{
\begin{array}{rl}
\frac{\varphi_{i,j}-\varphi_{i,j-1}}{\Delta{t}}  & \mbox{if } j >1,   \\ 
 0 & \mbox{if } j = 1.
\end{array} \right.
\end{equation}
Here, $\varphi\in{\mathbf{R}^{nr\times{nt}}}$ is the dual variable, which will be used to efficiently compute optimal transport distance with primal-dual method. The objective in equation (\ref{eq:OT_seismic}) can be written as 
\begin{equation}
\Phi = ||\mathbf{u}||_{1,1} + \frac{\epsilon}{2}||\mathbf{u}||_{2}^2,
\end{equation}
where 
\begin{equation}
||\mathbf{u}||_{1,1}=\sum_{i=1}^{nr}\sum_{j=1}^{nt}|{u}_{i,j}^x| + |{u}_{i,j}^t|,
\end{equation}
\begin{equation}
||\mathbf{u}||_{2}^2=\sum_{i=1}^{nr}\sum_{j=1}^{nt}({u}_{i,j}^x)^2 + ({u}_{i,j}^t)^2.
\end{equation}
Here $\epsilon$ is a small regularization parameter which is used  to ensure strict convexity \citep{li2017parallel}, and $\epsilon$ used in this paper is set as $1e^{-4}$.  Note that the EMD between $P_{cal}$ and $P_{obs}$ equals $\Phi\times{\Delta{x_r}\Delta{t}}$. Thus,
the discretized problem becomes an $L_1$-type convex optimization with linear
constraints. The Lagrangian of the convex optimization problem is given by
\begin{equation}
\label{Convex Problem}
\mathcal{L}(\mathbf{u},\varphi)=||\mathbf{u}||_{1,1} + \frac{\epsilon}{2}||\mathbf{u}||_{2}^2 + <{\varphi}, \nabla{\cdot}\mathbf{u}+P_{cal}-P_{obs}>,
\end{equation}
where $<\cdot,\cdot>$ is the inner product between $nr\times{nt}$ matrices treated as vectors, i.e.,
\begin{equation}
<A,B>=\sum_{i=1}^{nr}\sum_{j=1}^{nt}A_{i,j}B_{i,j}.
\end{equation}
Now the convex problem can be reformulated as the following saddle-point problem \citep{chambolle2011first}
\begin{equation}
\label{saddle-point problem}
f_{OT}(P_{cal},P_{obs}) = \min_{\mathbf{u}}\max_{\varphi}\mathcal{L}(\mathbf{u},\varphi).
\end{equation}
The saddle point of (\ref{saddle-point problem}) can be found by the primal-dual hybrid gradient method
\begin{equation}
\label{primal problem}
{\mathbf{u}}^{k+1} = \arg \min_{\mathbf{u}}||\mathbf{u}||_{1,1} + \frac{\epsilon}{2}||\mathbf{u}||_{2}^2 + <{\varphi}^k,\nabla{\cdot}\mathbf{u}> + \frac{1}{2\alpha}||\mathbf{u}-\mathbf{u}^k||_2^2 ,
\end{equation}
\begin{equation}
\label{dual problem}
\varphi^{k+1} = \arg \max_{\varphi}<{\varphi}, \nabla{\cdot}(\bar{\mathbf{u}}^{k+1})+P_{cal}-P_{obs}> - \frac{1}{2\beta}||\varphi-\varphi^k||_2^2,
\end{equation}
where $\bar{\mathbf{u}}^{k+1} = 2{\mathbf{u}}^{k+1}-{\mathbf{u}}^{k}$, $\alpha$ and $\beta$ are primal and dual stepsizes \citep{parikh2014proximal}. In particular, one has $-\nabla{\cdot}=\nabla^{*}$ ($*$ is conjugate operator) which is defined by the identity 
\begin{equation}
\label{gradient-divergence}
<\mathbf{u},\nabla{\varphi}> = - <\nabla{\cdot}{\mathbf{u}},\varphi>.
\end{equation}
Here, $<\cdot,\cdot>$ is inner product over the space $\mathbf{R}^{nr\times{nt}\times{2}}$. Applying the identity (\ref{gradient-divergence}) to the primal problem (\ref{primal problem}), we get
\begin{equation}
{\mathbf{u}}^{k+1} = \arg \min_{\mathbf{u}}||\mathbf{u}||_{1,1} + \frac{\epsilon}{2}||\mathbf{u}||_{2}^2 - <\nabla{\varphi}^k,\mathbf{u}> + \frac{1}{2\alpha}||\mathbf{u}-\mathbf{u}^k||_2^2 .
\end{equation}  
Then we have the explicit formulas to update the primal variables $u_{i,j}^x$ and  $u_{i,j}^t$  
\begin{equation}
\label{primal-update}
(u_{i,j}^x)^{k+1} = \frac{1}{1+\epsilon\alpha}\mbox{shrink}((u_{i,j}^x)^{k}+\alpha{(\partial_x{\varphi})_{i,j}^k},\alpha), \nonumber \\
\end{equation}
\begin{equation}
\label{primal-update}
(u_{i,j}^t)^{k+1} = \frac{1}{1+\epsilon\alpha}\mbox{shrink}((u_{i,j}^t)^{k}+\alpha{(\partial_t{\varphi})_{i,j}^k},\alpha),
\end{equation}
where shrink operation is defined as 
\begin{equation}
\mbox{shrink}(x,y)= \left\{
\begin{array}{rl}
\left(1-\frac{y}{|x|}\right)x  & \mbox{if } |x| \ge{y},   \\ 
 0 & \mbox{if } |x| <{y}.
\end{array} \right.
\end{equation}

Likewise, the explicit iteration to update the dual variable $\varphi$ can be written as 
\begin{equation}
\label{dual-update}
\varphi_{i,j}^{k+1} = \varphi_{i,j}^{k} + \beta((\nabla{\cdot}\bar{\mathbf{u}}^{k+1})_{i,j}+(P_{cal})_{i,j}-(P_{obs})_{i,j}).
\end{equation}
Note that $\alpha$ and $\beta$ need to satisfy $\alpha{\beta}\le{\frac{1}{\lambda_{max}}}$ to ensure stability \citep{esser2010general}, where $\lambda_{max}$ is the largest eigenvalue
of the discrete Laplacian operator $\nabla{\cdot}\nabla$. Since the first-order finite difference method is used to discretize the gradient operator and the divergence operator, combining with Gershgorin Circle Theorem \citep{chambolle2011first}, we get
$\lambda_{max}\le{\frac{4}{\Delta{x_r}^2}+\frac{4}{\Delta{t}^2}}.$
In general, the primal and dual stepsizes can be simply  chosen as $\alpha=\beta=\sqrt{\frac{4}{\Delta{x_r}^2}+\frac{4}{\Delta{t}^2}}$.

It is well-known that the convergence speed of PDHG is highly sensitive to stepsize choice \citep{goldstein2013adaptive}, we adopted the PDHG with linesearch scheme \citep{malitsky2018first} given in Algorithm (\ref{PD_line}) to accelerate speed of computing the EMD, which can automatically adjust the stepsize parameters for fast convergence without user inputs. For readers who are interested in the rigorous proof on the convergence of PDHG, we recommend these research papers \citep{esser2010general,chambolle2011first,malitsky2018first}. After several numerical tests, some parameters in Algorithm (\ref{PD_line}) are empirically chosen as $\xi=0.5$, $\zeta=0.99$ and $\delta=1e^5$ for all numerical examples. 
The primal and dual variables can be updated by (\ref{primal-update}) and (\ref{dual-update}), which is easy to implement and parallelize with GPU device. It is empirically noted that the convergence rate of PDHG method may depend on the size of the problem. Since seismic data is continuous signals, for large-scale application, multi-grid strategy can be used to save computation time \citep{metivier2016measuring}. We first compute EMD on a coarse grid, then implement primal-dual method on the original fine gird. For more detailed comparison of the existing numerical solution to EMD, please refer to \citep{li2017parallel}. In the following sections, we will focus on the application of EMD to FWI for mitigating local minima issues.

\begin{algorithm}[t]
\caption{PDHG algorithm with linesearch for EMD}  
\label{PD_line}
\hspace*{0.02in} {\bf Input:}  
$P_{cal}\in{\mathbf{R}^{nr\times{nt}}}$, $P_{obs}\in{\mathbf{R}^{nr\times{nt}}}$, $\epsilon$, $\alpha_1$, $\xi\in(0,1)$, $\zeta\in(0,1)$, $\delta>0$ and $\theta_1=1$ \\
\hspace*{0.02in} {\bf Output:}  
$\mathbf{u}\in{\mathbf{R}^{nr\times{nt}\times{2}}}$, $\varphi\in{\mathbf{R}^{nr\times{nt}}}$
\begin{algorithmic}[1]
\While{$Not$ $Converged$} 
\State {${\mathbf{u}}^{k+1} = \arg \min_{\mathbf{u}}||\mathbf{u}||_{1,1} + \frac{\epsilon}{2}||\mathbf{u}||_{2}^2 + <{\varphi}^k,\nabla{\cdot}\mathbf{u}> + \frac{1}{2\alpha_k}||\mathbf{u}-\mathbf{u}^k||_2^2$}
\State{Choose any $\alpha_{k+1}\in{[\alpha_{k},\alpha_{k}\sqrt{1+\theta_{k}}]}$} and run linesearch
\Repeat
\State {$\theta_{k+1}=\frac{\alpha_{k+1}}{\alpha_{k}}$}
\State {$\beta_{k+1}=\delta\alpha_{k+1}$}
\State {$\bar{\mathbf{u}}^{k+1} = {\mathbf{u}}^{k+1} + \theta_{k+1}({\mathbf{u}}^{k+1}-{\mathbf{u}}^{k})$}
\State {$\varphi^{k+1} = \arg \max_{\varphi}<{\varphi}, \nabla{\cdot}(\bar{\mathbf{u}}^{k+1})+P_{cal}-P_{obs}> - \frac{1}{2\beta_{k+1}}||\varphi-\varphi^k||_2^2$}
\State {$\alpha_{k+1}=\xi\alpha_{k+1}$}
\Until{$\sqrt{\delta}\alpha_{k+1}||\nabla{\varphi^{k+1}}-\nabla{\varphi^{k}}||_{1,1}\le{\xi\zeta||\varphi^{k+1}-\varphi^{k}||_{1}}$} 
\State{$\alpha_{k+1}=\alpha_{k+1}/\xi$} 
\State{$k=k+1$} 
\EndWhile
\State \Return  {$\varphi$,$\mathbf{u}$}
\end{algorithmic}
\end{algorithm}

\subsection{Application of the Earth Mover's Distance to FWI}
In this part, we consider  an application of EMD to FWI in acoustic media. The 2D acoustic equation with constant density can be expressed as
\begin{equation}
\frac{\partial^2{p}(x,z,t)}{\partial{t^2}}-v^2(x,z){\nabla}^2{p(x,z,t)} = {q}(x_s,z_s,t),
\end{equation} 
where $p$ is the pressure and $q$ is the seismic source. The model parameter here is the velocity $v$. In this paper, seismic wavefields are computed by finite difference method. In addition, an unsplit convolutional perfectly matched layer method is applied to suppress boundary reflection from the artficial boundary \citep{komatitsch2007unsplit}.

Seismic waveform inversion can be characterized as a PDE-constrained optimization problem:
\begin{equation}
\label{PED constrained}
\begin{aligned}
& \min_{{v}} f(P_{cal}[v](x_r,t),P_{obs}(x_r,t)) \\
& \mbox{  s.t.  } \frac{\partial^2{p}}{\partial{t^2}}-v^2{\nabla}^2{p} = {q} \mbox{ and } Rp-P_{cal}=0,
\end{aligned}
\end{equation}
where $R$ represents receiver sampling operator and $f(P_{cal},P_{obs})$ is the misfit function. Lagrange multiplier method is employed to formulate this constrained problem into an unconstrained problem, for which the  Lagrangian function is given by \citep{metivier2013full,metivier2016measuring}  
\begin{equation}
\begin{aligned}
\mathcal{J}(v,p,P_{cal},\phi,\psi)  &= f(P_{cal},P_{obs})+<\frac{\partial^2{p}}{\partial{t^2}}-v^2{\nabla}^2{p} -{q},\phi>_{\mathcal{W}} \\
				  & + <Rp-P_{cal},\psi>_{\mathcal{D}},
\end{aligned}
\end{equation}
where the scalar product in the wavefield space and the data space is denoted by $<\cdot,\cdot>_{\mathcal{W}}$ and  $<\cdot,\cdot>_{\mathcal{D}}$, respectively. Since seismic waveform inversion is a large-scale problem, all-at-once method is not feasible for seismic waveform inverse problem  \citep{van2015penalty}. The adjoint-state method  \citep{plessix2006review,fichtner2011hessian} is usually applied to reduce memory storage and computational cost. Let the derivative of the Lagrangian function $\mathcal{J}(v,p,P_{cal},\phi,\psi)$ with respect to the state variable $p$ equal to zero
\begin{equation} 
 \frac{\partial\mathcal{J}(v,p,P_{cal},\phi,\psi)}{\partial{p}}=0,
\end{equation}
we have the adjoint-state equation
\begin{equation}
\frac{\partial^2{\phi}(x,z,t)}{\partial{t^2}}-v^2(x,z){\nabla}^2{\phi(x,z,t)} = -R^T{\psi}(x_r,t), 
\end{equation} 
where ${\phi}(x,z,t)$ is the adjoint variable and ${\psi}(x_r,t)$ is the adjoint source. The adjoint-state equation represents back-propagating the data residuals (adjoint source) \citep{tarantola1984inversion}.

Similarly, let  
\begin{equation} 
 \frac{\partial\mathcal{J}(v,p,P_{cal},\phi,\psi)}{\partial{P_{cal}}}=0,
\end{equation}
the adjoint source can be computed by 
\begin{equation}
{\psi}(x_r,t)=\frac{\partial{f(P_{cal},P_{obs})}}{\partial{P_{cal}}} .
\end{equation} 
Using the adjoint-state approach, we have the gradient of the objective function
\begin{equation}
\begin{aligned}
\nabla{f}(v(x,z)) &=\frac{\partial\mathcal{J}(v,p,P_{cal},\phi,\psi)}{\partial{v}}  \\
                 &=-{2}{v(x,z)}\int_{0}^{T}{{\nabla}^2{p}(x,z,t)\phi(x,z,t)}dt.
\end{aligned}
\end{equation}

The gradient can be obtained by correlating the source wavefield ${p}(x,z,t)$ and the adjoint wavefield $\phi(x,z,t)$. 
Note that the source wavefield forward propagates, while the adjoint wavefield is backward-propagated. Since the computer can not store the source wavefields at all times, we first compute the source wavefield and save the boundary values, then back-propagates the source wavefield with the saved boundaries \citep{yang2015graphics}, at the same time,  we compute the back-forward adjoint wavefield. Thus, it requires sloving the wave equation three times to obtain the gradient for each shot.  

Now, take second-order Taylor expansion of $f(v)$ at the point $v^n$ with perturbation $\Delta{v}$, we have 
\begin{equation}
f(v^n+\Delta{v}) \approx f(v^n) + {\Delta{v}^T\nabla{f(v^n)} + \frac{1}{2}{\Delta{v}}^T{H^n}{\Delta{v}}}, 
\end{equation}
where $\nabla{f(v^n)}$ is the gradient of $f(v)$ at the point $v^n$ and $H^n$ is the Hessian matrix. The conventional quasi-Newton method \citep{nocedal2006sequential} can be applied to minimize the misfit function $f(v)$. The iteration can be expressed as
\begin{equation}
v^{n+1} = v^{n} + \gamma^{n}\Delta{v}^n,
\end{equation}
where $\gamma^{n}$ is a positive scalar parameter computed through a parabolic search strategy \citep{liu2017effects}, in which we at least need to compute the wave equation twice to obtain an optimal stepsize using a test step-length $\gamma$, which is chosen according to the following condition:
\begin{equation}
 \max(|\gamma\Delta{v}^n|)\le{\frac{1}{100}\max(v^n)}.
\end{equation}
When the optimum step-length is less than $0.1\gamma$, we force the step-length as $0.1\gamma$. $\Delta{v}^n$ is a model increment satisfying
\begin{equation}
\Delta{v}^n = -(\bar{H}^n)^{-1}\nabla{f}(v^n).
\end{equation}
Here, $(\bar{H}^n)^{-1}$ is an approximation of the inverse of the Hessian $H^n$ which is computed through the $\ell$-BFGS method. This approximation is based on  several  latest gradients and model increments \citep{ nocedal2006sequential,epanomeritakis2008newton,pan2017accelerating}. In addition, conjugate gradient method is also widely employed in FWI study, in which the current update direction can be constructed using the current gradient and the last update direction \citep{ nocedal2006sequential,liu2017effects}.   

When the conventional $L_2$ misfit function is used to measure the distance between the calculated data and  observed data, the adjoint source is given by 
\begin{equation}
\frac{\partial{f_{L_2}(P_{cal},P_{obs})}}{\partial{P_{cal}}}=P_{cal}(x_r,t)-P_{obs}(x_r,t). 
\end{equation} 
For the optimal transport distance, combine equation (\ref{eq:OT_seismic}) and equation (\ref{Convex Problem}), we have 
\begin{equation}
\frac{\partial{f_{OT}(P_{cal},P_{obs})}}{\partial{P_{cal}}}=\frac{\partial\mathcal{L}(\mathbf{u},\varphi,P_{cal})}{\partial{P_{cal}}},
\end{equation} 
where
\begin{equation}
\label{adj_EMD}
\frac{\partial\mathcal{L}(\mathbf{u},\varphi,P_{cal})}{\partial{P_{cal}}} =  \varphi(x_r,t) + \frac{\partial\mathcal{L}}{\partial{\mathbf{u}}}\times{\frac{\partial\mathbf{u}}{\partial{P_{cal}}}} + \frac{\partial\mathcal{L}}{\partial{\varphi}}\times{\frac{\partial\varphi}{\partial{P_{cal}}}}. 
\end{equation}
Base on equation (\ref{saddle-point problem}) and the definition of the EMD, we know  that  $\frac{\partial\mathcal{L}}{\partial{\varphi}}=0$ and $\frac{\partial\mathcal{L}}{\partial{\mathbf{u}}}=0$ are necessary for the point to be the solution of the optimization problem. Although in preatical
computation, the numerical solution can not strictly guarantee $\frac{\partial\mathcal{L}}{\partial{\varphi}}=0$ and $\frac{\partial\mathcal{L}}{\partial{\mathbf{u}}}=0$, when the EMD is obtained, $\frac{\partial\mathcal{L}}{\partial{\varphi}}$ and $\frac{\partial\mathcal{L}}{\partial{\mathbf{u}}}$ are close to zero  in the primal-dual method \citep{chambolle2011first,li2017parallel}.  Hence,  the second and the third terms on the right-hand side of equation (\ref{adj_EMD}) can be neglected, and the adjoint source of FWI using EMD is exactly the dual variable 
\begin{equation}
\frac{\partial{f_{OT}(P_{cal},P_{obs})}}{\partial{P_{cal}}}=\varphi(x_r,t).
\end{equation} 
This means that when the optimal transport problem has been solved using the proposed method,  one can obtain not only the EMD between the predicted and observed data but also the adjoint source for FWI.

\section{Numerical Examples}
\subsection{1D case study: sensitivity to time shift}
\begin{figure}
\centering
\begin{tabular}{cc}
\includegraphics[scale=.3]{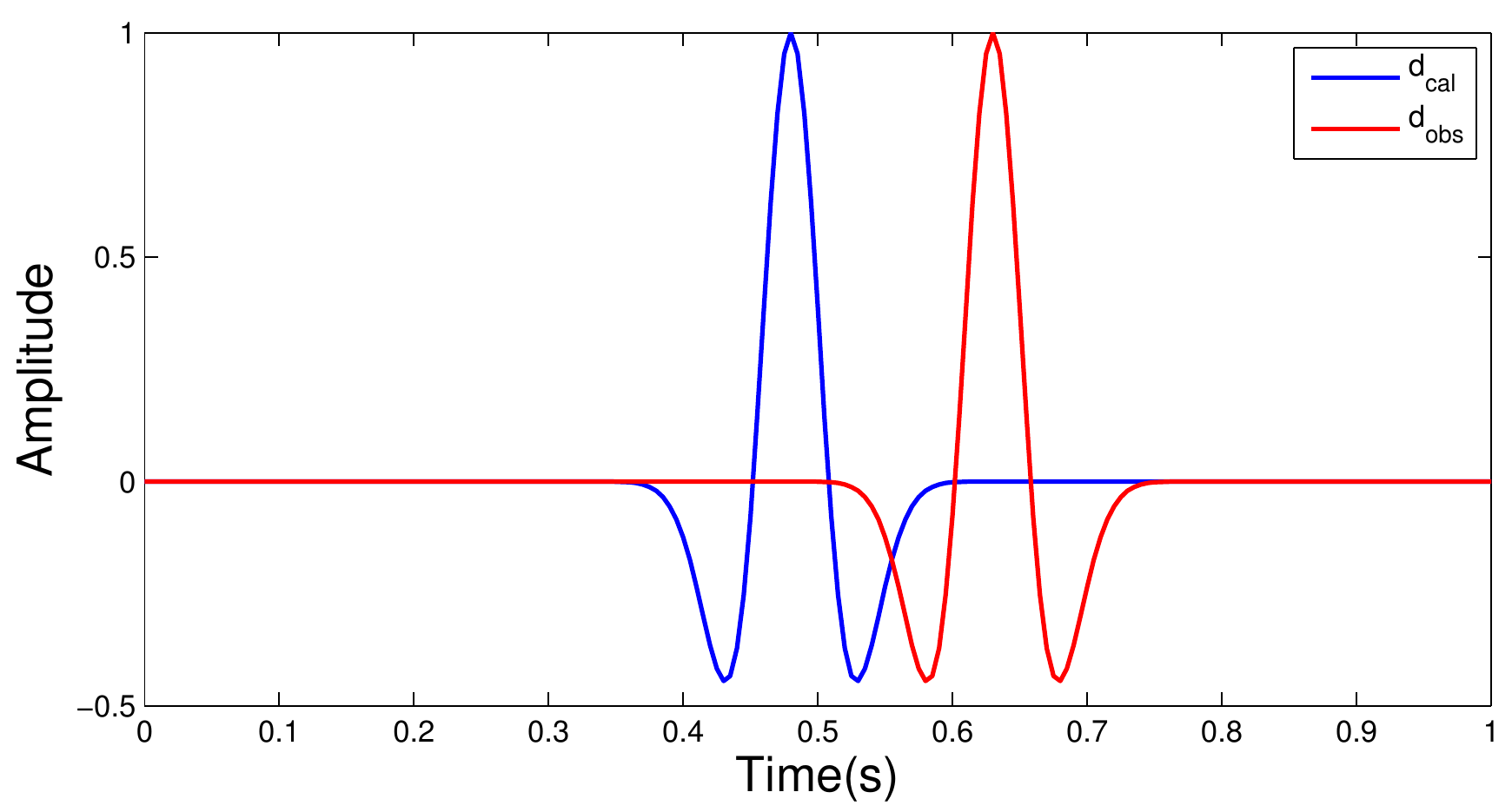}&
\includegraphics[scale=.3]{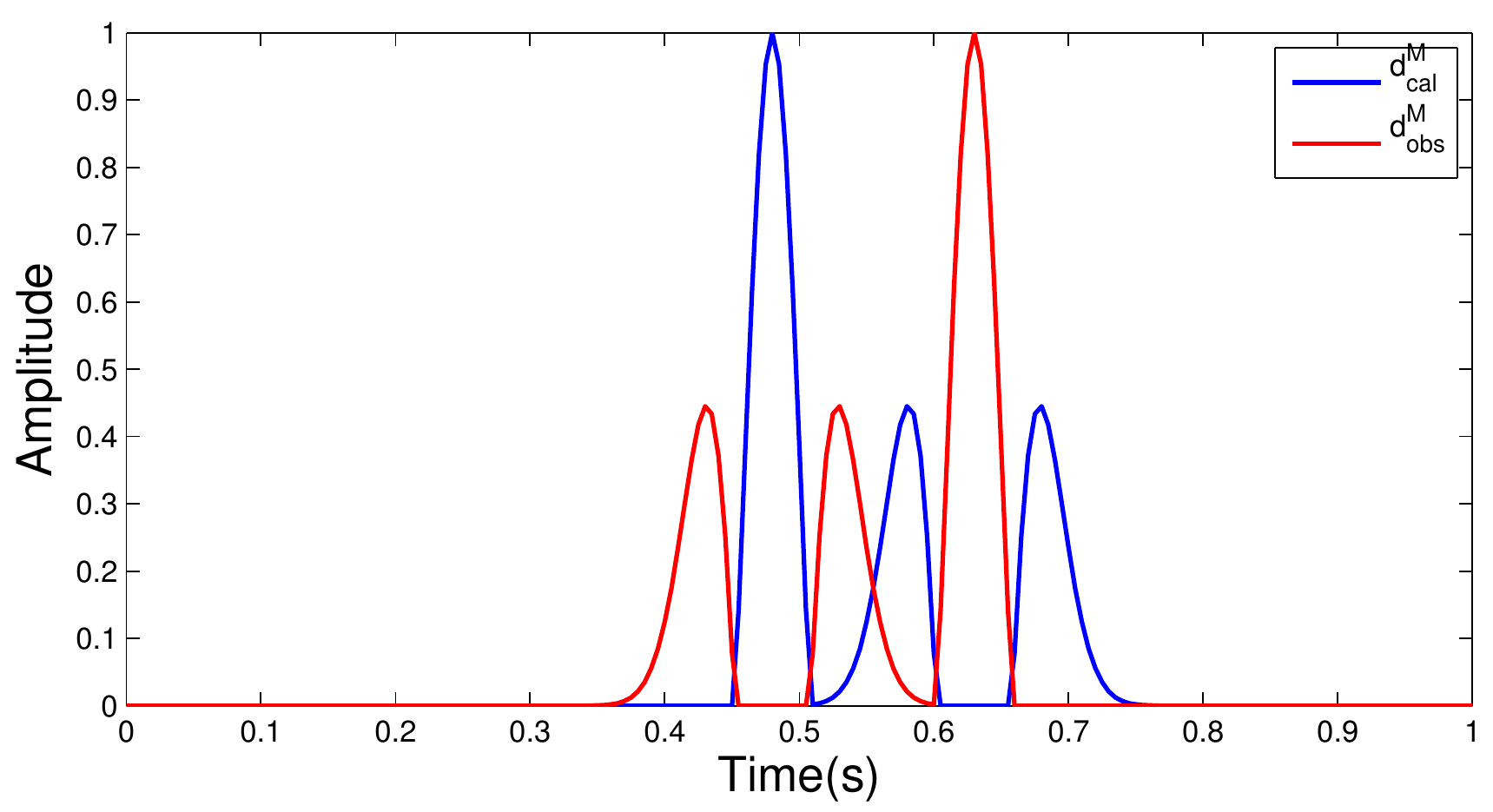}\\
{\small (a)}&{\small (b)}\\
\end{tabular} 
\caption{The original distribution, synthetic signal $d_{cal}$ and observed signal $d_{obs}$ (a).  The refactored distribution with the Mainini strategy, synthetic signal $d_{cal}^M$ and observed signal $d_{obs}^M$ (b). }
\label{fig:Initial_distribution and Refactored_distribution}
\end{figure}

\begin{figure} 
\centering
\begin{tabular}{cc}
\includegraphics[scale=.3]{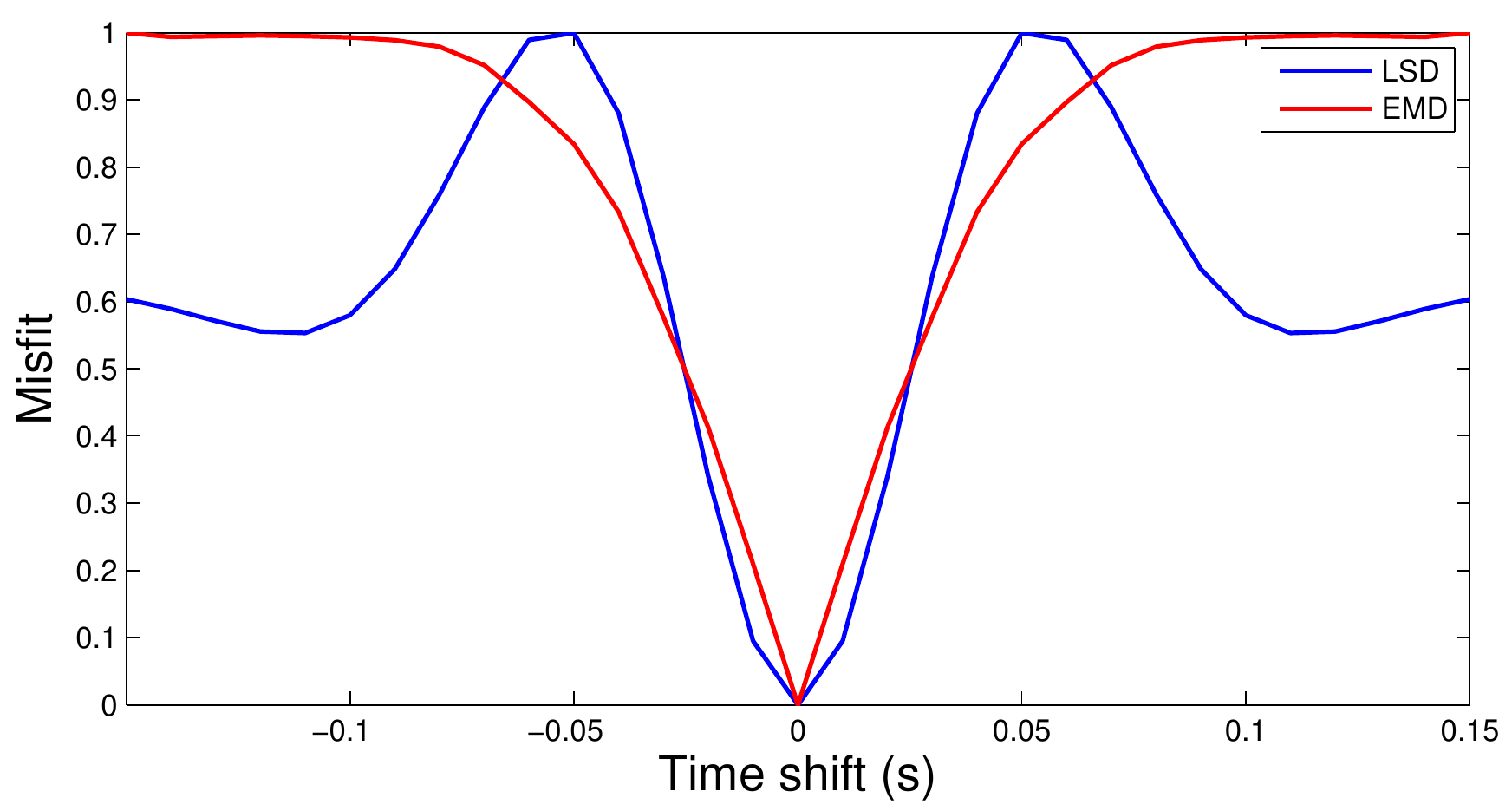}&
\includegraphics[scale=.3]{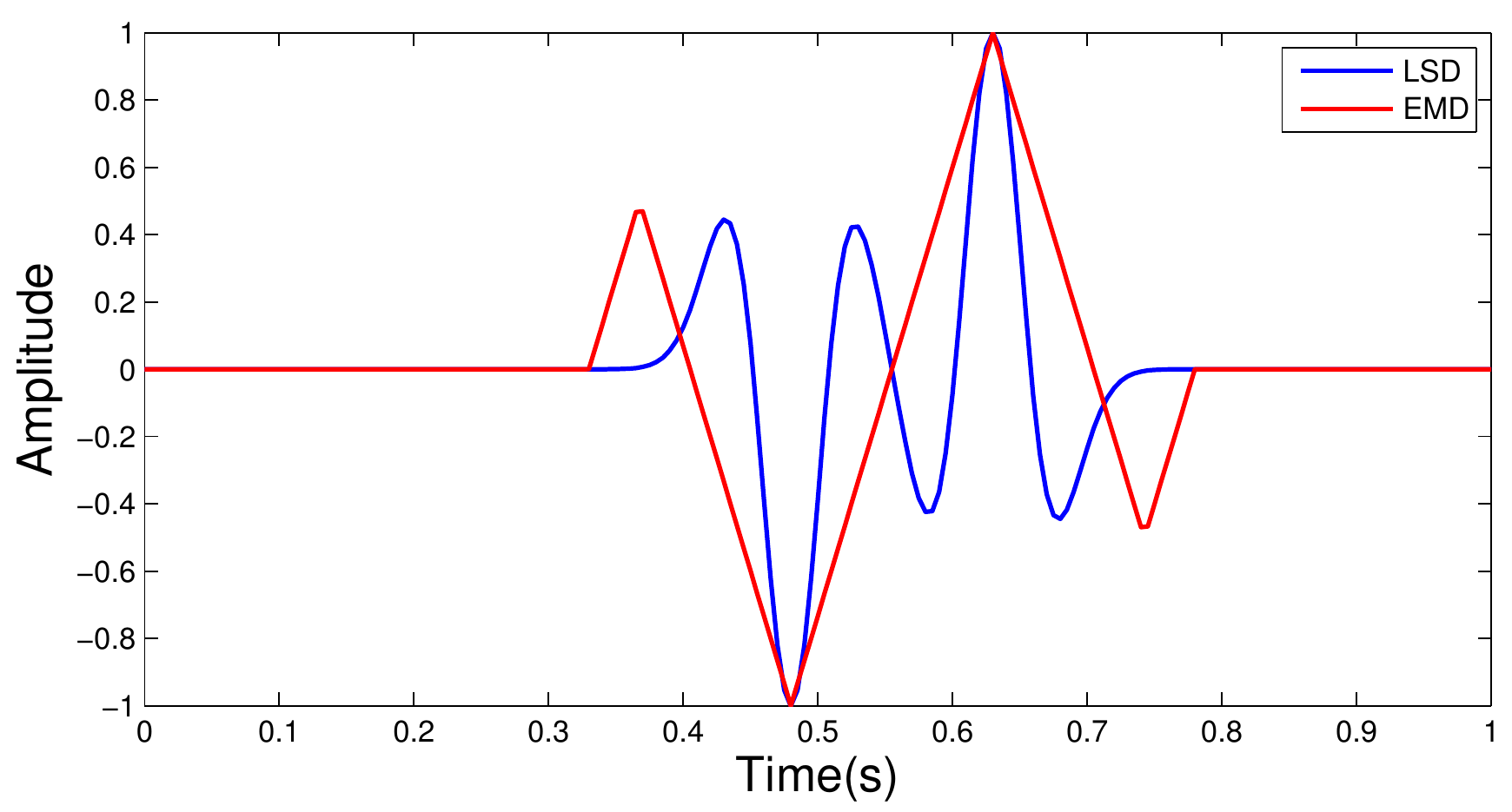}\\
{\small (a)}&{\small (b)}\\
\end{tabular} 
\caption{(a) Misfit function depending on the time shift of the Ricker signal, using the LSD
(blue) and the EMD(red). (b) Comparison between the LSD (blue) and EMD (red) adjoint sources for the two shifted
Ricker signals presented in Figure \ref{fig:Initial_distribution and Refactored_distribution}. }
\label{fig:misfit and Adjoint_source}
\end{figure}

We start the numerical example to investigate EMD's capability to detect shifted patterns in 1D case. The computation parameter $T$  in numerical solution to  Earth Mover’s Distance  is set as one here. In Figure \ref{fig:Initial_distribution and Refactored_distribution} (a),  a Ricker wavelet with the peak frequency of 8 Hz serves as the observed data and the calculated data corresponds to the same Ricker wavelet, shifted in time. With Mainini strategy, we have the refactored distribution shown in figure \ref{fig:Initial_distribution and Refactored_distribution} (b). The misfit function of different time shifts using LSD and EMD are presented in Figure \ref{fig:misfit and Adjoint_source} (a). It is observed that two local minima and a global minimum emerge in the misfit function with LSD, which is known as cycle skipping in seismic waveform inversion. The misfit function based on the EMD presents a single minimum, while it appears not to be a strictly convex function of the time-shift, it is still more capable of detecting time shift compared to LSD.  Considering the physical meaning of optimal transport that rearranges the measure $d_{cal}^M$ into the measure $d_{obs}^M$, it is well understandable that misfit function increases with the decrement of the overlap.

The adjoint sources corresponding to the original signals (Figure \ref{fig:Initial_distribution and Refactored_distribution}) are given in Figure \ref{fig:misfit and Adjoint_source} (b). The adjoint source with LSD is the difference between these two signals, while the EMD adjoint source appears as an envelope of the LSD adjoint source. This feature is similar to the study of \citep{metivier2016measuring,metivier2016optimal}, in which the dual Kantorovich formulation is used. Note that, compared with the study of \citep{metivier2016measuring,metivier2016optimal},  solving a Poisson's problem
is not required in our new method. In addition, EMD adjoint source presents an angular shape  $L_1$ norm. This non-smooth property may doubt the use of standard quasi-Newton solvers. However,  the numerical experiments presented in the next section demonstrated that this property does not preclude the use of these solvers to minimize the EMD misfit function, as previously reported in the study of \citep{metivier2016measuring,metivier2016optimal}.
\clearpage

\clearpage
\subsection{Application to 2D crosshole configuration}
\begin{figure}
\centering
\begin{tabular}{ccc}
\includegraphics[scale=.15]{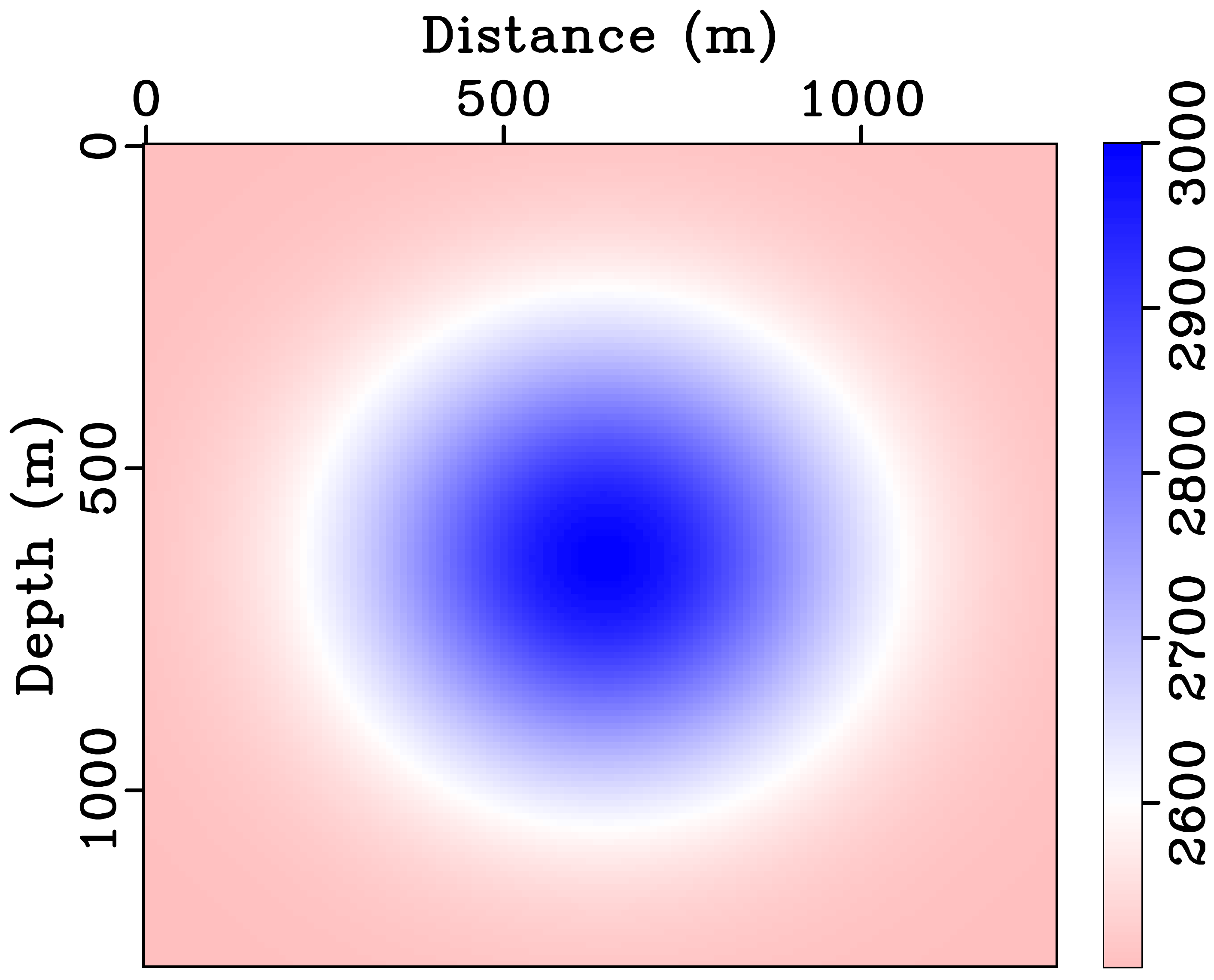}&
\includegraphics[scale=.15]{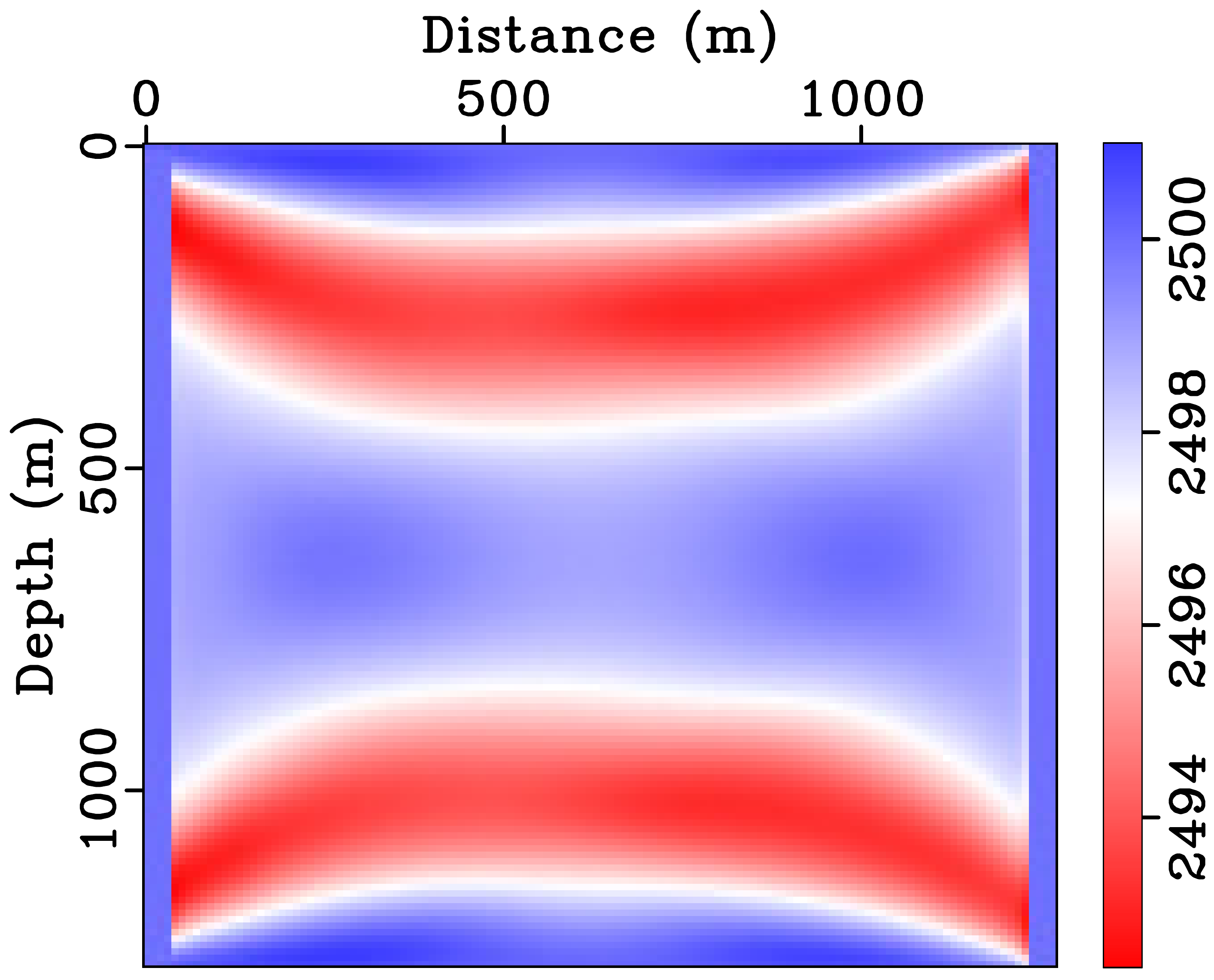}&
\includegraphics[scale=.15]{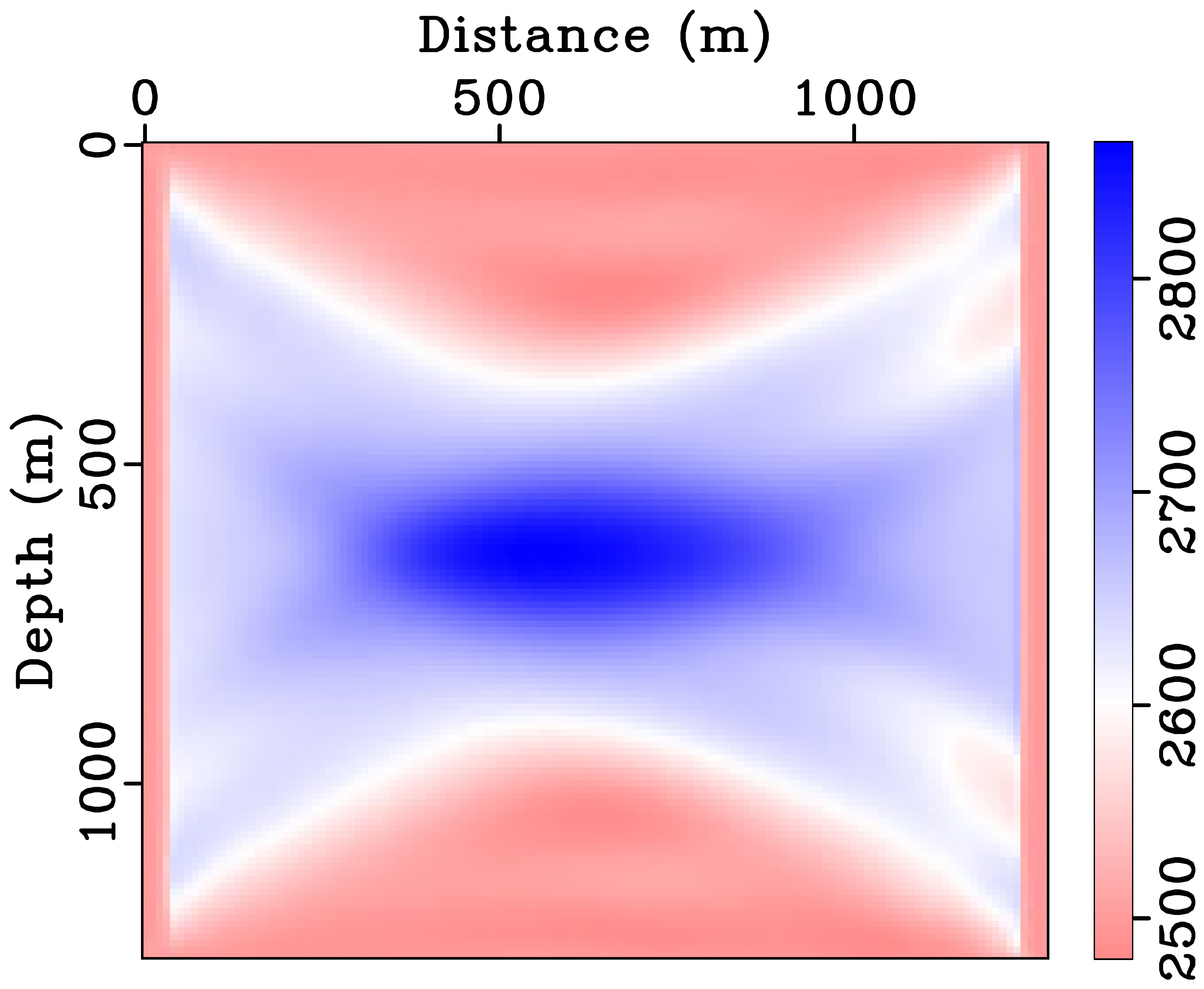}\\
{\small (a)}&{\small (b)} & {\small (c)}\\
\end{tabular} 
\caption{The true model used in the crosshole experiment (a). The inversion results after the 5th iteration by using LSD (b) and EMD (c), respectively. }
\label{fig:Camembert}
\end{figure}

In the 1D case, we have studied the capability of EMD to detect time shifted patterns. In this part, we perform full waveform inversion with two different distance measurements. The 2D crosshole configuration is used to investigate the effectiveness of EMD in 
mitigating local minima. In total 61 sources are equally spaced on the left side and 128 receivers on the right side with 10 m fixed acquisition. The true model is displayed in Figure \ref{fig:Camembert} (a). A Gauss-shaped inclusion is located in the centre of the rectangular velocity model. The background velocity is 2500 m/s and the maximum value is 3000 m/s. The velocity of the initial model is set as 2500 m/s. The synthetic data is generated using a Ricker source function centred on 10 Hz. The spatial discretization step is set to 10 m and the time discretization step is set to 0.001 s. The recording is performed over 1000 time steps for a total recording time 1.0 s. The maximal number of iterations in the primal-dual method with linesearch to obtain EMD is set to 100. Numerical experiments are carried out on the DELL workstation T7610 with Quadro M5000 8G video memory. The computational time of the gradient of all shots in the conventional LSD FWI formulation is 14.5 s, while the computational time used for EMD based FWI is 22.7 s. Moreover, $P_{cal}(x_r,t)$ and $P_{obs}(x_r,t)$ are defined on $\Omega=[0,1]\times[0,1]$ in this case study.

\begin{figure}
\centering
\begin{tabular}{cc}
\includegraphics[scale=.3]{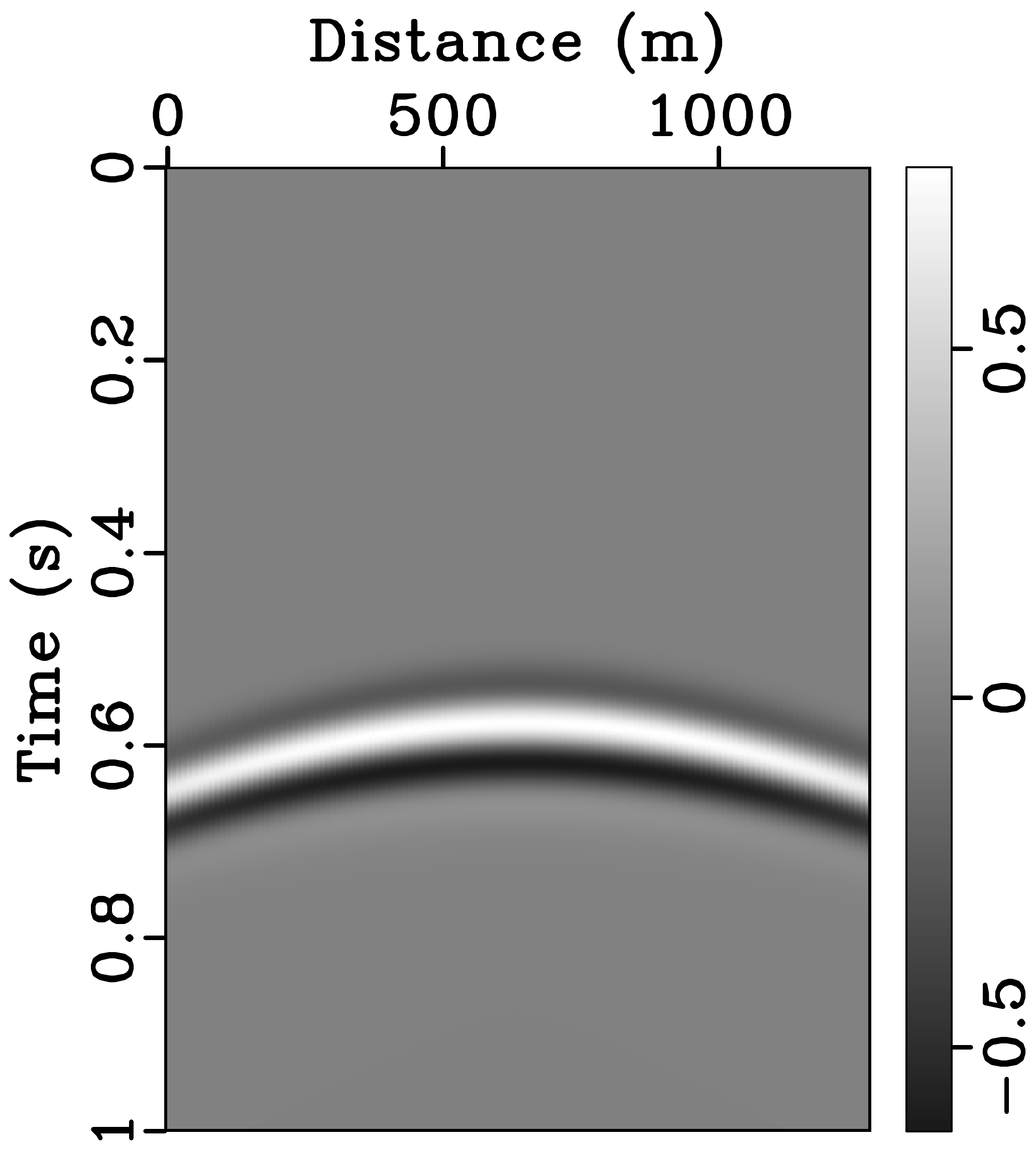}&
\includegraphics[scale=.3]{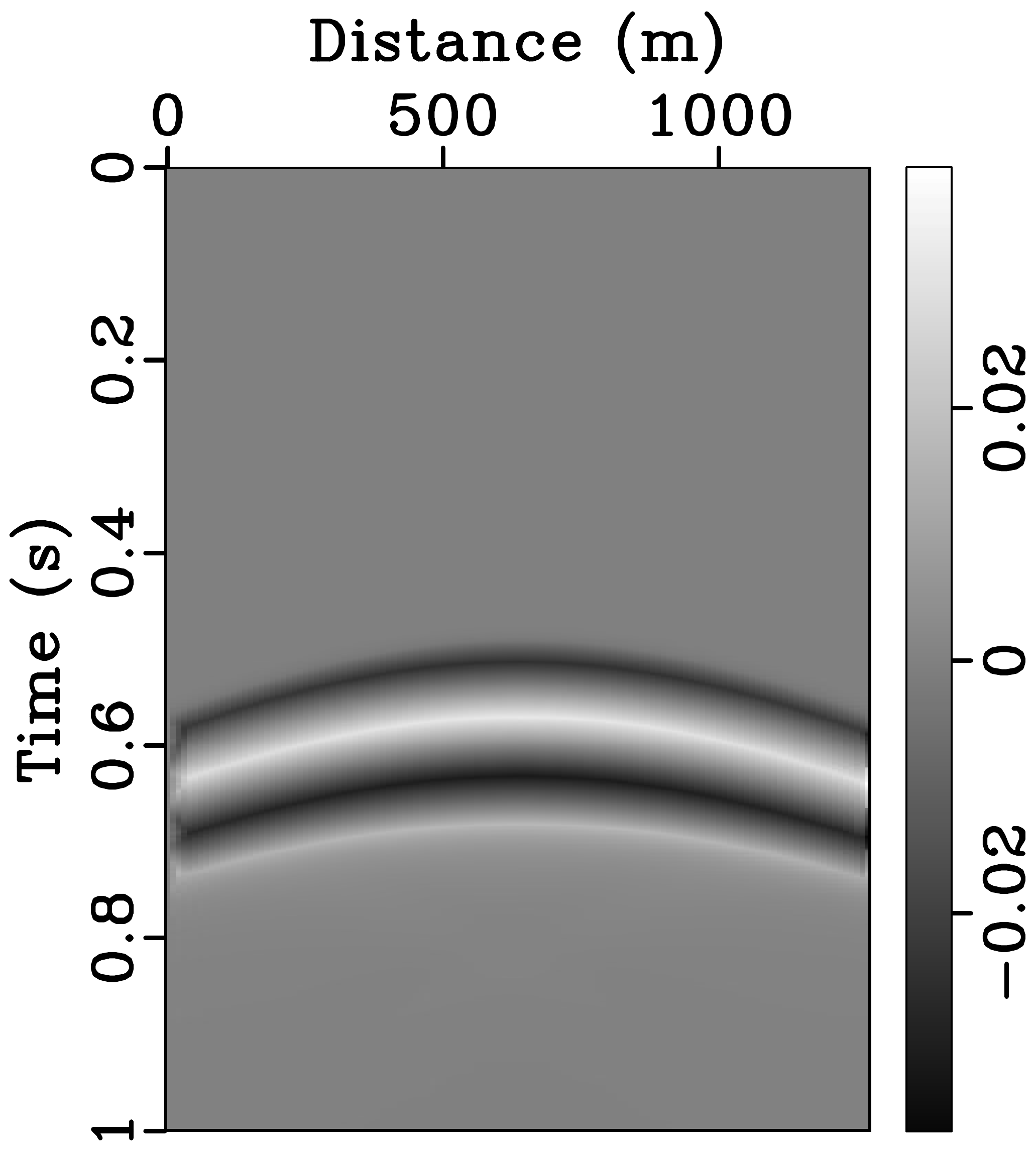}\\
{\small (a)}&{\small (b)}\\
\end{tabular}
\caption{The adjoint source of LSD (a) and the adjoint source of EMD (b) in the crosshole experiment.}
\label{fig:adjoint source}
\end{figure}

Figure \ref{fig:adjoint source} (a-b) show the adjoint sources of the first iteration with LSD and EMD, respectively. Figure \ref{fig:Camembert} (b) and Figure \ref{fig:Camembert} (c) display the inversion results after the 5th iteration with conventional LSD and EMD, respectively. The inversion results indicate that inversion using LSD suffers from cycle skipping and converges to local minima. The inversion result with EMD demonstrates that inversion converges in the correct direction. From this
experiment, we know that the EMD has the capability to reduce the risk of being trapped in a local minimum. 
\clearpage

\clearpage
\subsection{Application to 2D Marmousi 2 model}
In the previous subsection of numerical solution to EMD, we treat multidimensional seismic data as the general imaging data, and the 2D seismic data are defined on $\Omega=[0,T]\times[0,X]$. In this part, we will discuss the effect of weight distribution on time and receiver axes using synthetic offshore data in 2D reflection configuration.  This numerical test is implemented on the Dell
workstation with NVIDIA Quadro P5000 16GB GPU device. 

\begin{figure}
\centering
\begin{tabular}{cc}
\includegraphics[scale=.12]{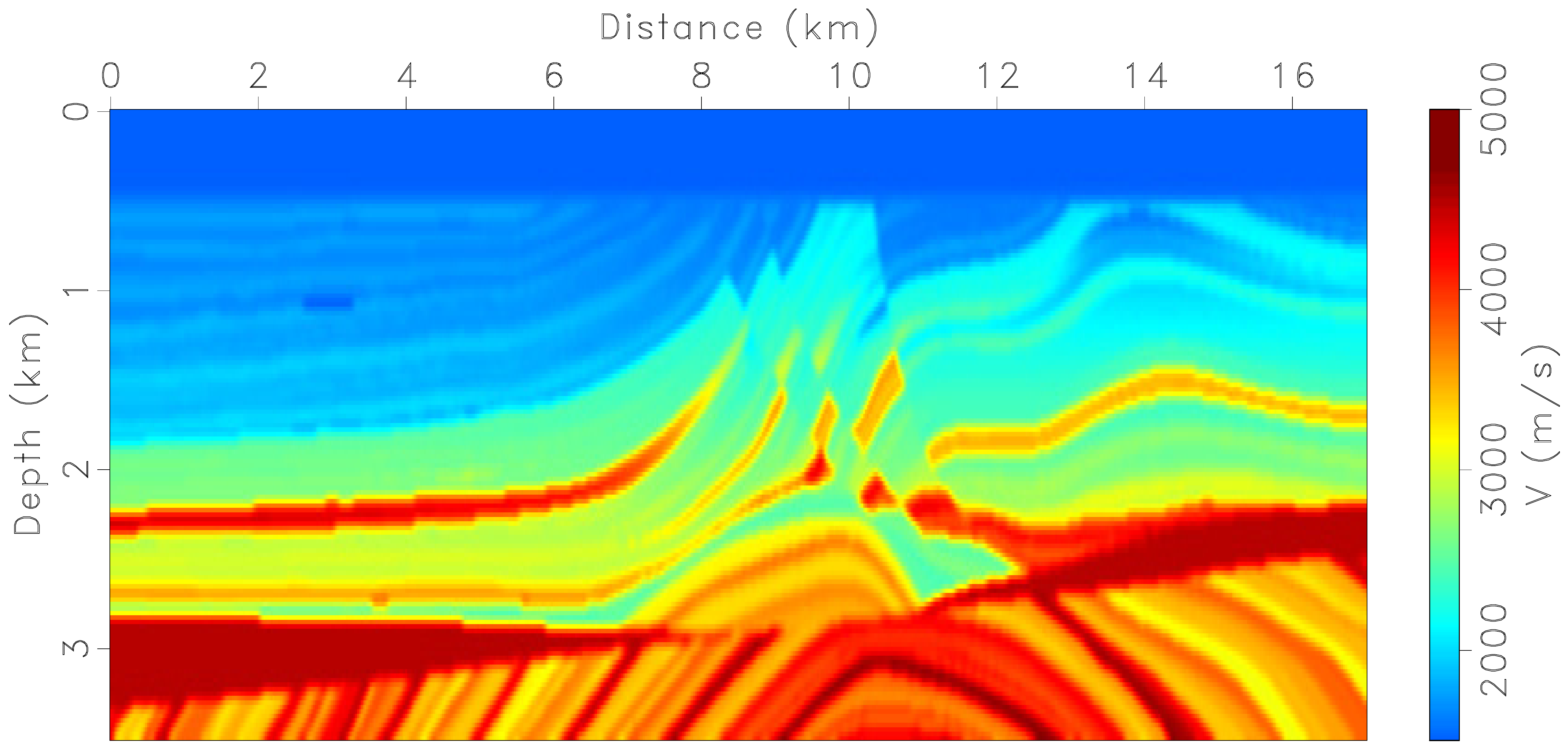}&
\includegraphics[scale=.12]{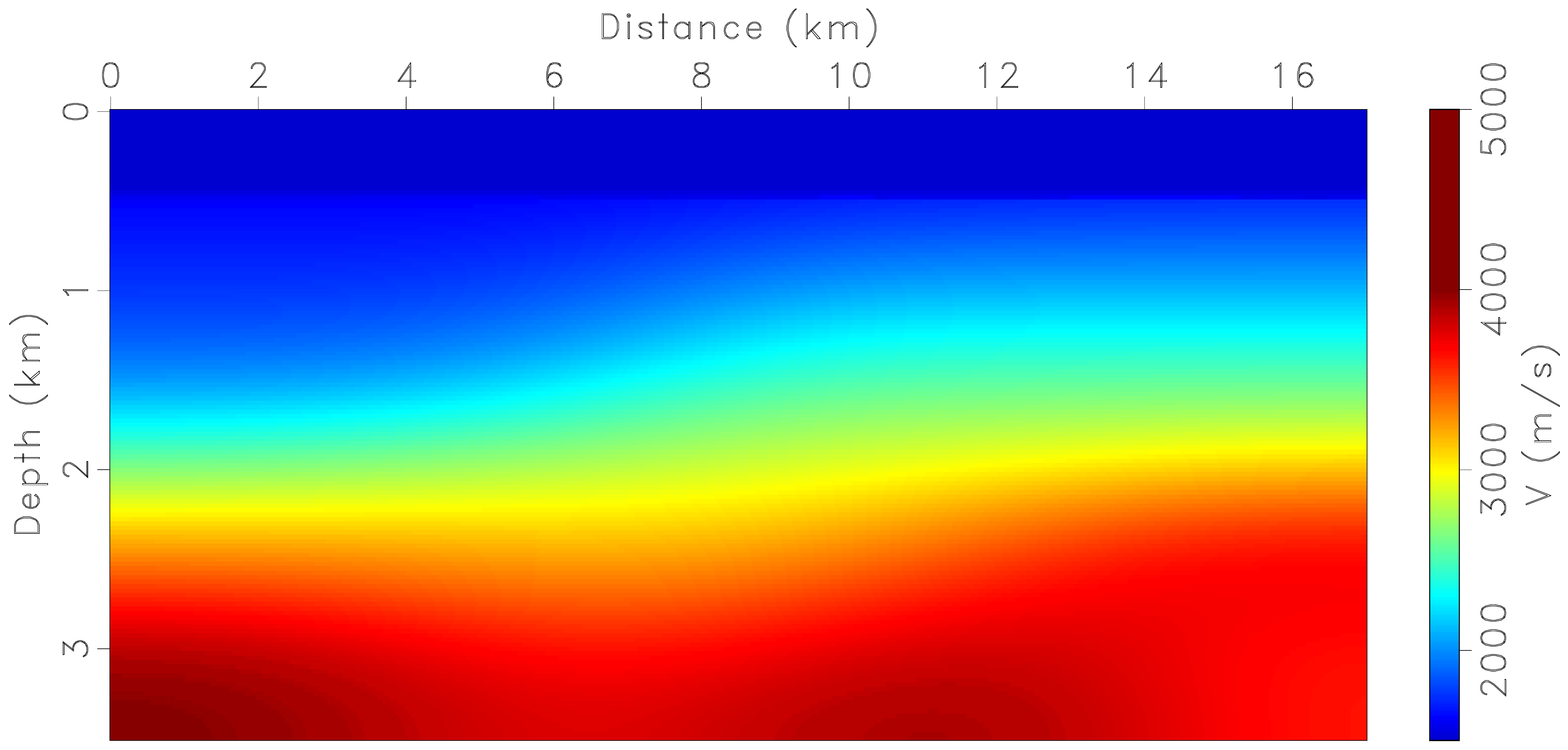}\\
{\small (a)}&{\small (b)}\\
\end{tabular}
\caption{Marmousi 2 P-wave velocity model used in the reflection experiment. Exact velocity model (a), initial model for FWI (b). }
\label{fig:mar_model}
\end{figure}

In the numerical tests, the misfit functions are minimized using the $\ell$-BFGS method with memory parameter $\ell$  as 5.  Figure \ref{fig:mar_model} (a) shows the P-wave velocity of the Marmousi 2 model, which is defined on a grid with the size of $681\times{141}$. The spatial discretization step is set to 25 m. A fixed-spread surface acquisition with 76 equally spaced sources  and 227 equally spaced receivers placed  at depth of 50 m is considered. The synthetic data is generated using a Ricker source function with peak frequency of 7 Hz. The time discretization step is set to 2 ms.   The recording is performed over 3000 time steps for a total  recording time 6.0 s. To improve the overall efficiency, a multi-grid strategy is used to solve the $L_1$ type optimization problem for the purpose of  EMD calculation. For each shot, we first resample the seismogram to the size of $301\times{227}$ to obtain EMD on the coarse grid. The iteration times of computing EMD is 150.  Then, sinc interpolation method is used to construct the primal variable $\mathbf{u}$ on the original grid. The computation time of forward modeling with 2-6 finite difference method \citep{yong2018forward} for each shot is 0.18 s 
and the computation time of calculating EMD for each shot is 0.09 s. The computation time for each iteration in FWI with EMD is 89.30 s, in which we usually need to solve the wave equation five times and compute the EMD three times. Compared with the computational time of 79.64 s in the classical FWI formulation,  the proposed method has an approximately 11.2$\%$ increase of the computational time.

\begin{figure}
\centering
\begin{tabular}{cc}
\includegraphics[scale=.3]{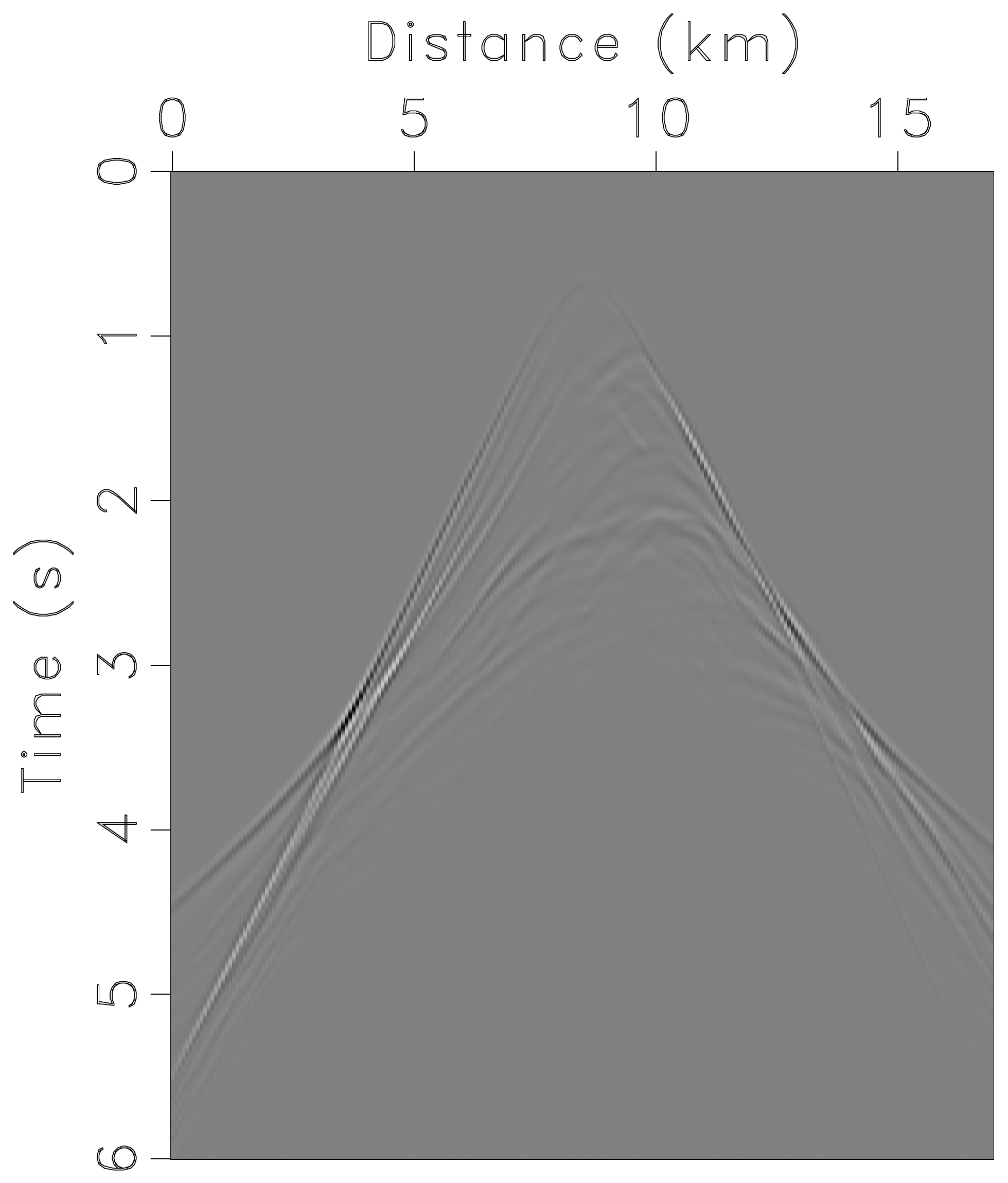}&
\includegraphics[scale=.3]{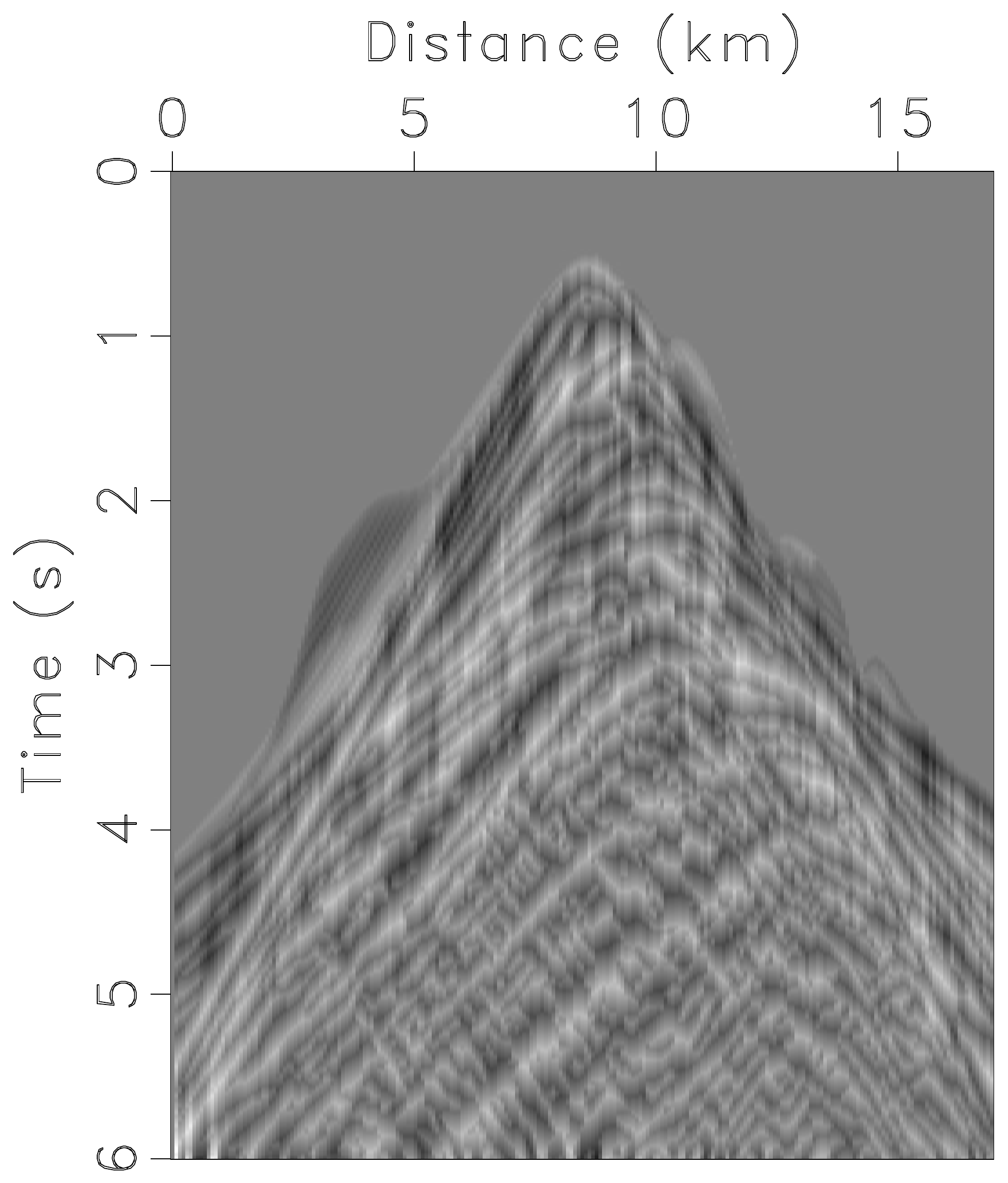}\\
{\small (a)}&{\small (b)}\\
\includegraphics[scale=.3]{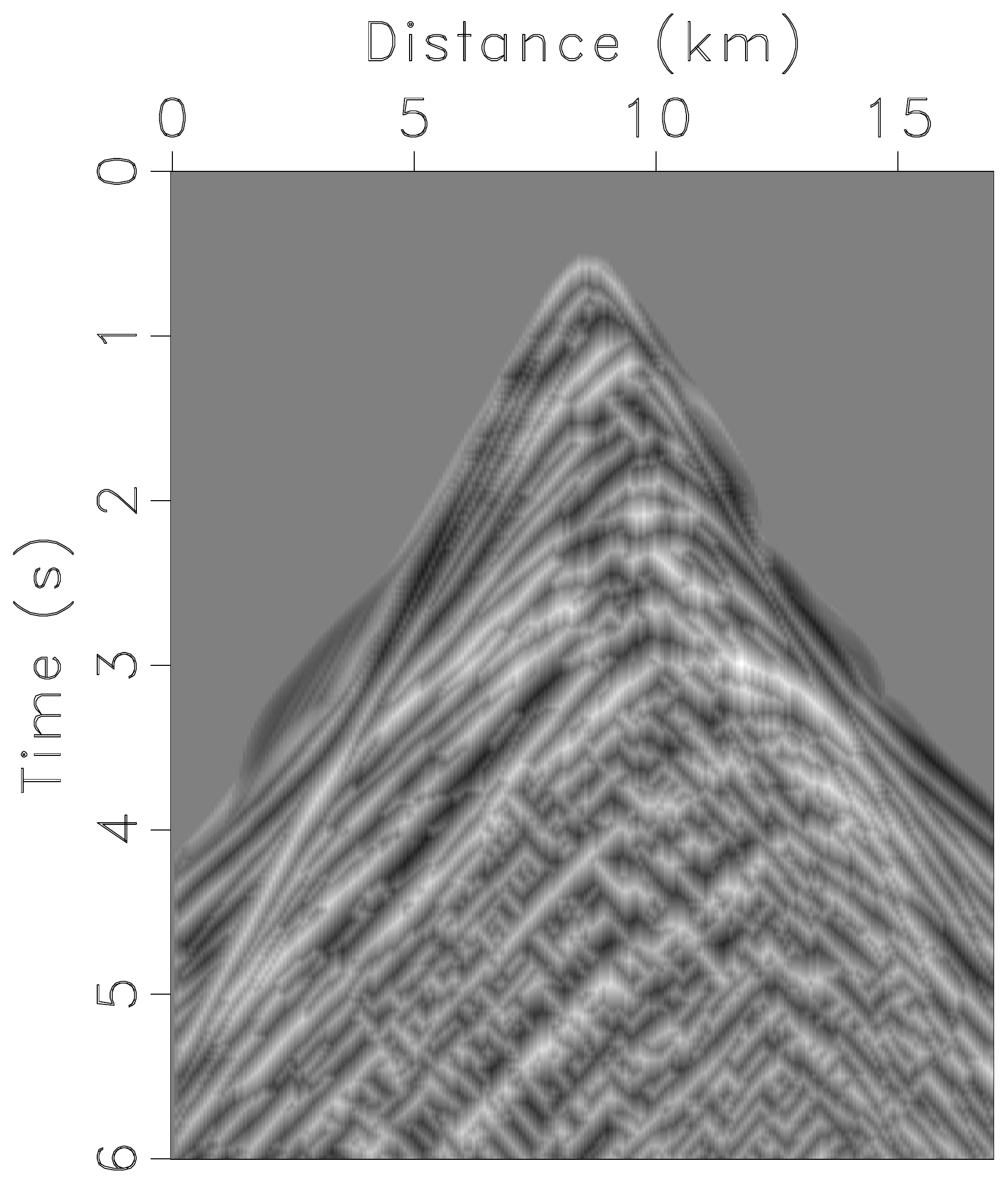}&
\includegraphics[scale=.3]{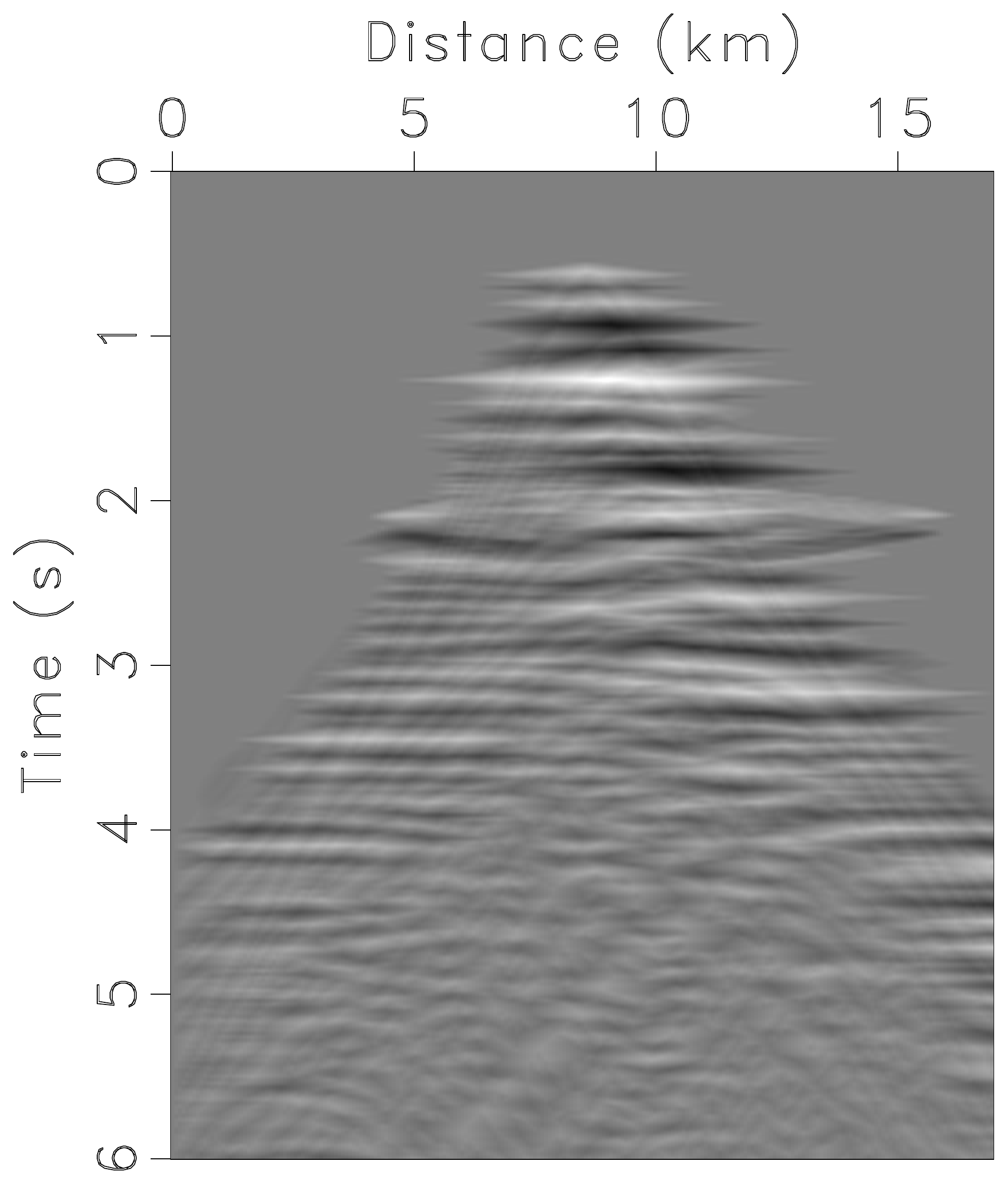}\\
{\small (c)}&{\small (d)}\\
\end{tabular}
\caption{ The adjoint sources of the first iteration. LSD (a), EMD with $\Omega=[0,0.1]\times[0,1]$ (b),  EMD with $\Omega=[0,1]\times[0,1]$ (c)  and EMD with $\Omega=[0,10]\times[0,1]$ (d).}
\label{fig:mar_adj}
\end{figure}

\begin{figure}
\centering
\begin{tabular}{cc}
\includegraphics[scale=.12]{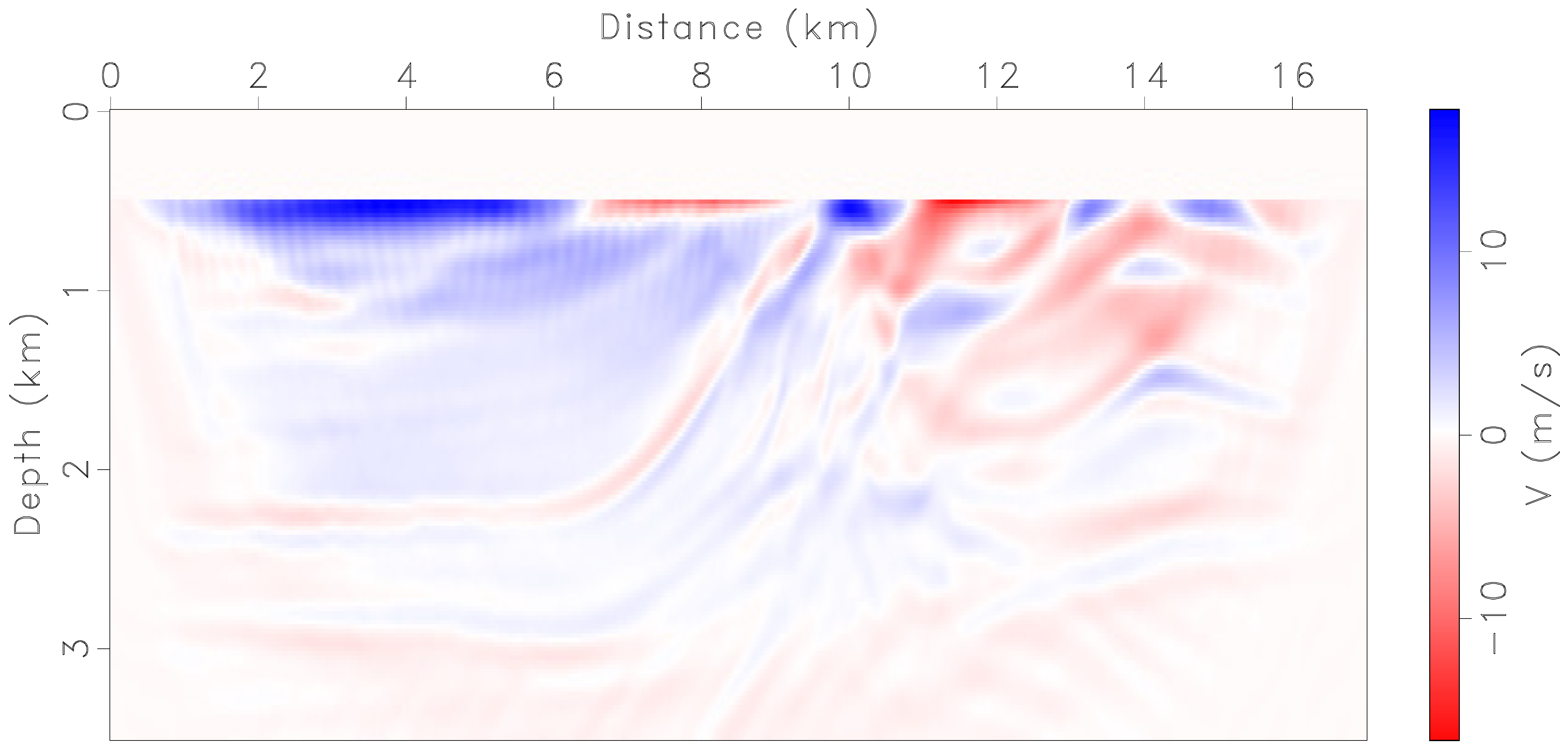}&
\includegraphics[scale=.12]{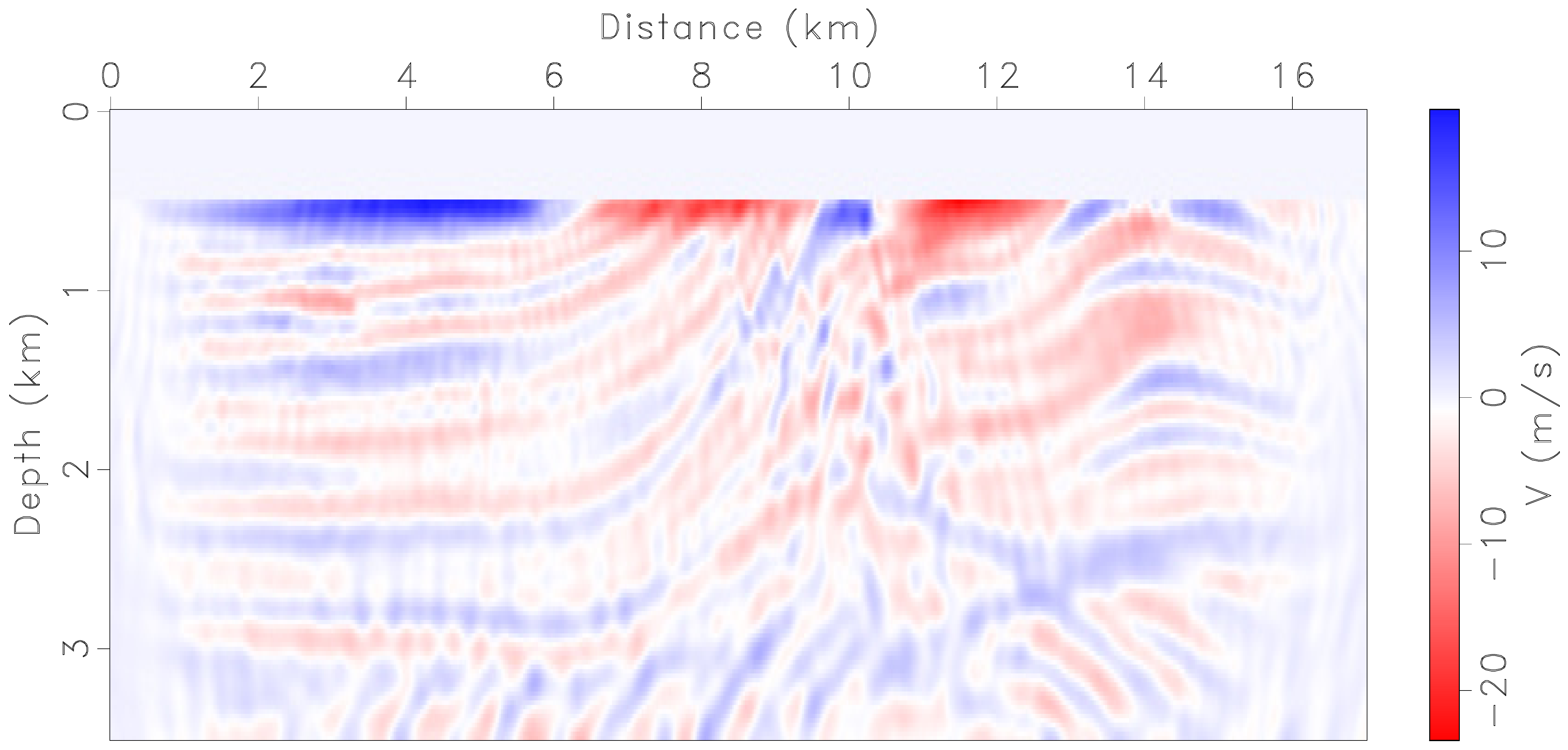}\\
{\small (a)}&{\small (b)}\\
\includegraphics[scale=.12]{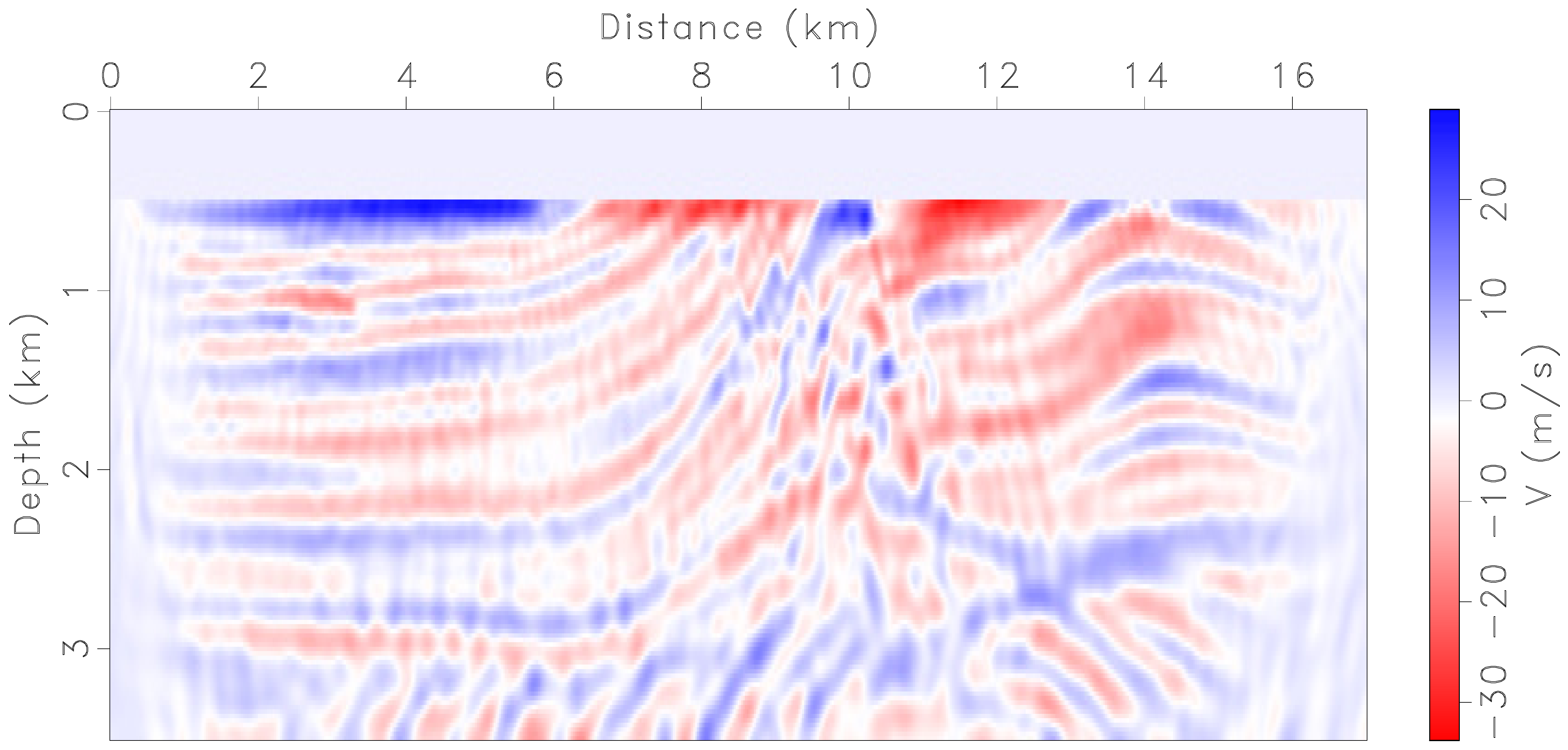}&
\includegraphics[scale=.12]{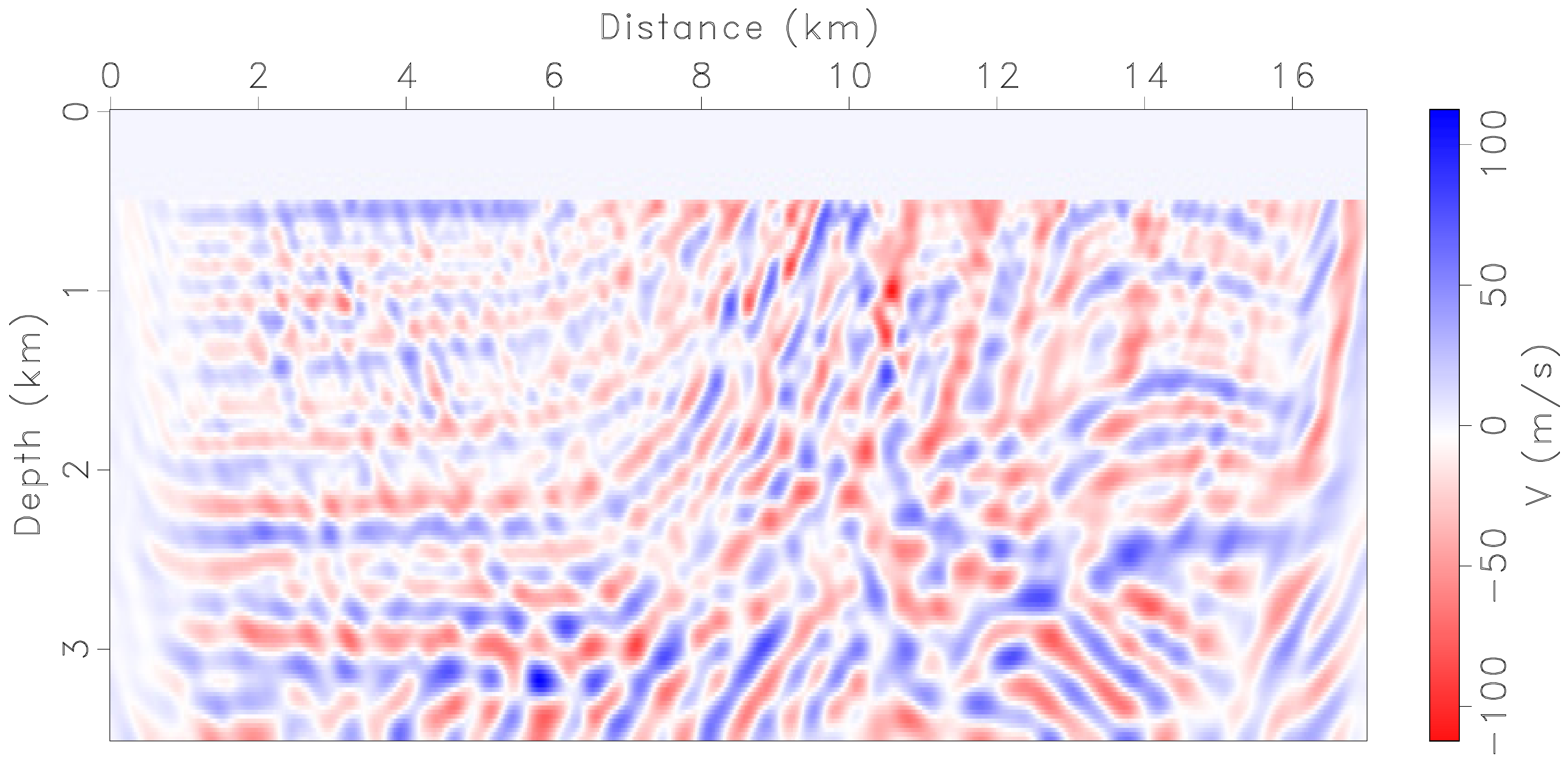}\\
{\small (c)}&{\small (d)}\\
\end{tabular}
\caption{The increments of the first iteration. LSD (a), EMD with $\Omega=[0,0.1]\times[0,1]$ (b),  EMD with $\Omega=[0,1]\times[0,1]$ (c)  and EMD with $\Omega=[0,10]\times[0,1]$ (d).}
\label{fig:mar_increment}
\end{figure}

\begin{figure}
\centering
\begin{tabular}{cc}
\includegraphics[scale=.12]{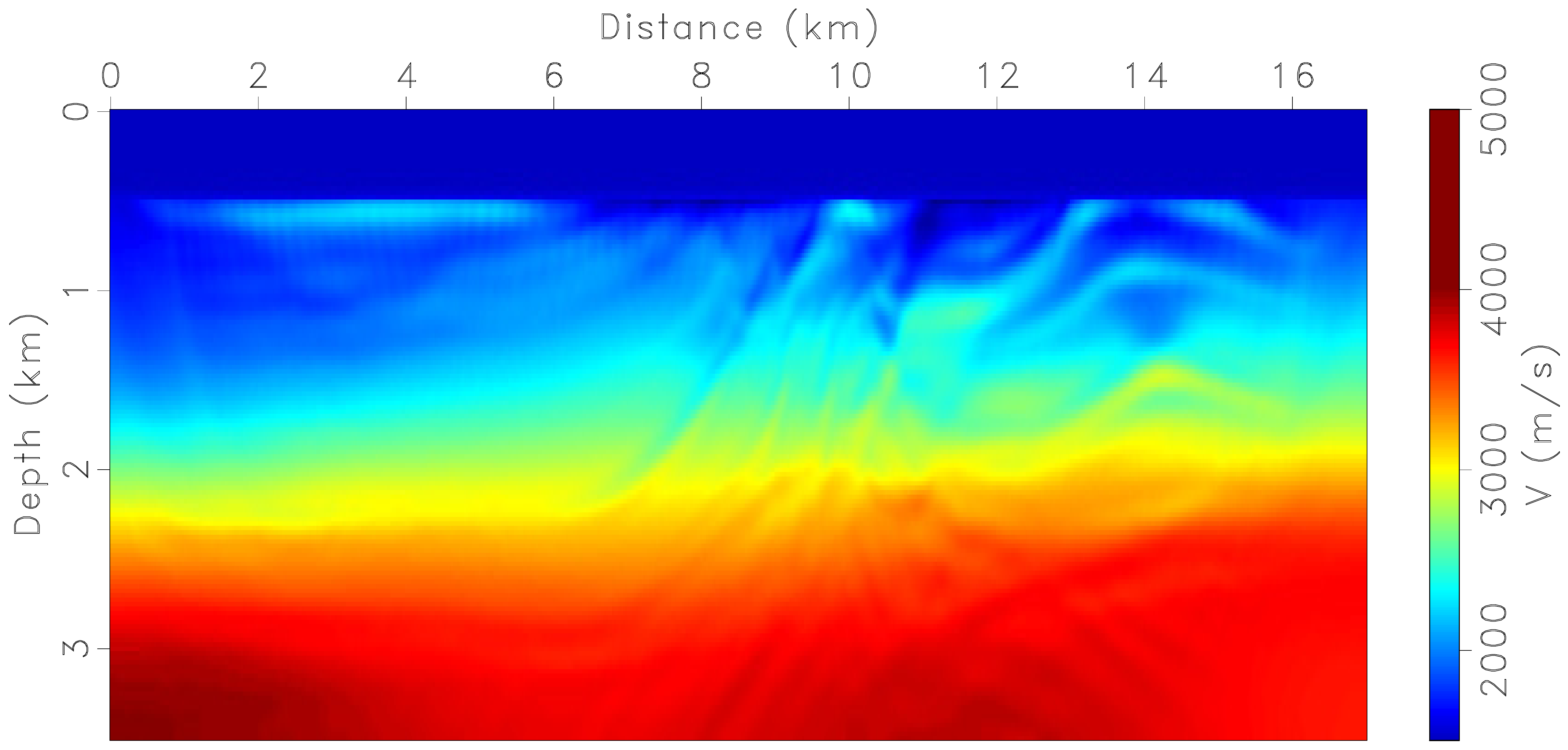}&
\includegraphics[scale=.12]{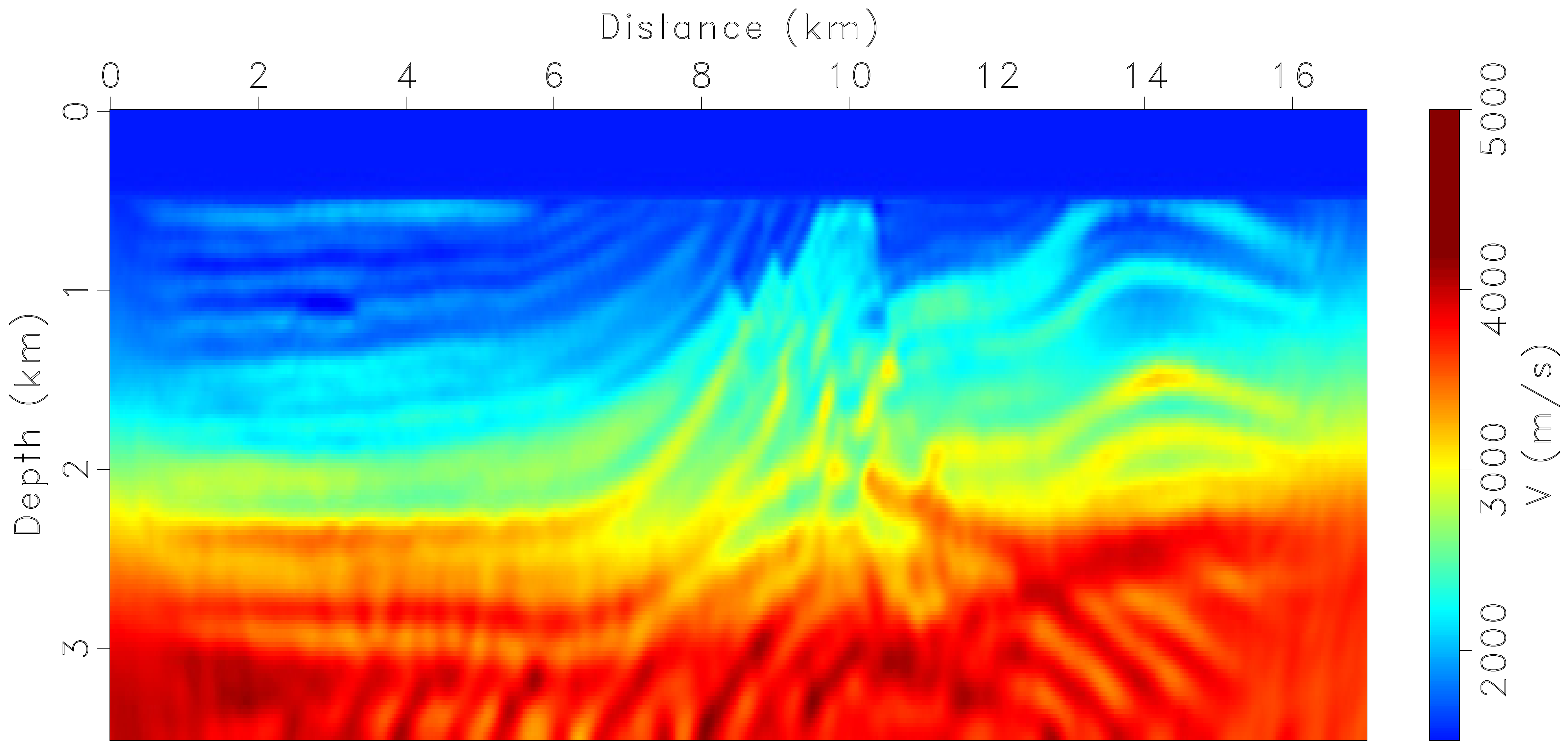}\\
{\small (a)}&{\small (b)}\\
\includegraphics[scale=.12]{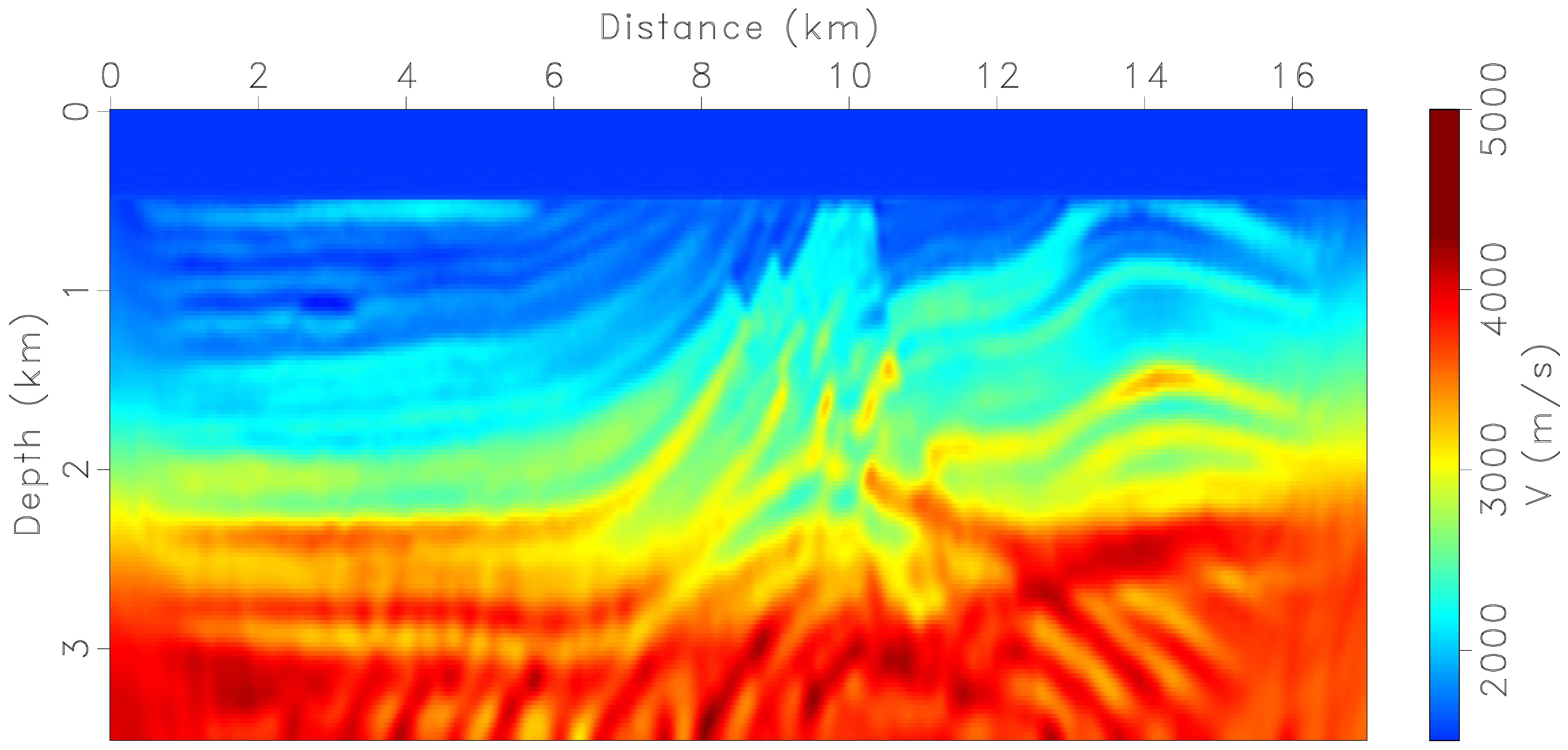}&
\includegraphics[scale=.12]{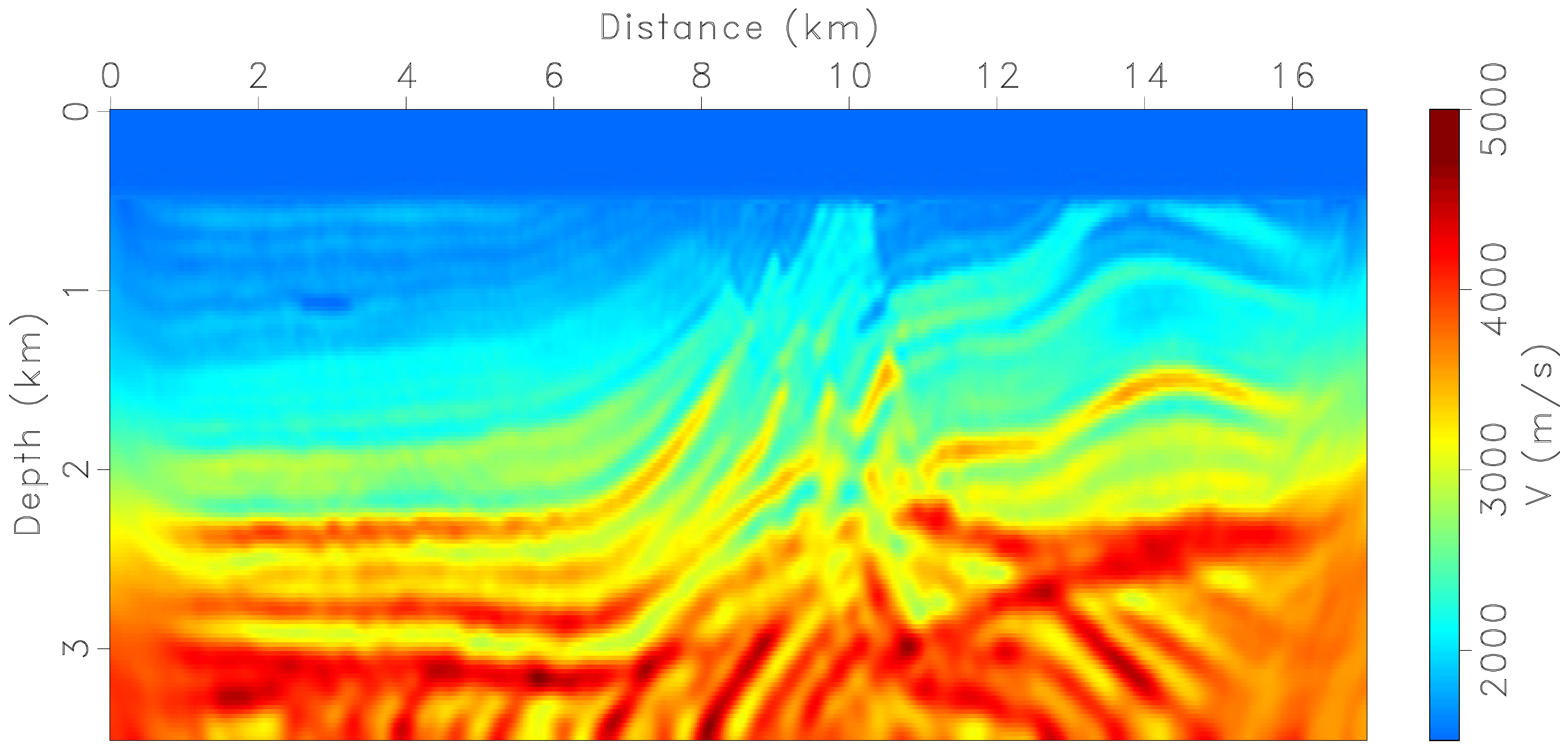}\\
{\small (c)}&{\small (d)}\\
\end{tabular}
\caption{The estimated P-wave velocity after the 10th iteration. LSD (a), EMD with $\Omega=[0,0.1]\times[0,1]$ (b),  EMD with $\Omega=[0,1]\times[0,1]$ (c)  and EMD with $\Omega=[0,10]\times[0,1]$ (d).}
\label{fig:mar_vel10}
\end{figure}

\begin{figure}
\centering
\begin{tabular}{cc}
\includegraphics[scale=.12]{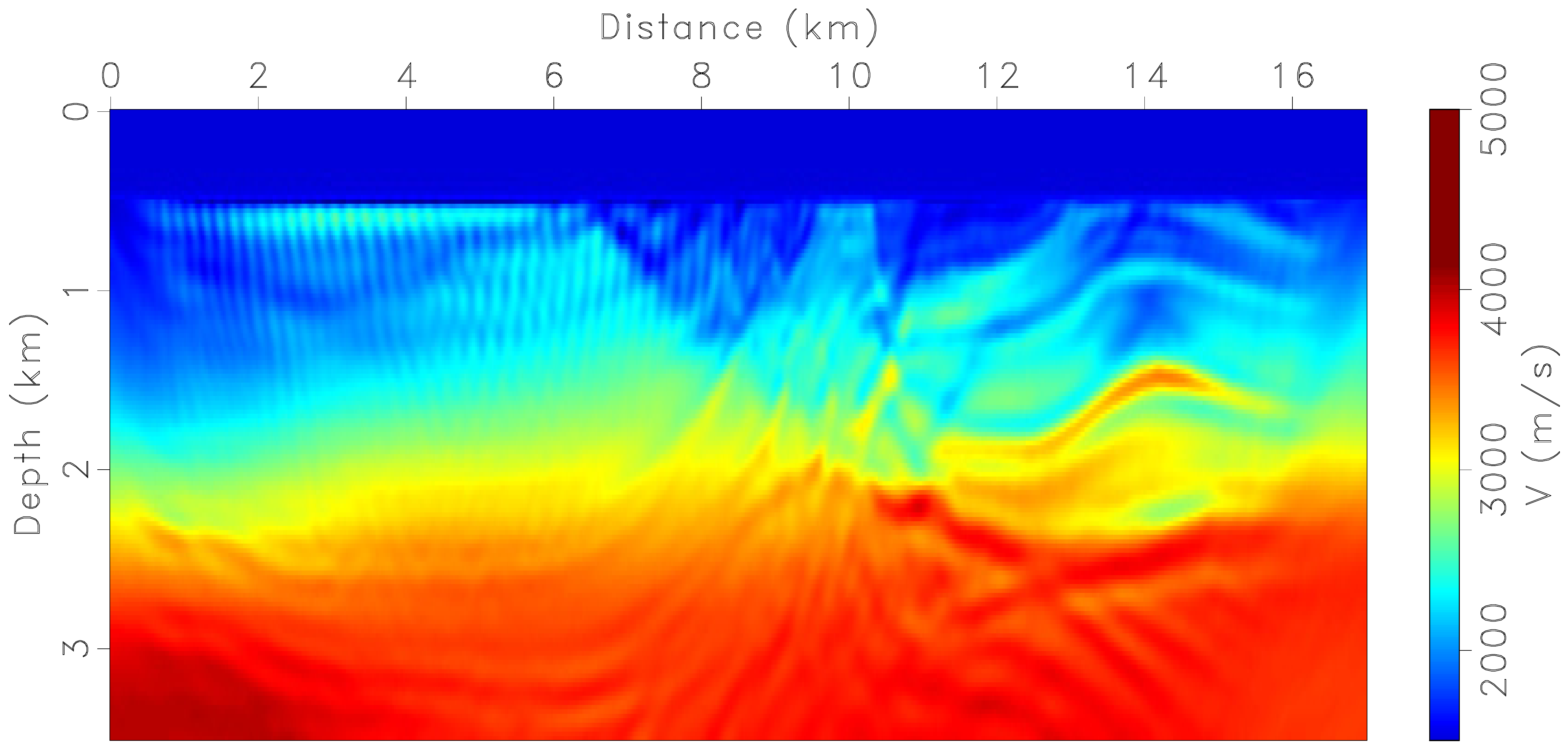}&
\includegraphics[scale=.12]{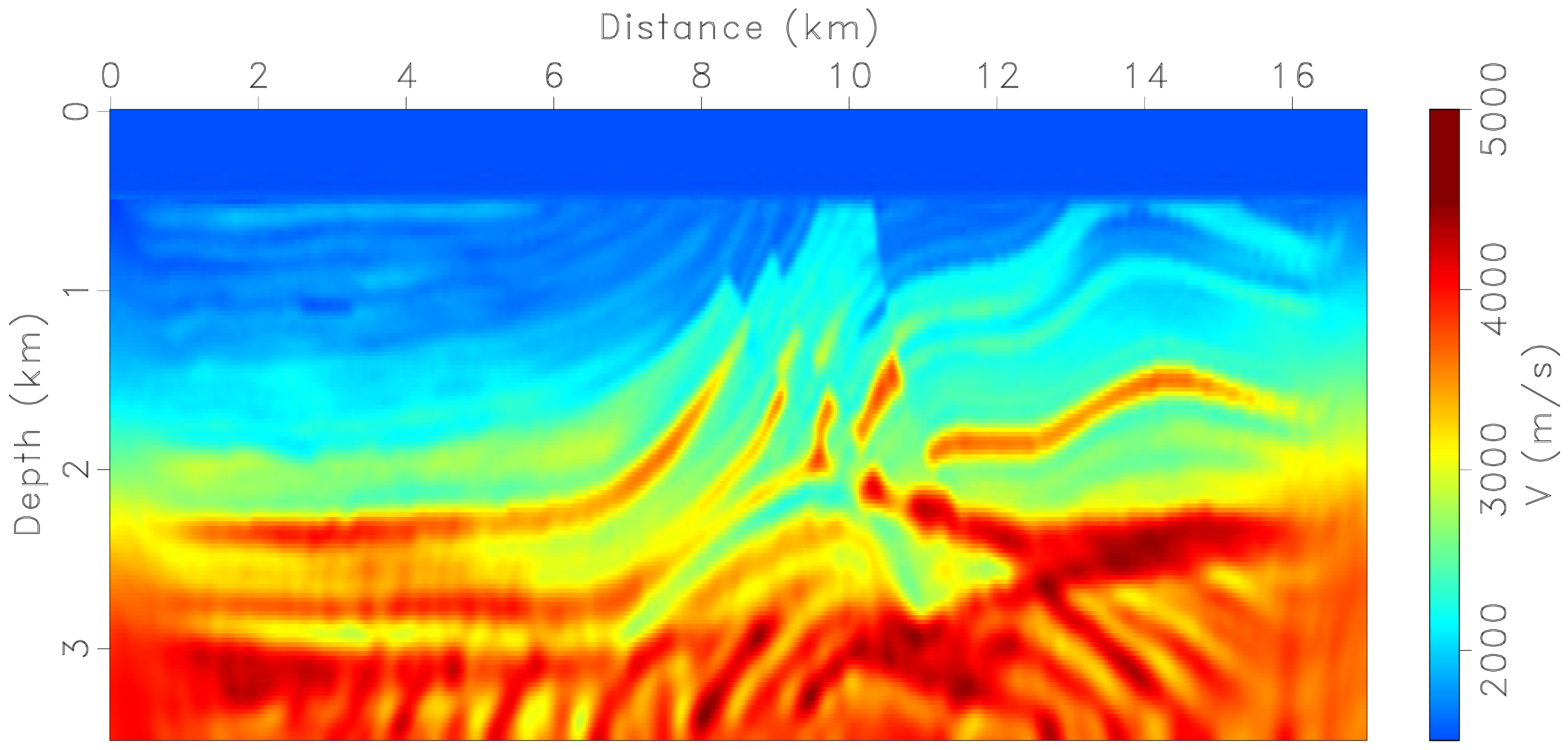}\\
{\small (a)}&{\small (b)}\\
\includegraphics[scale=.12]{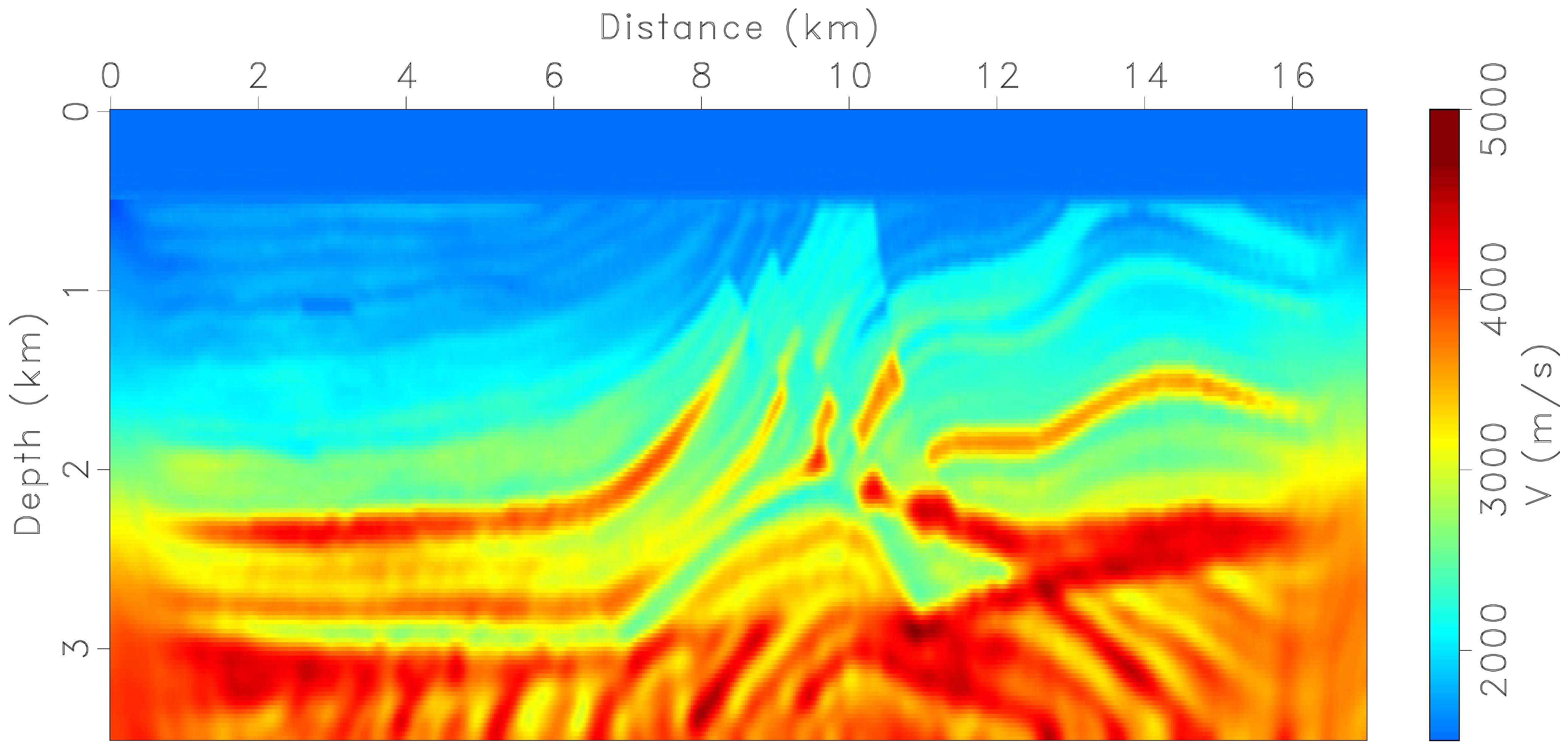}&
\includegraphics[scale=.12]{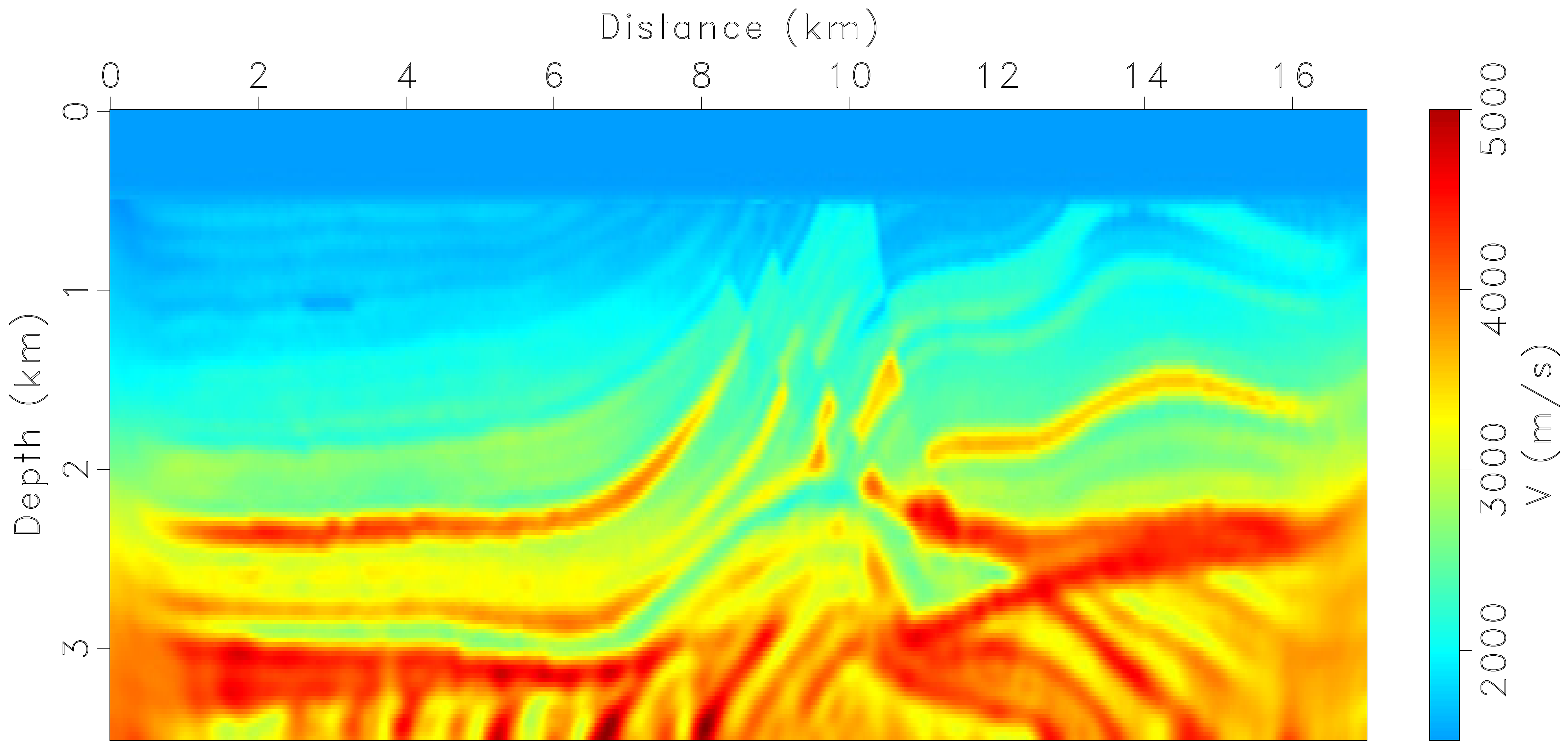}\\
{\small (c)}&{\small (d)}\\
\end{tabular}
\caption{The estimated P-wave velocity after the 30th iteration. LSD (a), EMD with $\Omega=[0,0.1]\times[0,1]$ (b),  EMD with $\Omega=[0,1]\times[0,1]$ (c)  and EMD with $\Omega=[0,10]\times[0,1]$ (d).}
\label{fig:mar_vel30}
\end{figure}

\begin{figure}
\centering
\begin{tabular}{cc}
\includegraphics[scale=.12]{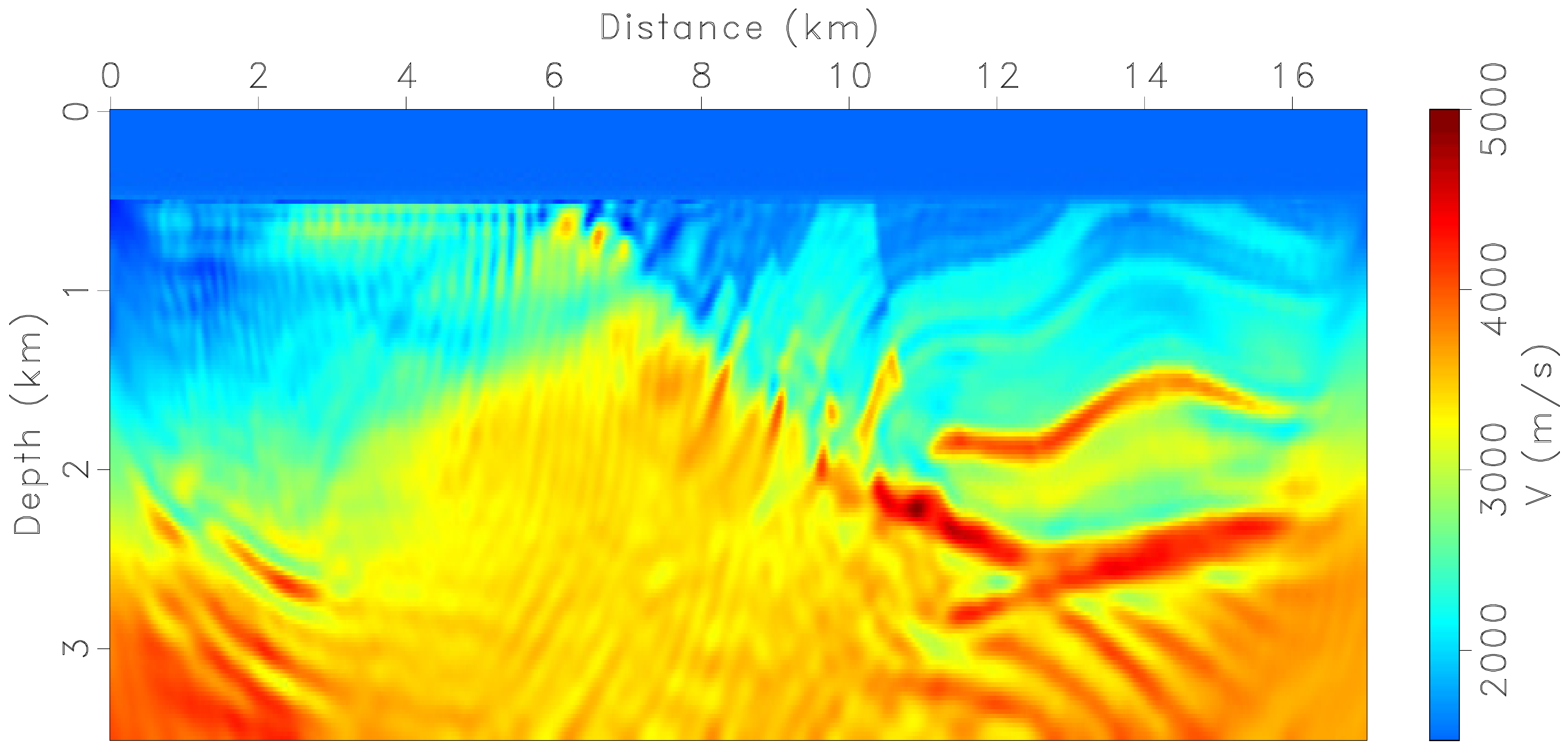}&
\includegraphics[scale=.12]{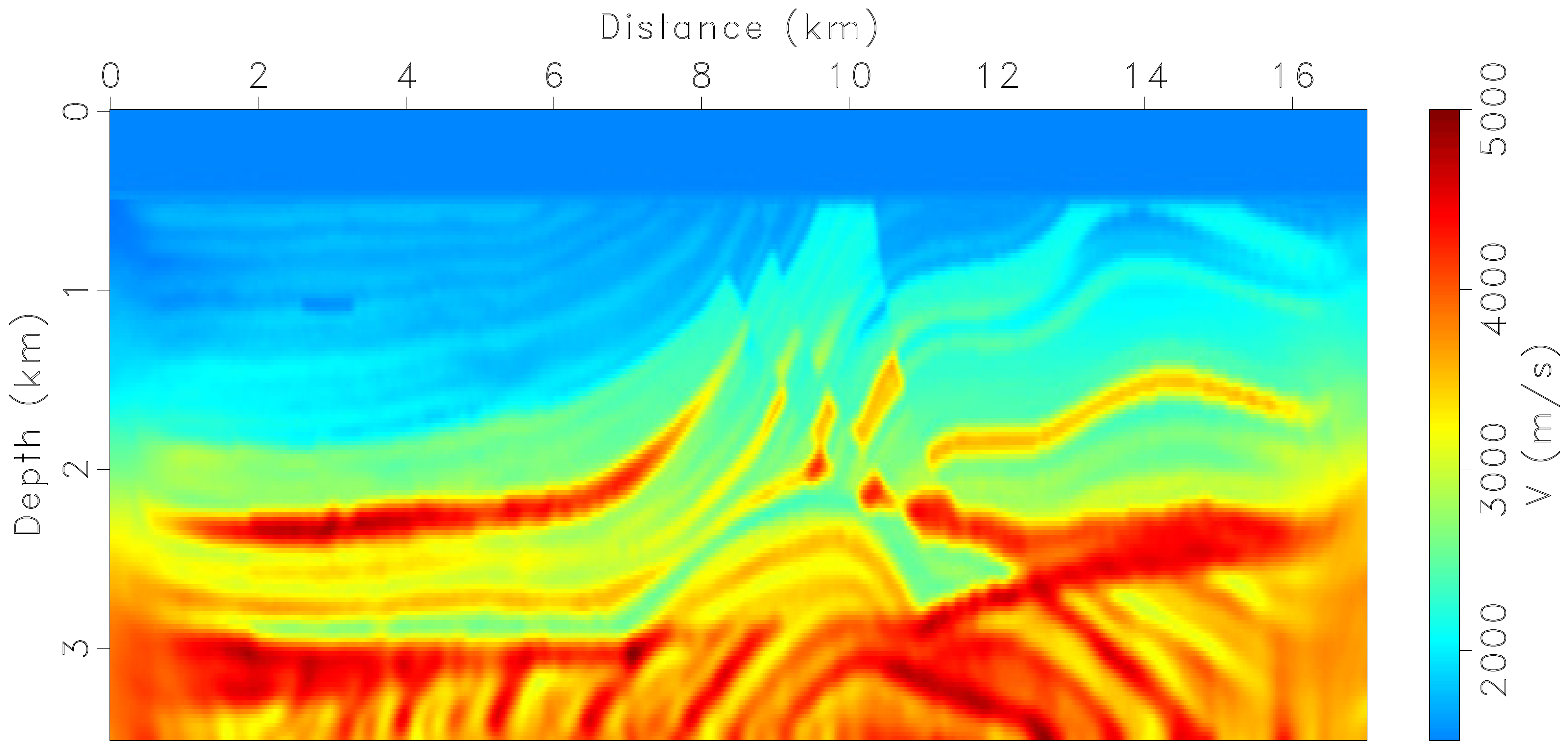}\\
{\small (a)}&{\small (b)}\\
\includegraphics[scale=.12]{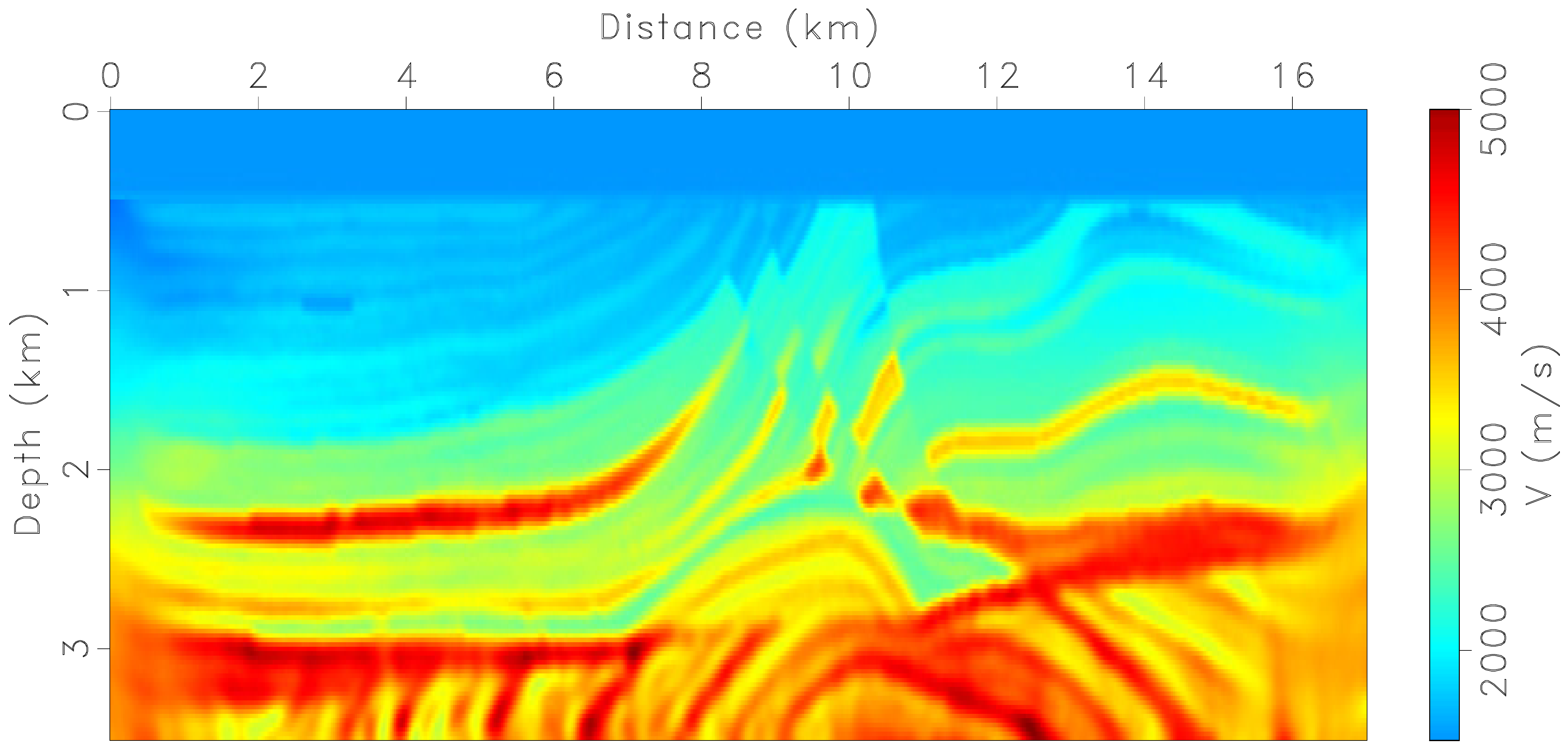}&
\includegraphics[scale=.12]{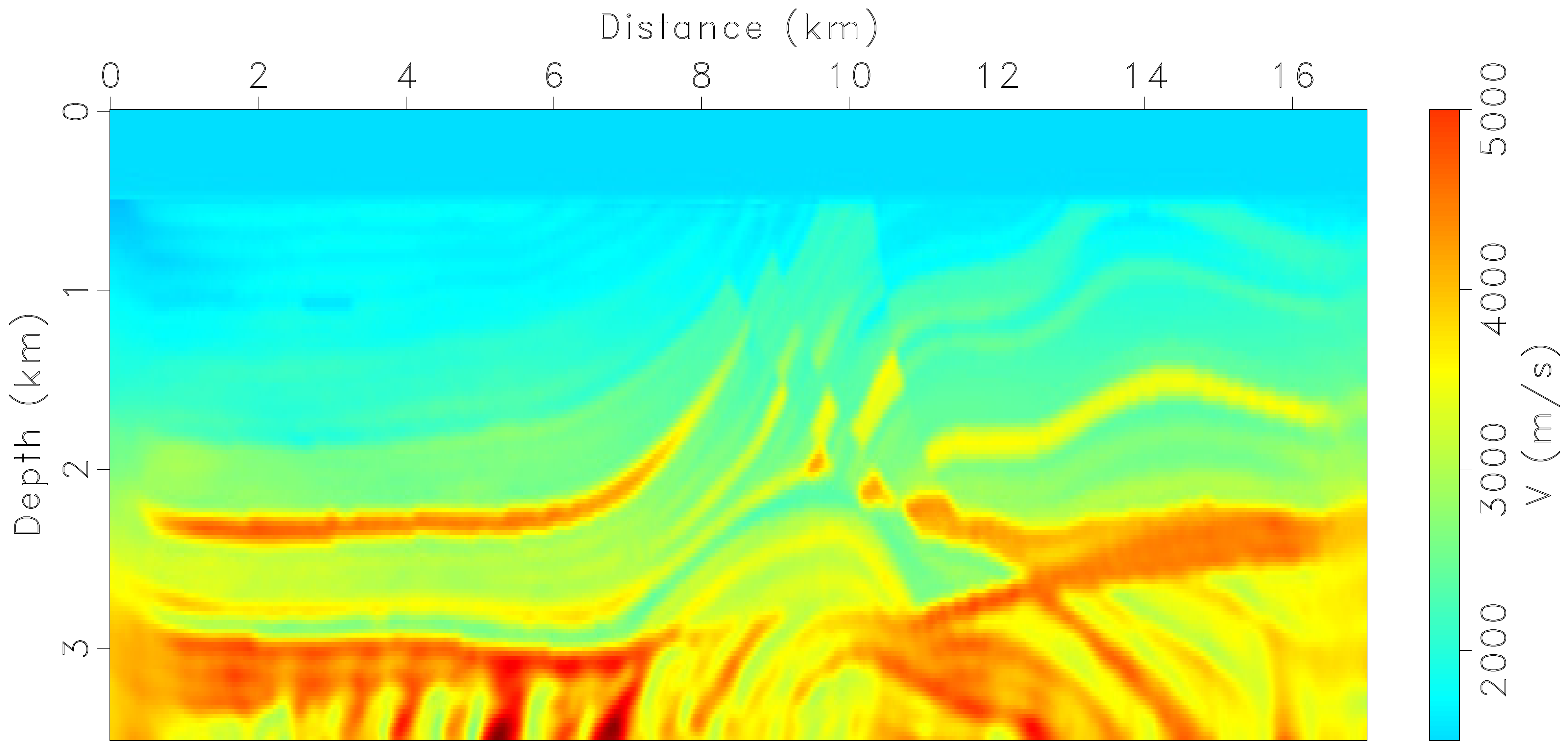}\\
{\small (c)}&{\small (d)}\\
\end{tabular}
\caption{The estimated P-wave velocity after the 100th iteration. LSD (a), EMD with $\Omega=[0,0.1]\times[0,1]$ (b),  EMD with $\Omega=[0,1]\times[0,1]$ (c)  and EMD with $\Omega=[0,10]\times[0,1]$ (d).}
\label{fig:mar_vel100}
\end{figure}

\begin{figure}
\centering
\includegraphics[scale=0.75]{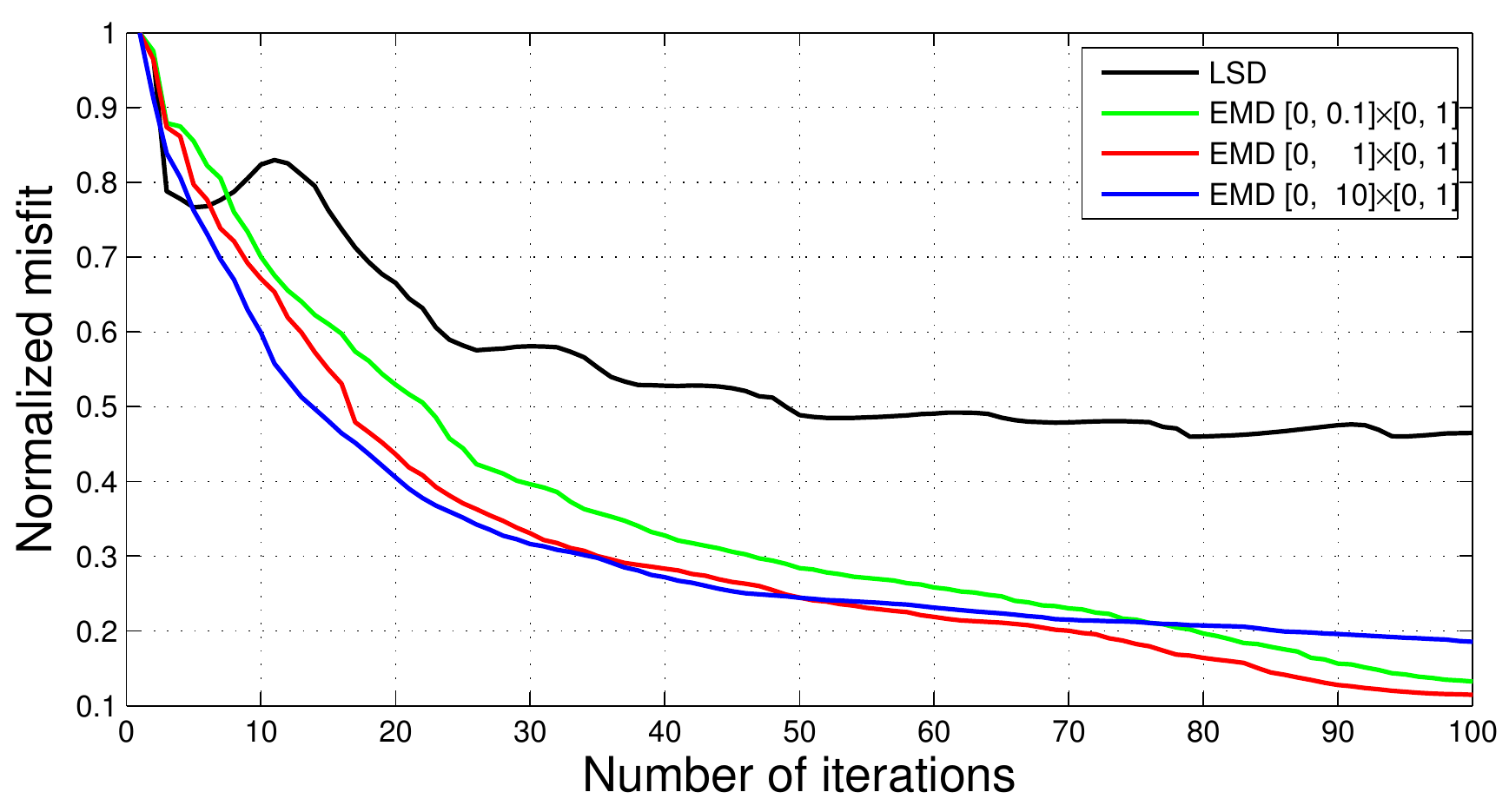}\\
\caption{The convergence rate of FWI using different objective functions. }
\label{fig:mar_objs}
\end{figure}

A strongly smoothened version of the exact model, presented in Figure \ref{fig:mar_model} (b), is considered as the initial guess. Figure \ref{fig:mar_adj} (a) displays the adjoint source of the 1st iteration of the classical FWI. Figures \ref{fig:mar_adj} (b-d) show the   adjoint sources of the 1st iteration using EMD with different weight distributions on time and receiver axes. We fix $X=1$ and use different values of $T$ to change the weight distributions. Comparison of the adjoint sources displayed in Figures \ref{fig:mar_adj} (b-d) suggests that, the continuity along receiver axis in adjoint source increase with the increase of the value of $T$. This can be well understood that, as the value of $T$ increases, the unit cost of transporting data along time axis increases and the transport will increase in the receiver direction. It can be seen from Figure \ref{fig:mar_adj} that, EMD can enhance the weighting of weaker amplitude seismic events. Since the interpretation of weak reflection data is enhanced, the corresponding deep structures can be seen in the updates of the first iteration shown in Figures \ref{fig:mar_increment} (a-d). Figures \ref{fig:mar_vel10}, \ref{fig:mar_vel30} and \ref{fig:mar_vel100}  display the inversions after 10th, 30th and 100th iteration, respectively. The classical FWI using LSD fails to invert velocity of shallow area at the beginning, thus the velocity of the deep area can not be updated validly. On the other hand, the estimation obtained with the EMD is significantly improved.  Compare the updates of the first iteration displayed in Figures \ref{fig:mar_increment} (b-d), we can find that, the capability of EMD to detect lateral variation of velocity model increases with the increase of the value of $T$. Since we have a long-offset recording geometry, the observed data contains abundant diving waves, which can bring lots of information about velocity variation in the lateral direction. Therefore, we can more effectively invert lateral variation of velocity model when a lager number is used for $T$. We also find that, when $T$ is chosen as a large number, the vertical resolution of the estimated velocity model will decrease.  The convergence rate displayed in Figure \ref{fig:mar_objs} has revealed that, the misfits of FWI using EMD decrease monotonously whatever parameters are chosen, but FWI using LSD suffers from local minima. Compare three convergence rate lines of FWI using EMD, we can find that, misfit decreases faster at the beginning when $T$ is larger, but the convergence speed gradually slows down and the misfit value converges to relatively larger number. In this case study, the dominant structures of the Marmousi 2 model are horizontal layers and the velocity variation in vertical direction is larger than that in horizontal direction, so it is supposed to give more weights on the time axis. Certainly,  developing systematic methods to distribute weight among different directions is an important work for a better application of EMD to FWI.        

\clearpage
\subsection{Application to SEG 2014 benchmark data}
\begin{figure}
\centering
\begin{tabular}{cc}
\includegraphics[scale=.3]{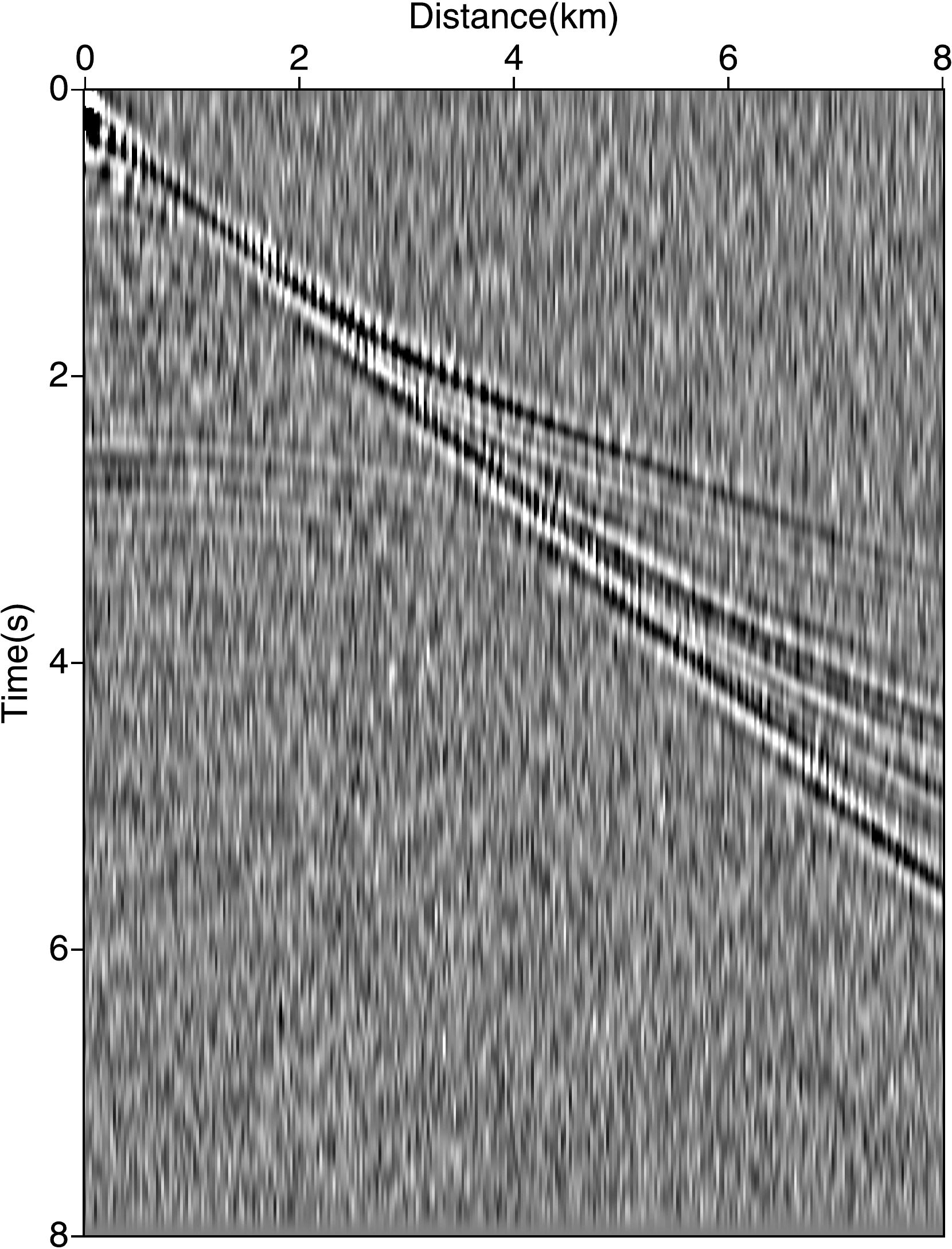}&
\includegraphics[scale=.3]{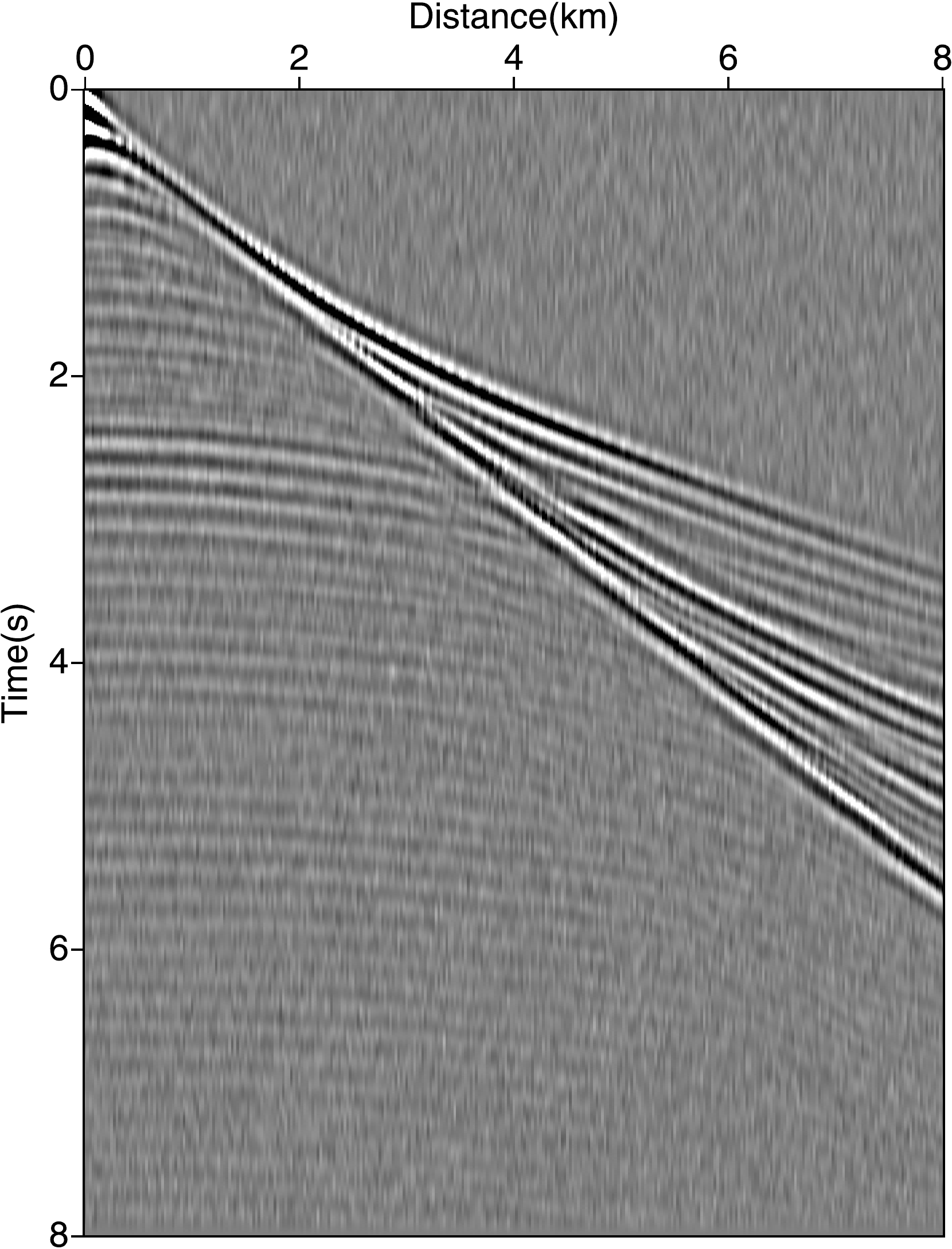}\\
{\small (a)}&{\small (b)}\\
\includegraphics[scale=.3]{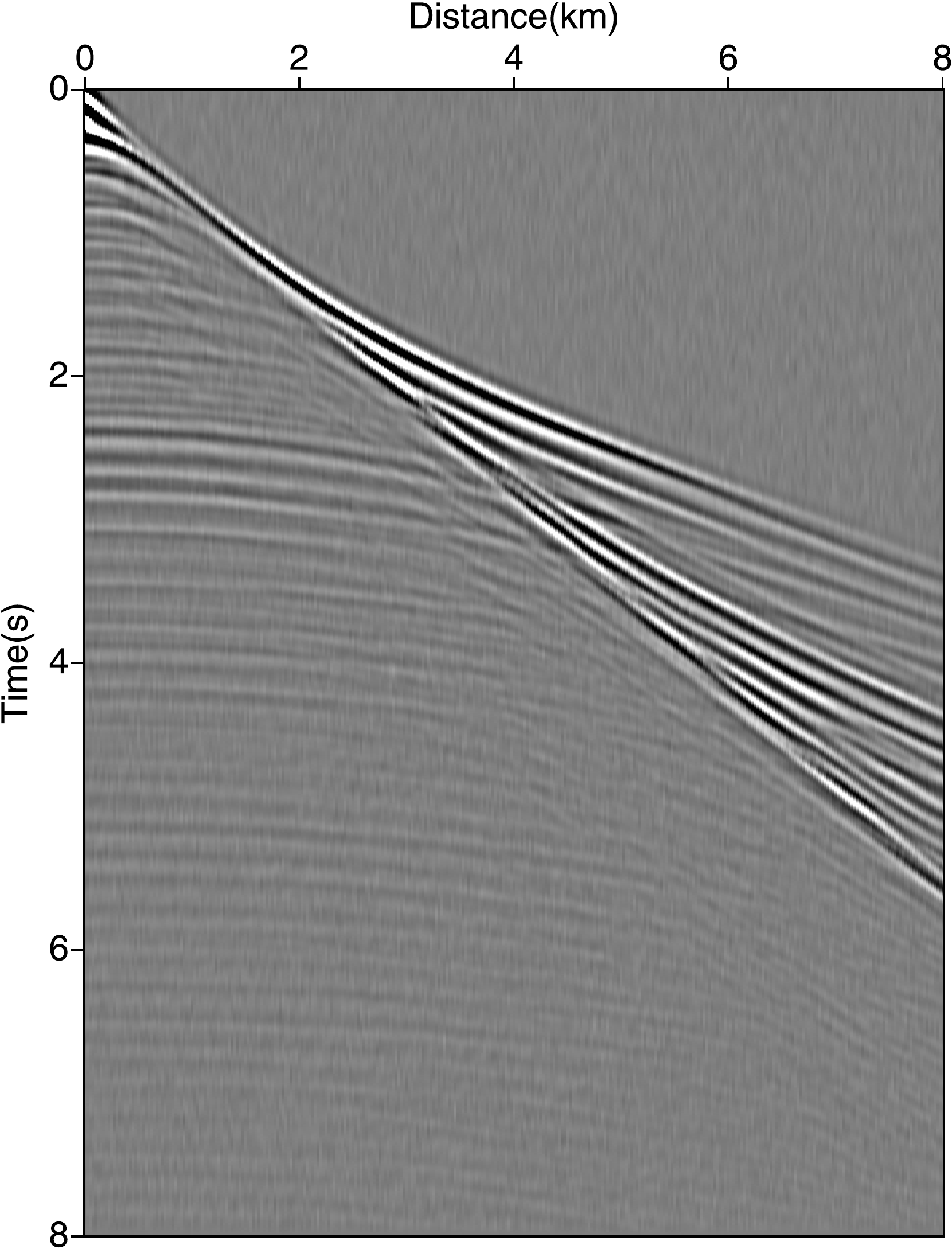}&
\includegraphics[scale=.3]{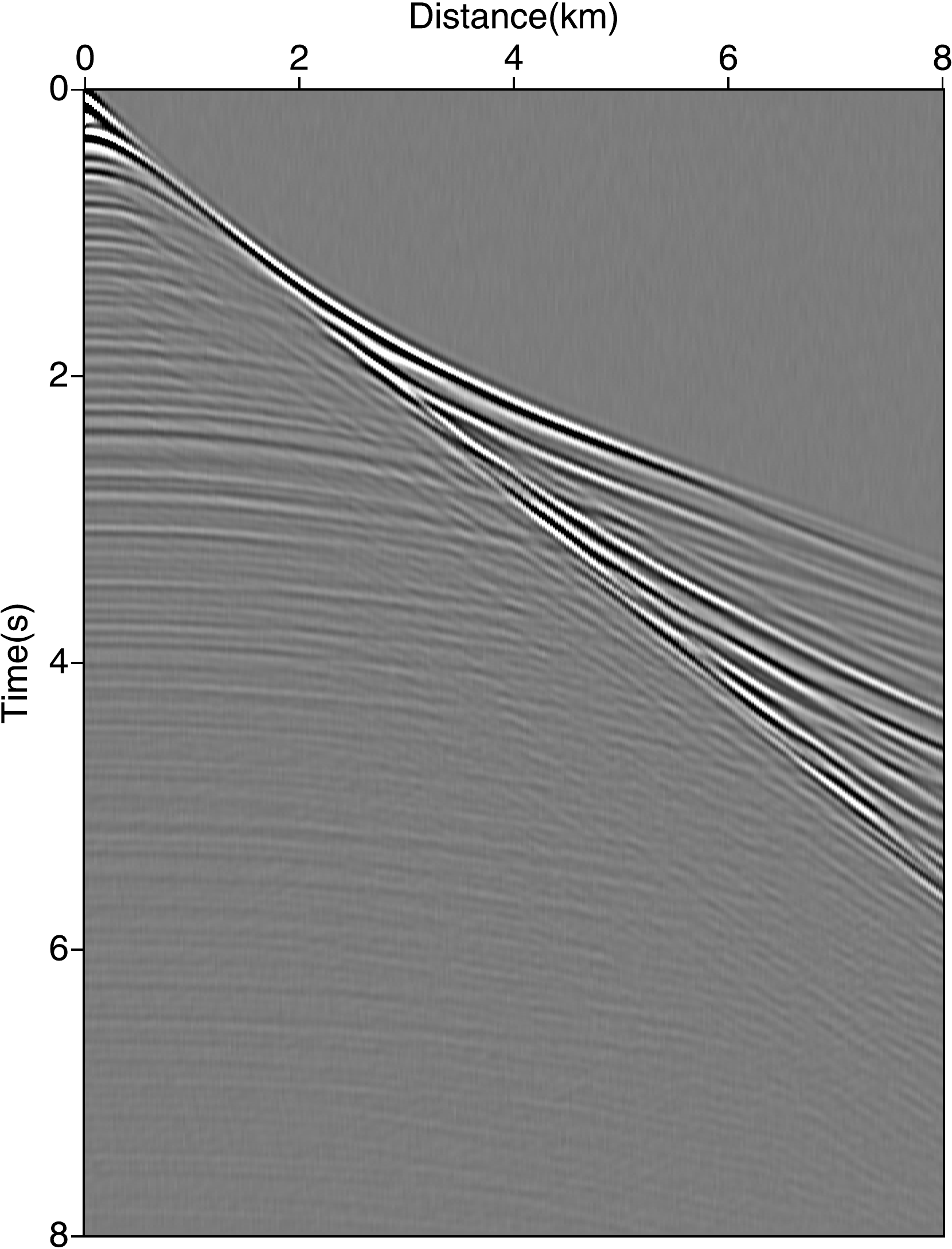}\\
{\small (c)}&{\small (d)}\\
\end{tabular}
\caption{  Chevron 2014  synthetic seismic benchmark. The 1st common-shot gather with the frequency bands 0-3 hz (a), 0-5 hz (b), 0-7 hz (c) and 0-10 hz (d).}
\label{fig:ChevronData}
\end{figure}

\begin{figure}
\centering
\includegraphics[scale=0.6]{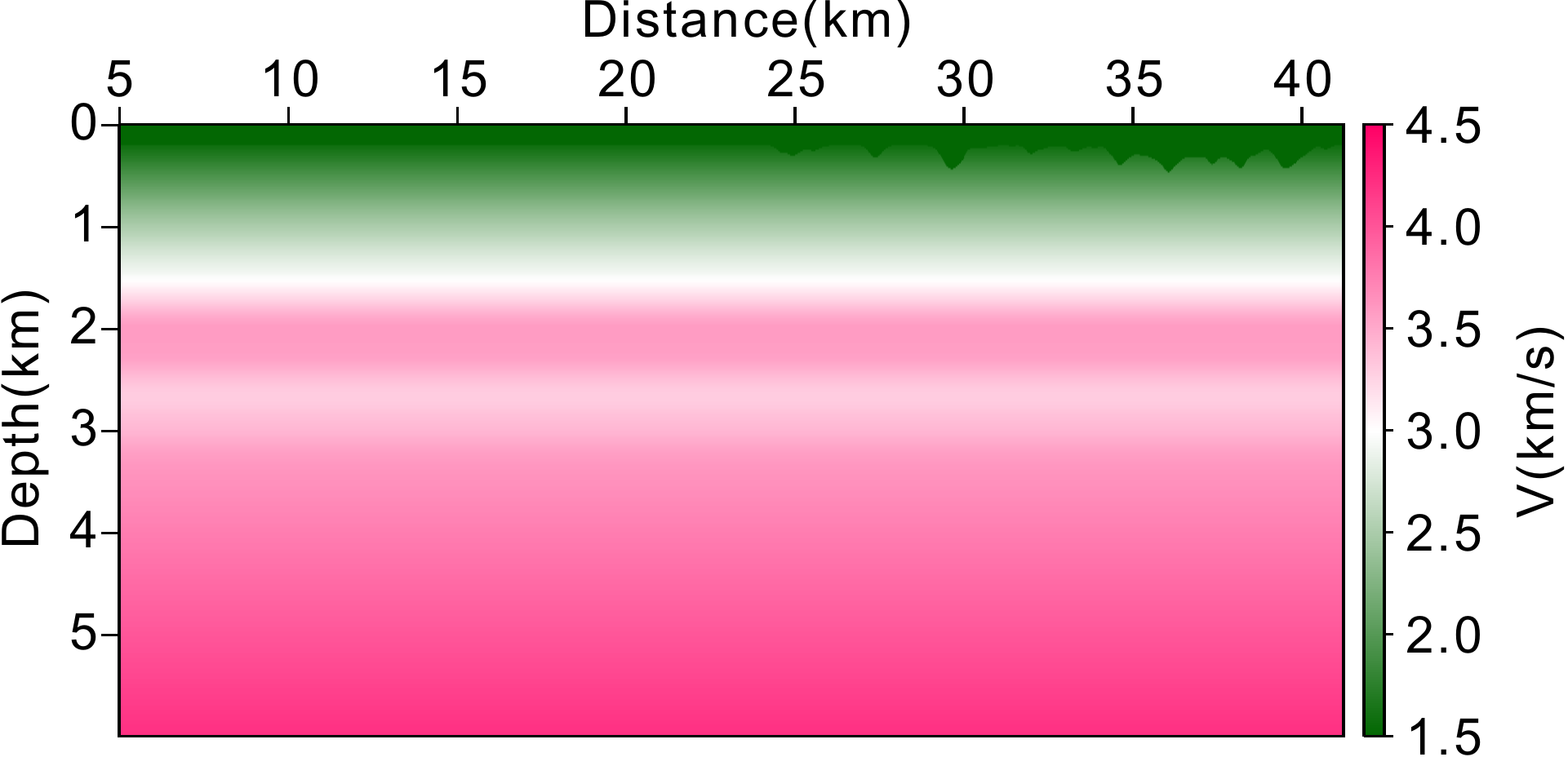}\\
\caption{Chevron 2014 initial P-wave velocity model. }
\label{fig:ChevronInitial}
\end{figure}

The SEG 2014 blind benchmark data set provided by Chevron oil company is a 2D marine isotropic elastic synthetic data with free surface multiples for FWI. The data set includes 1600 shots with an interval of 25 m at the depth of 15 m. Each shot has 321 receivers with 25 m sampling at the same depth. The observed data are plotted in Figure \ref{fig:ChevronData}, from which we can see that the data have a low signal-to-noise ratio (SNR) below 3 Hz and strong noise even in the 3-5 Hz frequency band. This is close to realistic exploration seismic data and makes the classical FWI easily converge to local minima. In addition, an initial P-wave velocity model, shown in Figure \ref{fig:ChevronInitial} is also provided. The true velocity model used to generate the data has not been released so far and only a single velocity profile of the true model at $x=39375$ m is given to verify the inversion results. 

Due to  the high computational cost, we select every seven shot gather, 229 shots in total for this study. The predicted data are generated with 2-8 finite difference modeling \citep{yong2018forward} with free-surface boundary condition on the top side of the domain and the PML boundary condition on the other three sides \citep{komatitsch2007unsplit}. The source wavelet for each frequency band is estimated by following  the frequency-domain strategy introduced by  \citep{pratt1999seismic}. We apply the proposed inversion
method with $15\times{25}$ m space sampling and 2 ms time sampling. Multi-scale inversion strategy \citep{bunks1995multiscale} is employed on  
frequency bands of 0-3 Hz, 0-5 Hz, 0-7 Hz, 0-10 Hz, 0-12 Hz and 0-15 Hz, sequentially. For each frequency bands, 20 iterations of a preconditioned conjugate gradient algorithm with parabolic search method \citep{nocedal2006sequential,liu2017effects} are performed.  $P_{cal}(x_r,t)$ and $P_{obs}(x_r,t)$ are also defined on $\Omega=[0,1]\times[0,1]$ in this case study. We compute the EMD between the observed and predicted data using 200 iterations of Algorithm (\ref{PD_line}) with the grid size of $401\times{321}$. This study was carried out on the DELL workstation T7610 with 16 G video memory using CUDA-C programming. The computation time of forward modeling for each shot is 0.305 s and the computation time of calculating EMD for each shot is 0.126 s. The computation time for each iteration in FWI using EMD is 443.85 s, in which we usually need to solve the wave equation five times and compute the EMD three times. Compared to FWI using LSD,  there is about 11.04$\%$ increase of the computational time of each iteration in this case.

\begin{figure}
\centering
\includegraphics[scale=0.6]{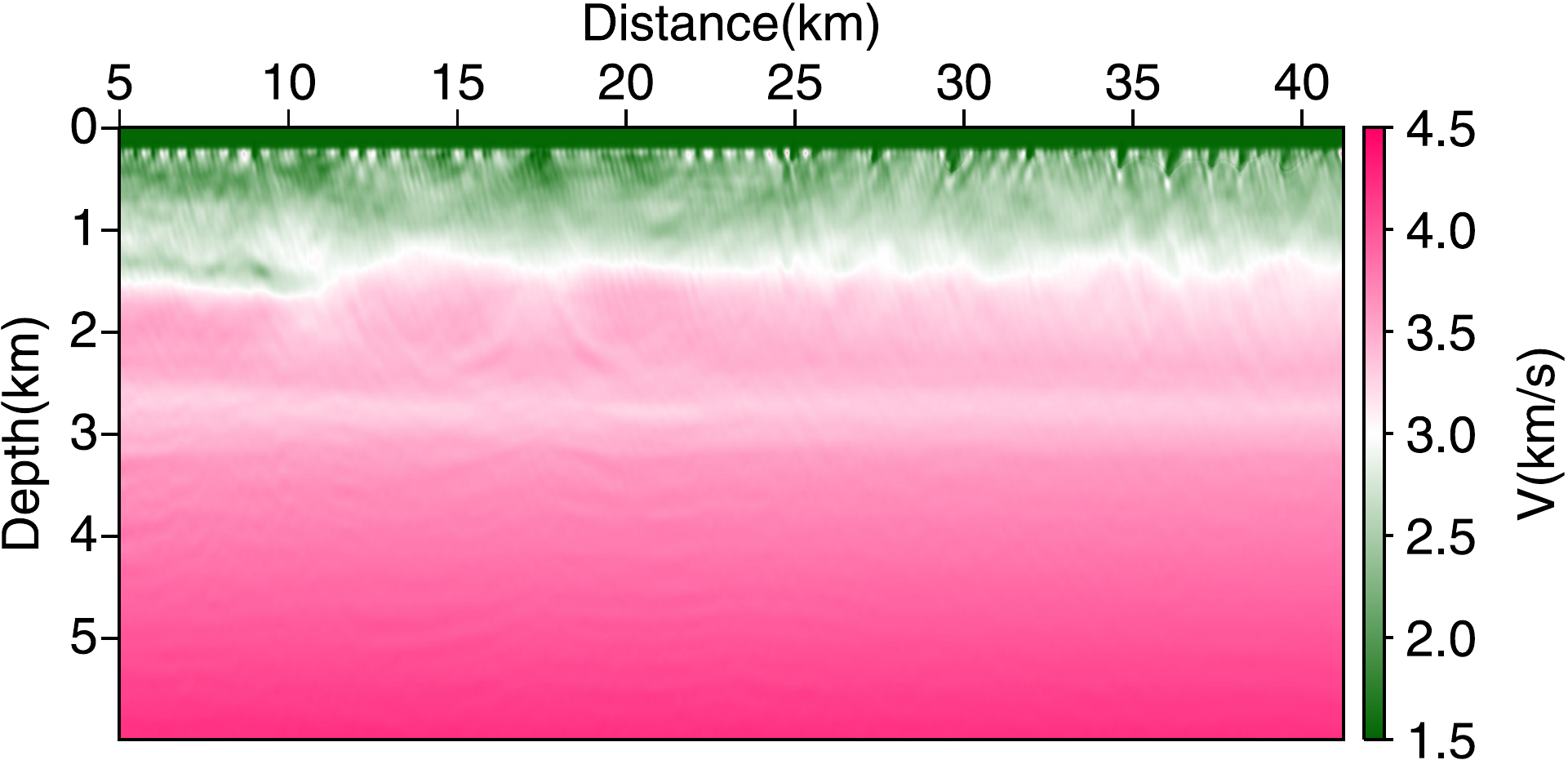}\\
\caption{The inversion results using the classical FWI with the least-square misfit function.}
\label{fig:ChevronFWI}
\end{figure}

\begin{figure}
\centering
\includegraphics[scale=0.6]{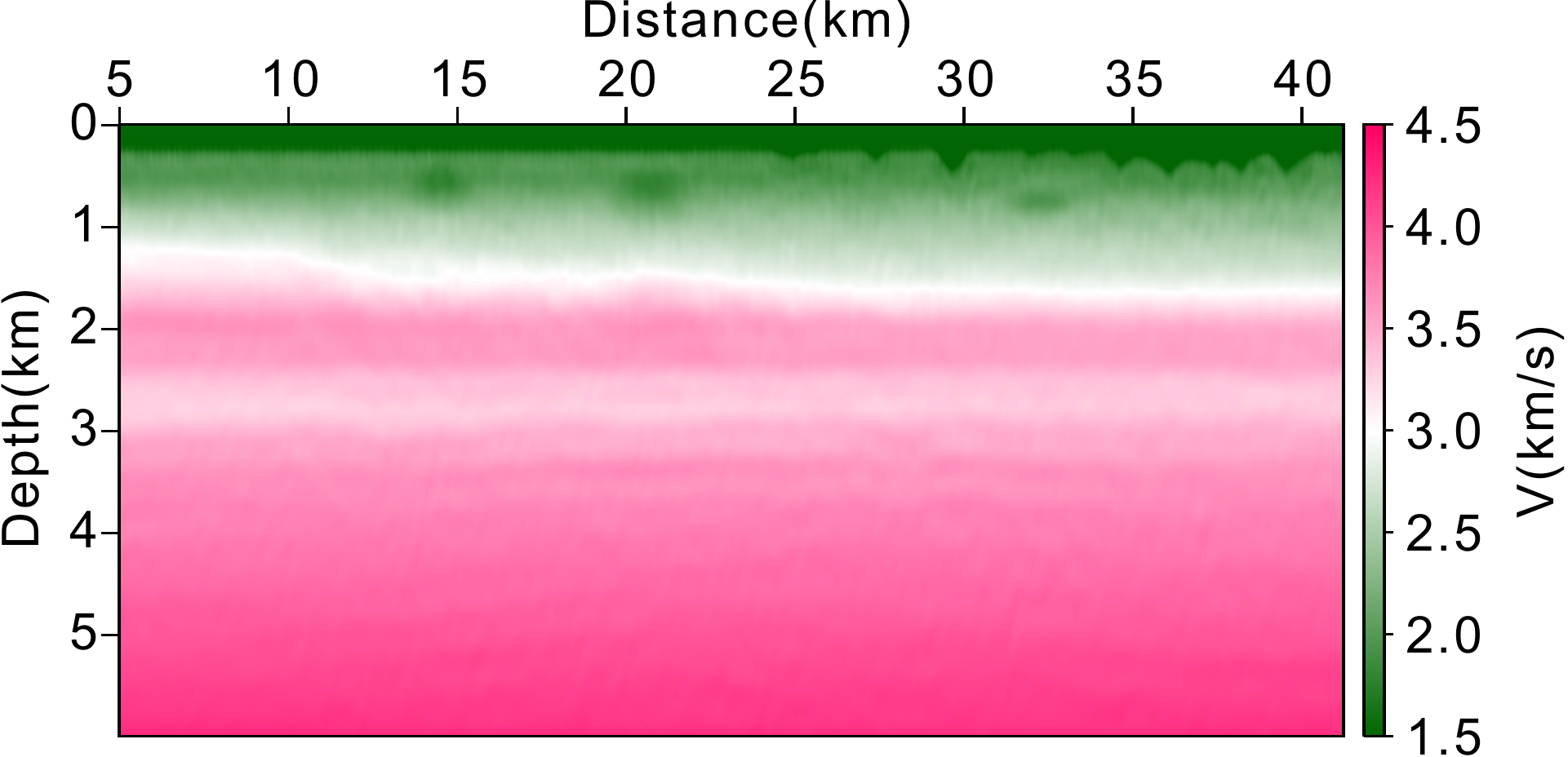}\\
{\small (a)}\\
\includegraphics[scale=0.6]{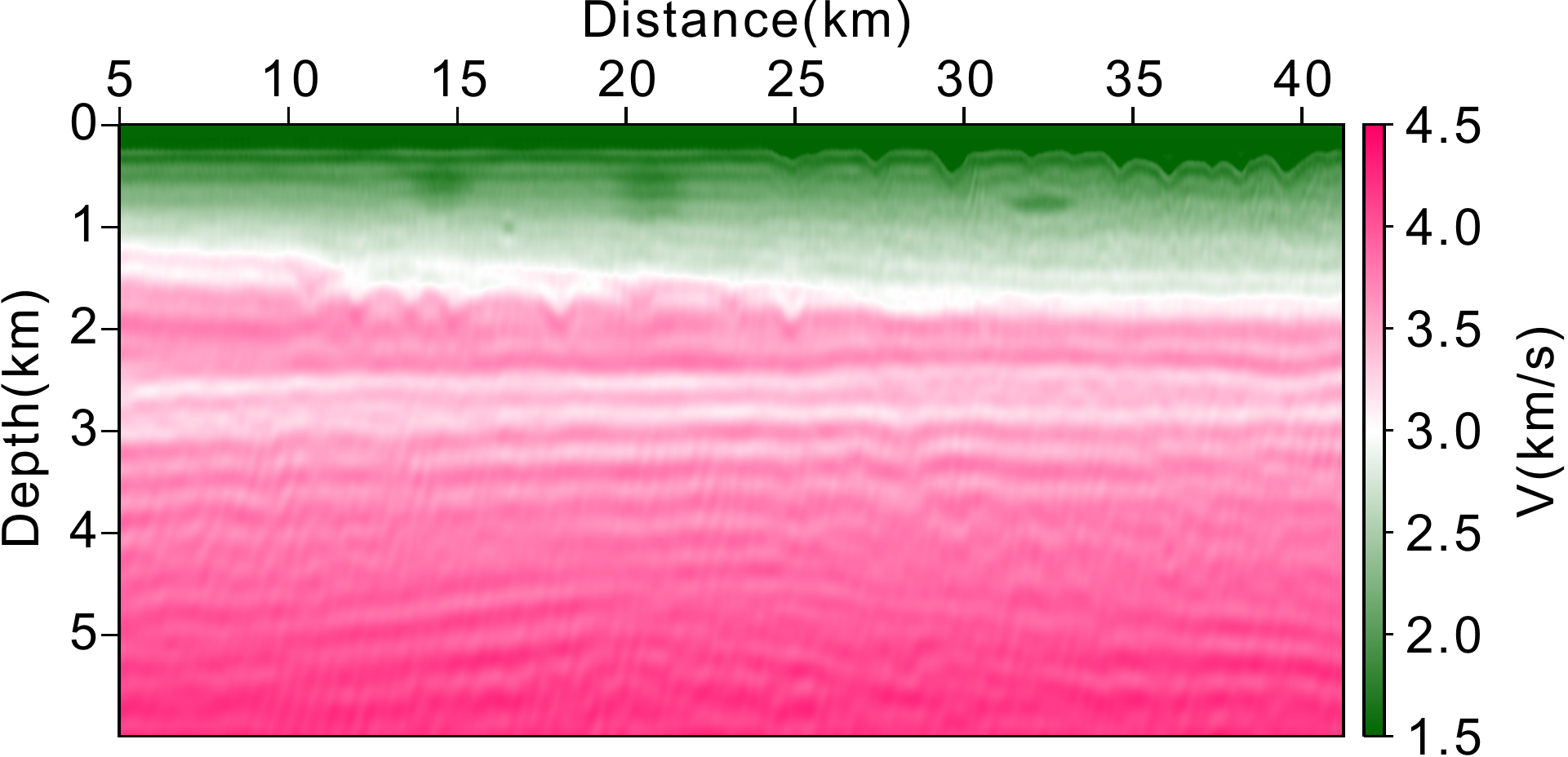}\\
{\small (b)}\\
\includegraphics[scale=0.6]{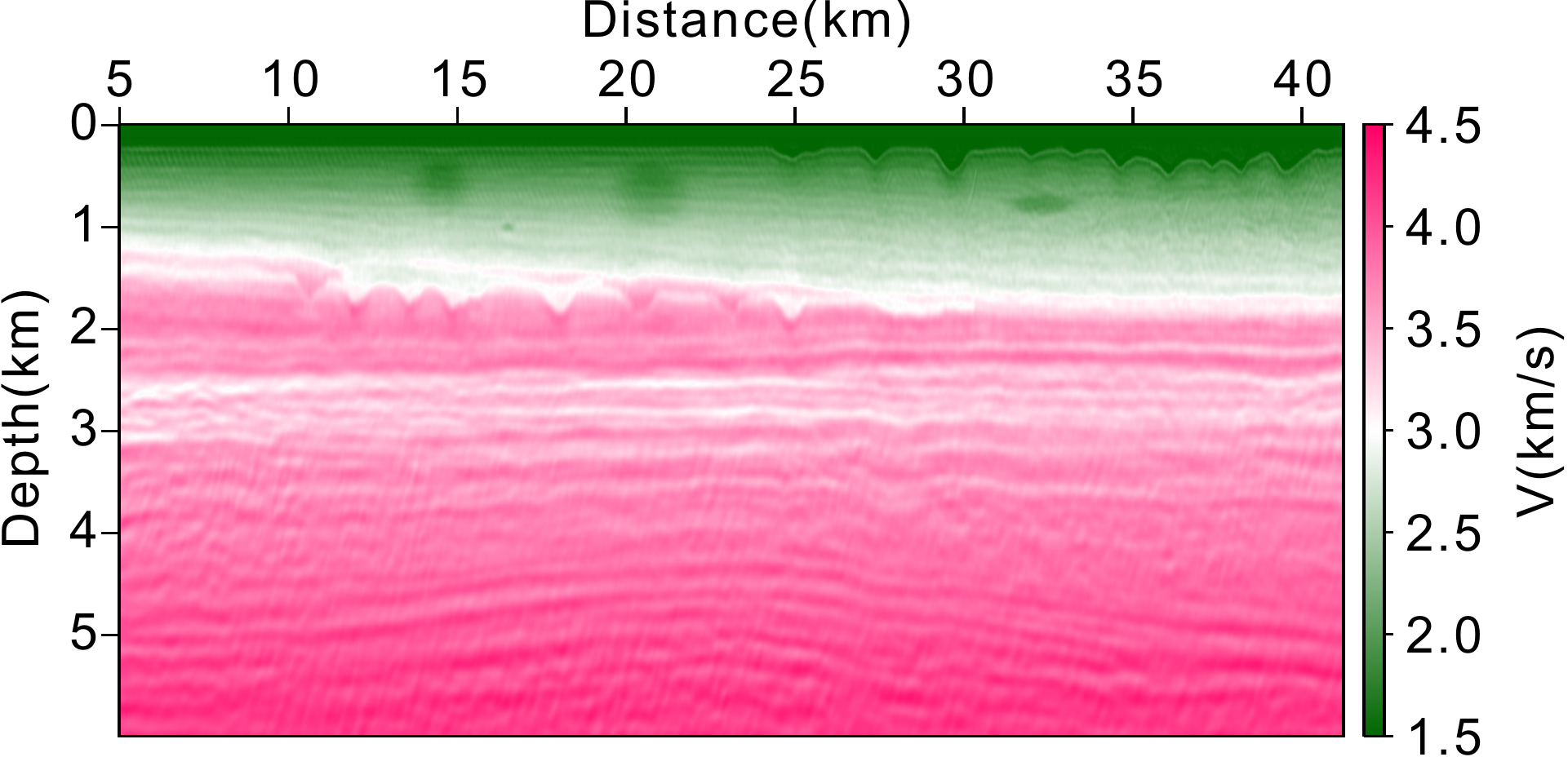}\\
{\small (c)}\\
\caption{The estimated P-wave velocity model at 3 Hz (a), 7 Hz (b) and 15 Hz (c).}
\label{fig:ChevronEMDFWI}
\end{figure}

\begin{figure}
\centering
\includegraphics[scale=0.25]{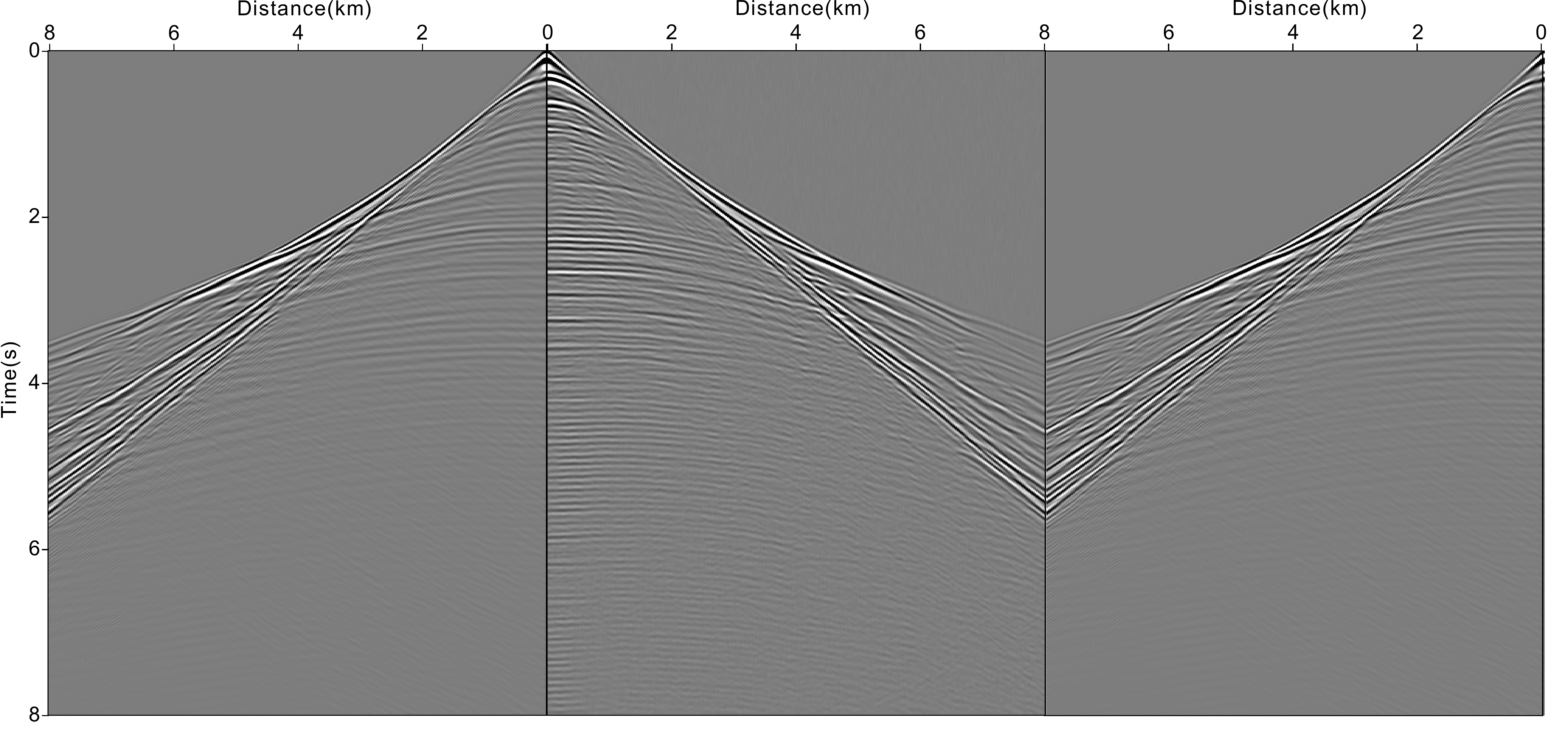}\\
\caption{Comparison of the observed (middle part) and predicted data (left and right part) after inversion at 15 Hz. The predicted data are mirrored and placed on both sides of the real data to better compare the match of reflection and transmission. }
\label{fig:seismogram}
\end{figure}

\begin{figure}
\centering
\includegraphics[scale=0.8]{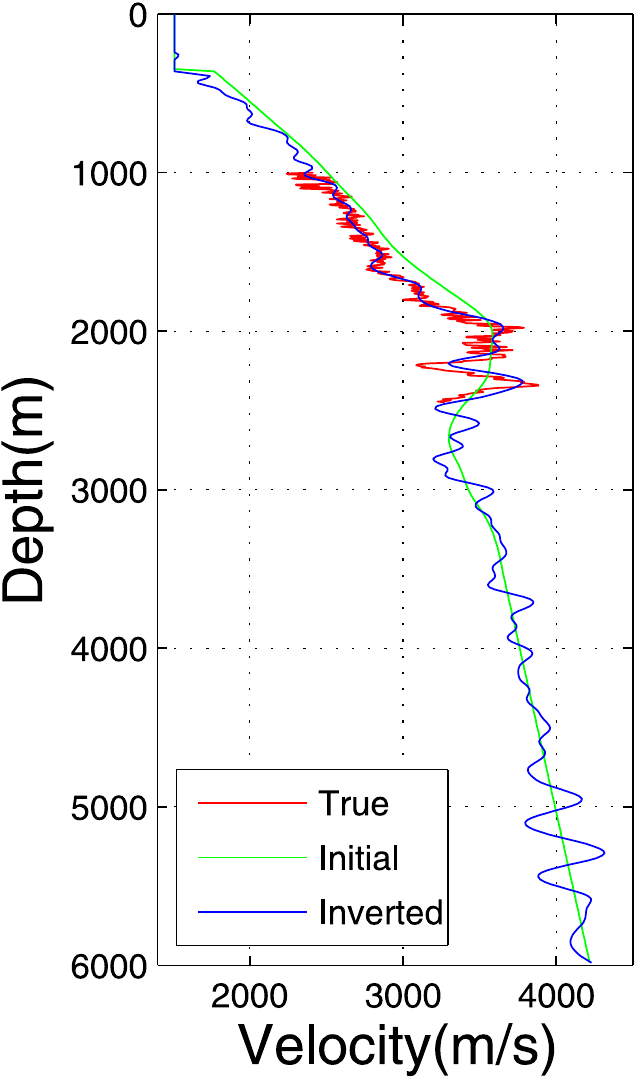}\\
\caption{Comparison the well logs taken at $x=39375$ m of the inverted model (blue line), the  initial model (green line) and given true one (red line).}
\label{fig:welllog}
\end{figure}

Figure \ref{fig:ChevronFWI} shows the inverted result using the classical FWI,  from which barely no  geology information can be obtained. The inverted velocity models using the proposed method at 3 Hz, 7 Hz and 15 Hz are shown in Figure \ref{fig:ChevronEMDFWI} (a-c). The detailed subsurface structures are gradually inverted via the application of more high-frequency components.  We also
compare the observed data of 115th shot with the predicted data from the inverted model at 15 Hz. Figure \ref{fig:seismogram} shows that the predicted data have a similar kinematic features in both diving waves and reflected waves.  Since the observed data contain some converted P-wave in elastic media, which can not be generated by acoustic modeling, it is reasonable that, there are some events in observed data which have not been well matched in phase.  In addition, developing FWI method using elastic modeling is important for a better amplitude match. Figure \ref{fig:welllog} shows the comparison of well logs at 39,375 m.  We can see from Figure \ref{fig:welllog} that, the initial velocity  is far away from the true one and the inverted velocity matches the true one well from the depth of  1000 to 2500 m.

\clearpage

\clearpage

\clearpage

\section{Discussion}

OT related methods and applications have recently begun to be employed in a wide variety of problems related
to signal and image analysis and pattern recognition \citep{villani2008optimal,santambrogio2015optimal}. Mathematically, four formulations of the OT problem mentioned in the theory section are equivalent \citep{kolouri2016transport}. The numerical solutions to these OT problems are different and the corresponding algorithms have different computational complexities \citep{cuturi2013sinkhorn,benamou2014numerical,metivier2016measuring,li2017parallel}. For large-scale seismic data, we have applied primal-dual method to efficiently compute the EMD between two seismograms based on the dynamic formulation of the OT problem. Compared to other methods used in the FWI application, the new method  is very easy to implement and has high computation efficiency on GPU device \citep{li2017parallel}. The applications of FWI using EMD to Marmousi 2 model and Chevron data  have shown that, the extra computation time of the proposed method is about 
11\%  higher than that of the classical FWI method. Note that the computation time of wave propagation is related to the model size of $nz\times{nx}$ and numerical modeling methods, and the computation time of calculating EMD depends on the seismogram size of $nt\times{nr}$.   The percentage of the extra computation time will change with different case studies. Overall, This extra computation time is acceptable for further 2D and 3D large size application of multi-dimensional EMD to real seismic data.    

The strategy proposed by Mainini \citep{mainini2012description} has been applied to satisfy non-negativity and conservation of mass for signed data in this paper. The first numerical example has shown that,  for time-shift signals,  EMD based misfit function increases with the decrement of the overlap of two refactored distributions. Compared to the conventional LSD,  EMD makes the misfit function more convex, therefore, FWI using EMD can reduce the risk of converging towards a local minimum. On the other hand, it is noted that when there is no overlap between two refactored distributions, the misfit will not increase with increment of time-shifts, and the optimal map will transport $P_{cal}^+$ to $P_{cal}^-$ and $P_{obs}^+$ to $P_{obs}^-$. This limitation of using Mainini strategy in seismic inversion has also been observed in the previous studies \citep{metivier2016optimal}. Other strategies like separately transporting  positive and negative part of the signal \citep{engquist2013application}, adding a constant mass \citep{yang2018application} and  exponentially encoding \citep{qiu2017full} can   guarantee the non-negativity and may provide a better capability. However, it requires normalizing the data to satisfy mass balances. Therefore, it is an important future work to design a robust strategy that can  better capture the time shift between two signed signals.

In this paper, multidimensional seismic data are treated as the general imaging data of pixels, similar with the previous EMD of multidimensional  seismic data \citep{metivier2016measuring,metivier2016optimal}. Intuitively, the apparent velocity can be used to build the
connection between space and time axes in the seismogram. In Marmousi case study, if we define the computational area of EMD on the $\Omega=[0,1]\times[0,1]$, the weight factor is $16km/6s=2.67km/s$. The value of the number is close to the realistic P-wave velocity. However,  waves from different propagation directions have different apparent velocities. It is not trivial to find one reasonable number to address all waves. 
Since the interval along the time direction is smaller than that along the space direction, the transport will be mainly in the time direction. In the  our study cases, the dominant structures of the sediment are horizontal layers, velocity variation mainly arises in the vertical direction. Therefore, it is reasonable to put more weight on the time direction in the seismogram. For a better application of EMD to multidimensional seismic data, it is necessary to find systematic ways to distribute weight among different directions.    

High-contrast salt inversion is a  major focus and challenge problem in oil and gas exploration. Although numerical examples have shown that EMD can mitigate local minima to some extend,  the relation between the waveform and the velocity perturbation is strongly nonlinear due to the high velocity contrast between salt and sediment. The previous studies  have indicated that, combining the layer stripping  strategy \citep{wang2009reflection}, Gaussian smooth filter and EMD has a potential to invert salt dome \citep{metivier2016measuring}. However, it requires thousands of iterations, therefore, combining  it with total variation regularization techniques may accelerate automatic salt inversion \citep{esser2018total,yong2018total}. In addition, developing effective strategies to capture the time shift of signed seismic data may help to efficiently invert salt dome inversion \citep{yang2018application}. To further develop this method as a viable alternative to conventional FWI, it is necessary to test it on realistic field data in the future. 

\section{Conclusions}

FWI is a powerful technique that solves the inverse problem as a non-linear data-fitting problem. However, it suffers from several issues such as local minima, due to the lack of  low-frequency component in data and the limited accuracy of the starting model. In this work, we have investigated the application of EMD to measure the misfit for FWI.  Mainini strategy is employed to satisfy the two assumptions of EMD  for signed seismic data.  The computation of the EMD between two compared seismograms is cast as a large-scale convex optimization problem, which has been efficiently solved by a simple-to-code and memory efficient PDHG algorithm with linesearch. Numerical study has indicated that the application of EMD to measure the misfit between two seismograms can effectively mitigate local minima in FWI. Numerical result on the 1D case study has demonstrated that, compared to LSD, EMD used in this paper is more effective in capturing time shifts and makes the objective function more convex.  Hence,  it is helpful to mitigate cycle skipping issues related to the use of the conventional LSD. This is illustrated on a simple transmission from the crosshole experiment, as well as on the Marmousi 2 case study. Starting with a poor initial model, FWI using EMD is able to obtain more reliable estimations of the velocity model. For reflection configuration, the FWI with EMD can produce a higher resolution image of the deep velocity structures. Application to the SEG 2014 blind data set has demonstrated the potential of the proposed method.

\section{Acknowledgments}

The authors greatly appreciate the financial support jointly provided by the National Basic Research Program of China (2014CB239006), the National Gas and Oil project (2016ZX05002-005), the Fundamental Research Funds for the Central Universities (17CX06033). The first author is also grateful for China Scholarship Council to support his visit to the University of Calgary. The work of the second author is supported by NSERC DG grant. We would thank Wenyong Pan from Los Alamos National Laboratory for valuable suggestions and comments on this paper. We also thank Wuchen Li from UCLA and Da Li from the University of Calgary for inspiring discussions on the optimal transport distance.

\section*{References}

\bibliography{mybibfile}

\end{document}